\documentclass[desactivate]{aa}  
\usepackage{natbib}
\usepackage{graphicx}
\usepackage{txfonts}
\usepackage{ulem}
\usepackage{threeparttable}
\usepackage{mathrsfs}
\usepackage{amsmath}
\usepackage{subfig}
\usepackage{float}
\usepackage{comment}
\usepackage{mathtools}
\usepackage{hyperref}
\hypersetup{
    colorlinks=true,
    linkcolor=blue,
    citecolor=blue,
    urlcolor=green,
    }
    \usepackage{placeins}
\usepackage{afterpage}
\usepackage{arydshln}

\begin{document}

\title{Odd radio circle J1555+2726: Observational evidence for shock re-acceleration of fossil plasma }

   \author{L. Bruno 
          \inst{1},
          B. Koribalski
          \inst{2,3}, 
            R. P. Norris
          \inst{2,3}, 
          A. Ivleva
          \inst{4},
          T. Venturi
          \inst{1,5},
          A. Botteon
          \inst{1},  
          C. J. Riseley
          \inst{6,7},
          K. Dolag
          \inst{4,8},
          J. R. Callingham
          \inst{9,10},
          T. W. Shimwell
          \inst{9,11}
          }

   \institute{
    Istituto Nazionale di Astrofisica - Istituto di Radioastronomia, via Gobetti 101, 40129, Bologna, Italy
             \and
     Australia Telescope National Facility, CSIRO, Space and Astronomy, P.O. Box 76, Epping, NSW 1710, Australia   
     \and 
    Western Sydney University, Locked Bag 1797, Penrith South DC, NSW 2751, Australia
    \and
      Universit\"ats-Sternwarte, Fakult\"at f\"ur Physik, Ludwig-Maximilians-Universit\"at M\"unchen, Scheinerstr. 1, 81679, M\"unchen, Germany  
             \and
    Center for Radio Astronomy Techniques and Technologies, Rhodes University, Grahamstown 6140, South Africa
      \and 
    Astronomisches Institut der Ruhr-Universit\"at Bochum (AIRUB), Universit\"atsstraße 150, 44801, Bochum, Germany
    \and 
    Ruhr Astroparticle and Plasma Physics Center, D-44780 Bochum, Germany
    \and
      Max-Planck-Institut f\"ur Astrophysik, Karl-Schwarzschild-Str. 1, 85748, Garching, Germany
    \and 
    Netherlands Institute for Radio Astronomy (ASTRON), Oude Hoogeveensedijk 4, NL-7990 Dwingeloo, the Netherlands
    \and
    Anton Pannekoek Institute for Astronomy, University of Amsterdam, Science Park 904, NL-1098 XH Amsterdam, the Netherlands
    \and 
    Leiden Observatory, Leiden University, PO Box 9513, 2300 RA, Leiden, The Netherlands
             \\
\email{l.bruno@ira.inaf.it}
}


 
  \abstract
   {Odd radio circles (ORCs) are peculiar extragalactic sources recently discovered. They are characterised by one or more bright rings extending on scales of hundreds of kiloparsecs, and are filled with fainter emission. A massive elliptical galaxy lies at the centre of ORCs. While their formation mechanism is debated, a shock is plausibly involved.    }
   {We aim to constrain the nature of ORC J1555+2726 (ORC4) through the analysis of its morphological and  spectral properties.  }
   {We utilised Low Frequency ARray (LOFAR), upgraded Giant Metrewave Radio Telescope (uGMRT), and MeerKAT radio data at 50-1284 MHz to obtain images and spectral index maps at high resolution and sensitivity. } 
   {Both the resolution and observing frequency affect the observed morphology. In particular, the brightness contrast between the ring and the filling emission becomes progressively smoother from higher to lower frequencies. We measured an integrated spectrum of slope $\alpha=1.06\pm 0.03$, a radio power of $P_{\rm 150}= (5.4\pm 0.5)\times 10^{25} \; {\rm W \; Hz^{-1}}$ at 150 MHz, and a radial spectral flattening down to $\alpha\sim0.5$ at the ring location. However, we inferred a negative (convex) spectral curvature and a peculiar spectral shape in radio colour-colour diagrams, which imply spectral flattening towards high frequencies. These properties cannot be explained by shock acceleration of thermal particles and standard ageing. }
   {The observed features suggest that a mixture of electrons with different spectra coexist within ORC4 volume, likely associated with shock re-acceleration of a fossil relativistic reservoir. The ring emission is plausibly dominated by higher-energy electrons re-accelerated at the shock front, while lower-energy re-accelerated electrons are more uniformly distributed across the whole source (possibly ellipsoidal) volume.}


\titlerunning{ORC J1555+2726}
\authorrunning{Bruno et al. 2026}
   \maketitle
%

\section{Introduction}

Odd radio circles (ORCs) are peculiar extragalactic sources defined by a bright ring surrounding a massive central elliptical galaxy at radii of hundreds of kiloparsecs \citep{Norris21,Norris21b,Koribalski21}. The region inside the ring is not empty, but filled with faint radio emission that can be detected with adequate sensitivity. Since the discovery of the first ORCs in 2021, several others have been reported \citep[e.g.][]{Lochner23,Koribalski24b, Koribalski24c,Gupta25,Norris25}, outlining important observational features. First, ORCs can exhibit either one or multiple rings \citep[e.g.][]{riseley24,Hota25}. Second, their integrated spectrum is steep, with spectral indices of $\alpha\sim1-1.5$ \citep[e.g.][]{DeGasperin26}. Third, the ring can be highly polarised by $\sim 20-30\%$, and the magnetic field lines are tangential to it \citep[e.g.][]{norris22,Taziaux25}. Moreover, with the growing number of ORCs, a major caveat is emerging: sensitive and high-resolution data are necessary to distinguish genuine ORCs from sources resembling ORCs. For instance, ORC2 and ORC3 from the discovery paper by \cite{Norris21} have been afterwards associated with diffuse radio components of active galactic nuclei (AGN) emission (\citealt{Rupke24,Norris25}; MacGregor et al., in preparation), and have to be excluded from the population of ORCs with characteristic edge-brightened rings.

So far, multi-wavelength observations have not reported detections of possible ORC counterparts. However, optical and X-ray data provided information on the galaxies and environments associated with ORCs. The central elliptical galaxies have high stellar masses ($M_{*}\sim 10^{11} \; M_{\odot}$) and old ($\sim1$ Gyr) stellar populations \citep[e.g.][]{Coil24,Rupke24,Norris25}. They seem to favour relatively dense environments, such as small groups, and can have close companions \citep[e.g.][]{Norris21b,Bulbul24,Rupke24}.

Several scenarios have been proposed to explain the nature of ORCs. Current data suggest that the ring traces particle acceleration from a shock, but the triggering event and acceleration mechanism are debated. For instance, shocks may arise from a merger of supermassive black holes \citep{norris22} or galaxies \citep{Dolag23,Ivleva26}, accretion at the virial radius \citep{Yamasaki24}, intense starburst outflows \citep{Coil24}, and strong tidal stellar disruption around the central black hole \citep{Omar22}. Possibly, shock re-acceleration of fossil relativistic plasma, rather than acceleration of fresh particles, may be involved \citep{Koribalski24b,Koribalski24c,Shabala24,Wang&Heinz26}.

In this work we present new deep multi-frequency radio observations of ORC J1555+2726, best known as ORC4 \citep{Norris21}. We investigate the origin of ORC4 through Low Frequency ARray (LOFAR), upgraded Giant Metrewave Radio Telescope (uGMRT), and MeerKAT data covering the 50-1284 MHz frequency range, and auxiliary images from the Rapid ASKAP Continuum Survey (RACS; \citealt{Mcconnell20racs,hale21racs,Duchesne25racs}) and the K. Jansky Very Large Array Sky Survey (VLASS; \citealt{lacyVLASS20}). Throughout this paper, we adopted a standard $\Lambda$CDM cosmology with $H_0=70\;\mathrm{km\; s^{-1}\; Mpc^{-1}}$, $\Omega_{\rm M}=0.3$, and $\Omega_{\rm \Lambda}=0.7$. At the redshift of $z=0.4512$ for ORC4 \citep{Coil24}, $1''=5.768$ kpc and the luminosity distance is $D_{\rm L}=2.51$ Gpc. We adopted the convention on the spectral index $\alpha$ as defined from the flux density $S_\nu \propto \nu^{-\alpha}$. 

This paper is organised as follows. In Sect. \ref{sect: ORC4 info}, we summarise the literature on ORC4. In Sect. \ref{sect: Observations and data reduction}, we present the radio observations. In Sect. \ref{sect: Results}, we report images and results of the spectral analysis. In Sect. \ref{sect: Discussion} we discuss the formation of the target based on our results. In Sect. \ref{sect: Summary and conclusions} we summarise our work and discuss prospects to test formation scenarios.

\section{ORC4}  
\label{sect: ORC4 info}

\begin{figure}
        \centering
        \includegraphics[width=0.4\textwidth]{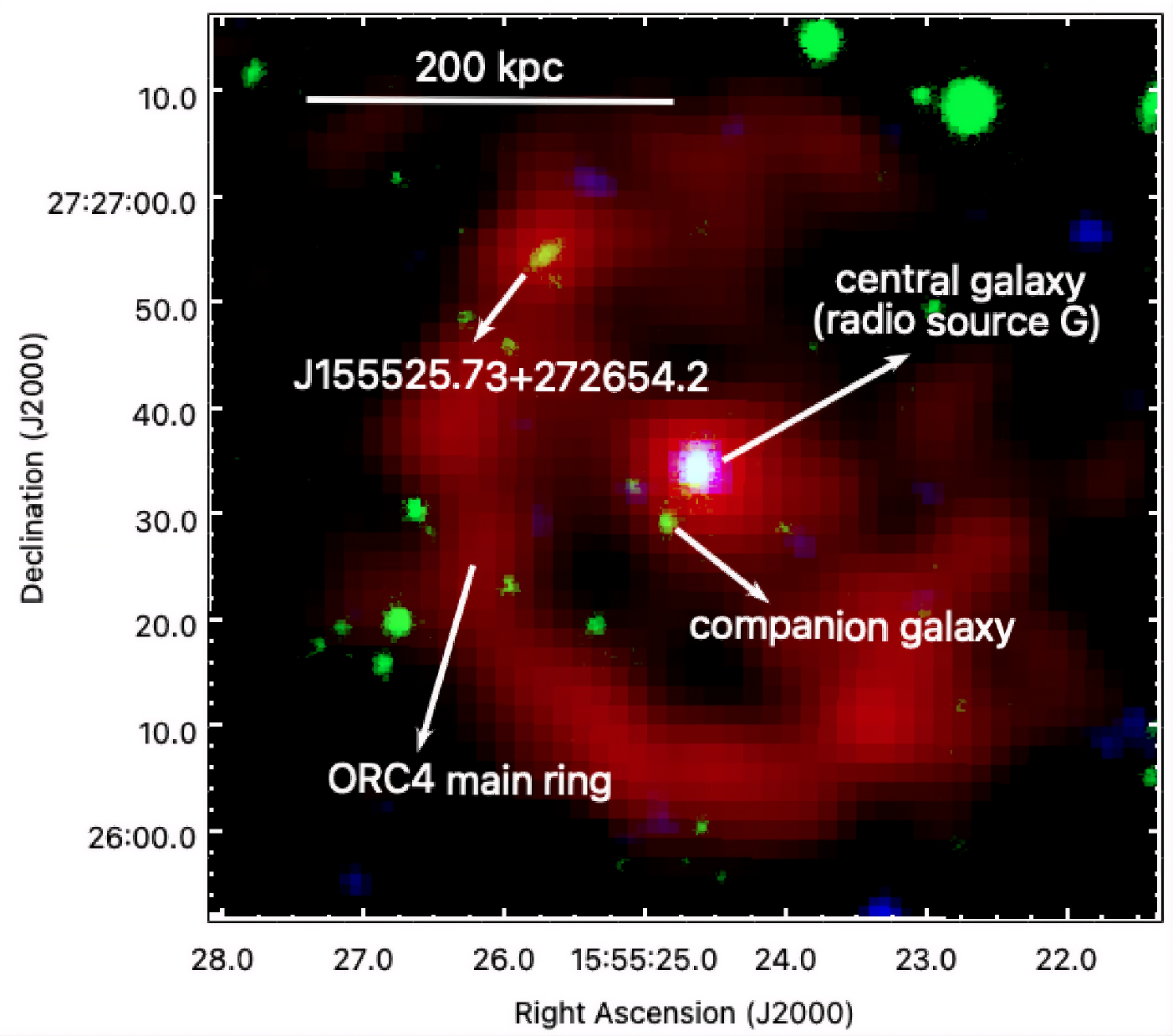}
        \smallskip
        
        \caption{Optical and radio overlay of ORC4 field. Green, red, and blue coloured images are DESI-DR10 (i, r, and g filters), GMRT (323 MHz), and VLASS (2988 MHz), respectively. }
        \label{fig: radio-ottico}
\end{figure}

ORC4 (${\rm RA}=15^{\rm h}55^{\rm m}24.6^{\rm s}$, ${\rm Dec}=  27^{\rm o}26'34.4''$) is one of the targets reported in the discovery paper by \cite{Norris21}. It was serendipitously discovered by H. Intema by visual inspection of GMRT images at 323 MHz of the nearby ($z\sim 0.09$) galaxy cluster Abell 2142 (A2142; ${\rm RA}=15^{\rm h}58^{\rm m}20^{\rm s}$, ${\rm Dec}=  27^{\rm o}14'00''$), which were first presented in \cite{venturi17}. ORC4 is located at $\sim40'$ west of A2142 and is not physically associated with this cluster. A radio-optical overview of the field of ORC4, which includes images from GMRT, VLASS, and the Dark Energy Spectroscopic Instrument Legacy Surveys \citep{Dey19_DESI} Data Release 10 (DESI-DR10), is shown in Fig.~\ref{fig: radio-ottico}. This region is not identified as overdense \citep{Norris21b}.  

ORC4 consists of a compact radio source at its centre (hereafter source G), which is surrounded by a bright circular ring; fainter radio emission fills the space between source G and the ring \citep{Norris21,Norris21b}. Deep MeerKAT images at 1284 MHz revealed a faint secondary ring, which lies slightly offset to the east of the main ring \citep{riseley24}. Using images from LOFAR surveys \citep{shimwell22LOTSSDR2,degasperin23,Shimwell26LOTSS}, a steep integrated spectral index of $\alpha=1.3\pm0.2$ was measured by \cite{DeGasperin26} between 54 and 144 MHz, suggesting the presence of aged electrons. 

The radio emission of source G arises from a low-power ($P_{1.4}\sim 3\times 10^{23} \; {\rm W \; Hz^{-1}}$ at 1.4 GHz) radio-quiet AGN \citep{Rupke24}. Its host is WISEA J155524.65+272633.7 \citep{Norris21,Norris21b}, a massive ($M_{*}\sim 1.9\times10^{11} \; M_{\odot}$) elliptical galaxy at high spectroscopic redshift ($z=0.4512$; \citealt{Coil24}). At such redshift, the ring has a radius of $\sim 200$ kpc. This galaxy shows no evidence of recent star formation, with the last burst estimated to be occurred $\sim 1$ Gyr ago \citep{Coil24}. At a projected distance of $\sim 35$ kpc, a less massive ($M_{*}\sim 4\times10^{10} \; M_{\odot}$) elliptical galaxy may form a close interacting system with WISEA J155524.65+272633.7 \citep{Norris21b,Rupke24}. Furthermore, \cite{Coil24} detected a high concentration of [O II] extending on a scale of $\sim 40$ kpc, which is uncommon for early-type galaxies and could trace gas from the host that has been ionised by a shock.

\section{Observations and data reduction}
\label{sect: Observations and data reduction}

\begin{table*}
 \centering
 \caption[]{Details of the radio data analysed in this work.}
 \label{table: datiRADIO}
\resizebox{\textwidth}{!}{
 \begin{tabular}{ccccccc}
\hline
        \noalign{\smallskip}
        Instrument & Band &  Frequency &  Observation date & On-source time & Project code & PI \\
        &  &  (MHz) & & (h) & & \\
        \noalign{\smallskip}
        \hline
        \noalign{\smallskip}
         LOFAR  & LBA  & 30-78 & 08,17,23-Dec.-2021 & 16.0  &  LC17\_012 & L. Bruno \\
         LOFAR & HBA & 120-168 & 15-Sep.-2018; 25,31-Oct.-2020; 13-Nov.-2020 & 32.0  &  LT10\_010; LC14\_018 & F. Vazza \\
    uGMRT  &  3 & 300-500 & 15-Mar.-2018 & 3.0  &  33\_052  & T. Venturi \\
  uGMRT &  4  & 550-950 & 21-Jun.-2019 & 2.0 & 36\_004 & H. Intema  \\
    uGMRT &  5  & 1060-1460 & 2,8,21-Mar.-2026 & 13.5 & 49\_011 & B. Koribalski  \\
    MeerKAT & L  & 872-1712 & 10-Oct.-2021; 12-Nov.-2021  & 5.5 & SCI-20210212-CR-01 & C. Riseley  \\

        \noalign{\smallskip}
        \hline
 
 \end{tabular}
}
\end{table*}

In this section we describe the radio data utilised in this work, which include observations pointed either on A2142 or ORC4. A summary of all observations is reported in Table \ref{table: datiRADIO}. For details on the observations of A2142 and the corresponding data processing we refer to \cite{bruno23b,riseley24}.

\subsection{Observations of A2142}

The field of A2142 was observed with LOFAR for 16 and 32 h with the Low Band Antenna (LBA) and High Band Antenna (HBA) arrays operating at 30-70 MHz and 120-168 MHz, respectively. These data were presented in \cite{bruno23b} and used to characterise the radio sources of the cluster. In the present work, we used the direction-dependent (DD) calibrated visibilities of the full field of view and carried out the extraction and self-calibration procedure \citep{vanweeren21} towards the region of ORC4. Specifically, sources outside areas of $21'\times 21'$ for LBA and $18'\times 18'$ for HBA in the direction of ORC4 were subtracted out from the visibilities, and cycles of direction-independent (DI) self-calibration of phase and amplitude were performed, improving the image fidelity towards the target. Recently, \cite{DeGasperin26} have presented images of ORC4 obtained from extracted and  self-calibrated pointings of 8 h of LOFAR surveys \citep{shimwell22LOTSSDR2,degasperin23,Shimwell26LOTSS}. Our deeper data provided higher image quality for HBA, while we did not reach significant improvements for LBA.

A2142 was also observed with the uGMRT at 300-500 MHz (band 3) for 3 h. These data were presented in \cite{bruno23b}. In the present work, we re-imaged the DI-calibrated data with {\tt WSClean} \citep{offringa14,offringa17} v. 3.10. We note that the primary beam model currently available within {\tt WSClean} (via the {\tt EveryBeam} library) for uGMRT becomes progressively less accurate at large radii ($\gtrsim 30'$) from the pointing centre, leading to unreliable primary beam correction factors towards the region of ORC4. Therefore, we corrected band 3 images for the primary beam attenuation through {\tt Finalflash}\footnote{\url{https://github.com/arpan-52/Finalflash}} v. 0.3.2, which provides more accurate models of the uGMRT primary beam. Despite this, residual inaccuracy of the beam models are still present at large radii (A. Pal, private communication). This is consistent with flux densities at 407 MHz biased low by $\sim 20-30\%$, as we found by measuring the multi-frequency spectra of compact sources around ORC4. To overcome this issue, we will apply a further correction factor of 1.25 for the 407 MHz flux densities, as empirically derived on average from the compact sources.

Finally, A2142 was observed with MeerKAT at 872-1712 MHz (L-band) for 5.5 h. These data were presented by \cite{riseley24} for the analysis of the cluster, and they additionally reported the first high-quality image of ORC4 at gigahertz frequencies. In the present work, we used a publicly available image\footnote{\url{https://cdsarc.cds.unistra.fr/viz-bin/cat/J/A+A/686/A44\#}} of the whole field at a resolution of $10''\times10''$.

\subsection{Observations of ORC4}
\label{sect: Data from pointed observations of ORC4}

New observations of ORC4 were acquired during three sessions in 2026 with the uGMRT at 1060-1460 MHz (band 5) for a total of 13.5 h. The total bandwidth consists of 4096 channels, each of $\sim98$ kHz width. The sources 3C286 and 1609+266 were used as absolute flux density and gain calibrators, respectively. We performed a standard data reduction with {\tt CASA} \citep{CASAteam22} v. 6.7, which includes excision of Radio Frequency Interference (RFI), and calibration of bandpass, amplitude, and phase for each observing run. We then performed rounds of phase and amplitude DI self-calibration, by producing model images with {\tt WSClean}. Finally, the visibilities of each session were concatenated into a single dataset for deeper imaging.

The target was also observed in 2019 with the uGMRT at 550-950 MHz (band 4) for 2 h. These archival observations have not been published thus far. The data were acquired with 2048 channels, each of $\sim$195 kHz width. We split the data into eight sub-bands (50 MHz each) to process each of them independently through {\tt SPAM} \citep{intema09}. The source 3C286 was used to derive DI gain solutions, which were transferred to the target. Afterwards, rounds of DD phase calibration were carried out to correct for ionospheric errors. To improve the DD calibration, we produced an accurate catalogue of the sources in the field, which was obtained with {\tt pyBDSF} \citep{Mohan&Rafferty15} based on the high-resolution ($\sim 3''$) images from the band 5 data. This catalogue provided {\tt SPAM} with solid initial phase solutions across the field, guaranteeing accuracy and stability for subsequent calibration cycles and consistency among the different sub-bands. After processing, the sub-bands were recombined for imaging with {\tt WSClean}.

\section{Results}
\label{sect: Results}

In this section we present the results of our analysis. These include radio images (Sect. \ref{sect: Radio morphology}), surface brightness profiles (Sect. \ref{sect: Surface brightness profiles}), integrated spectrum (Sect. \ref{sect: integrated spectrum}), and spatial distribution of the spectral properties (Sect. \ref{sect: Spatial spectral distribution}).

\subsection{Radio morphology}
\label{sect: Radio morphology}

\begin{figure*}
        \centering
        \includegraphics[width=0.40\textwidth]{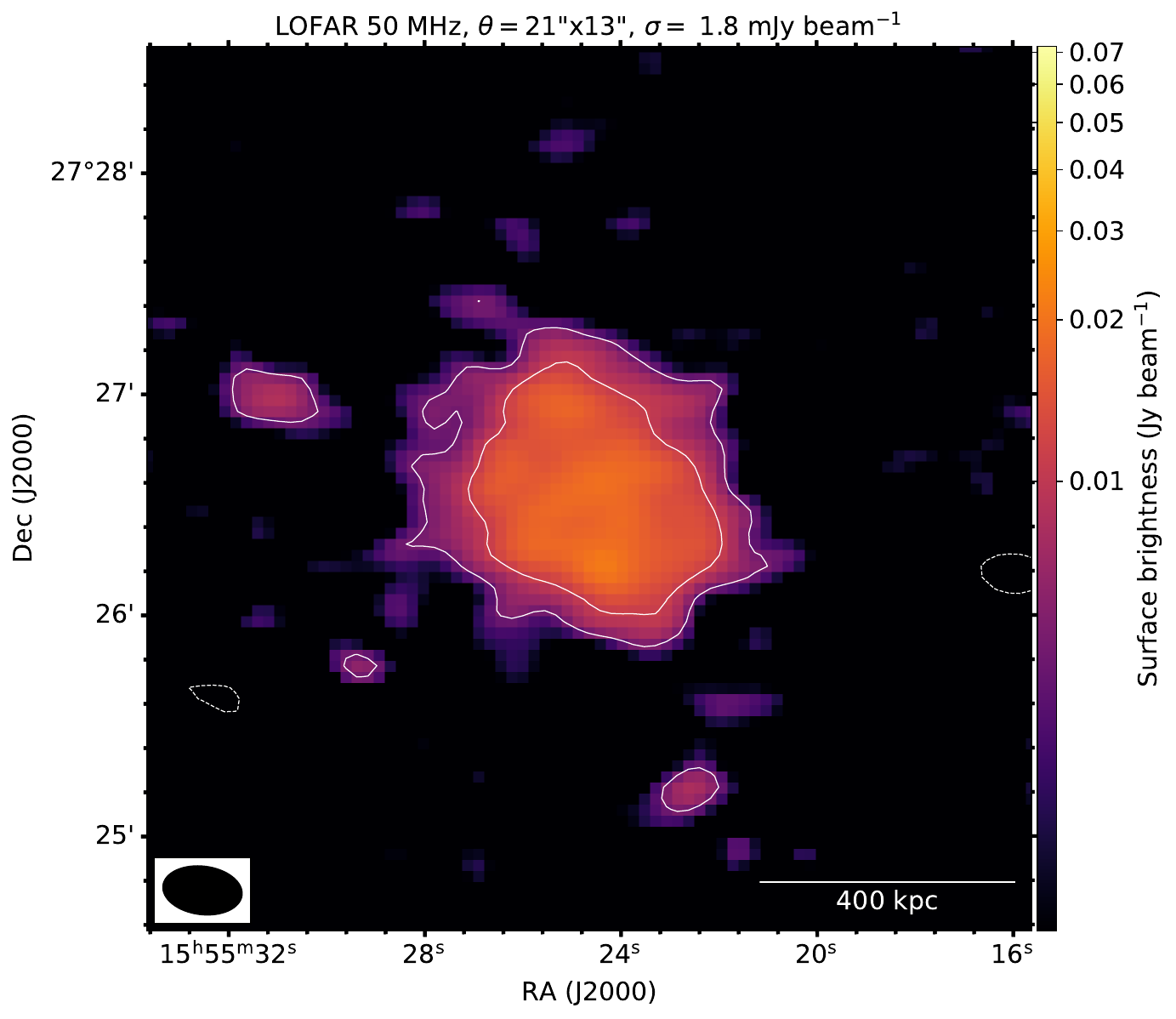}
        \includegraphics[width=0.40\textwidth]{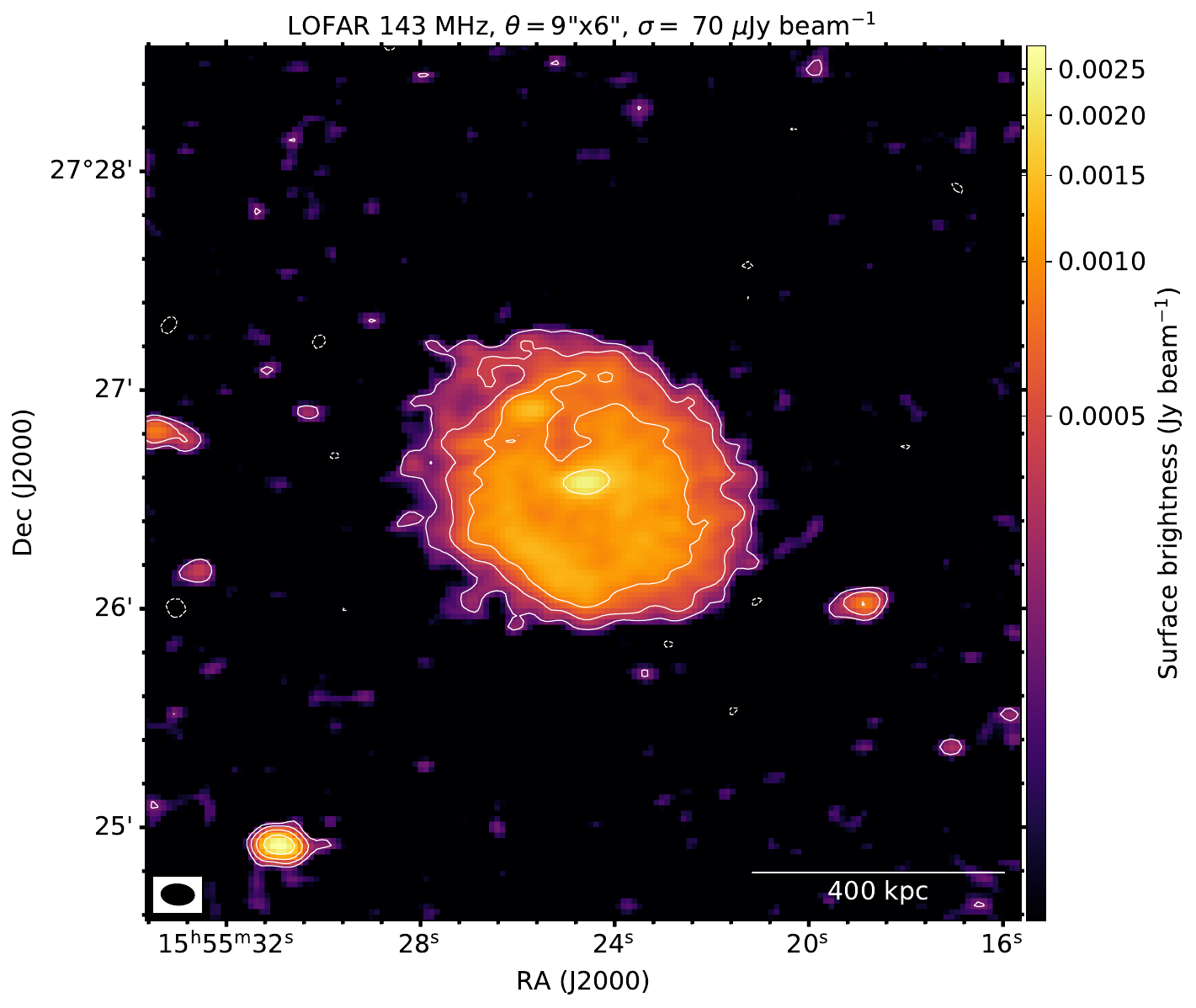}
        \includegraphics[width=0.40\textwidth]{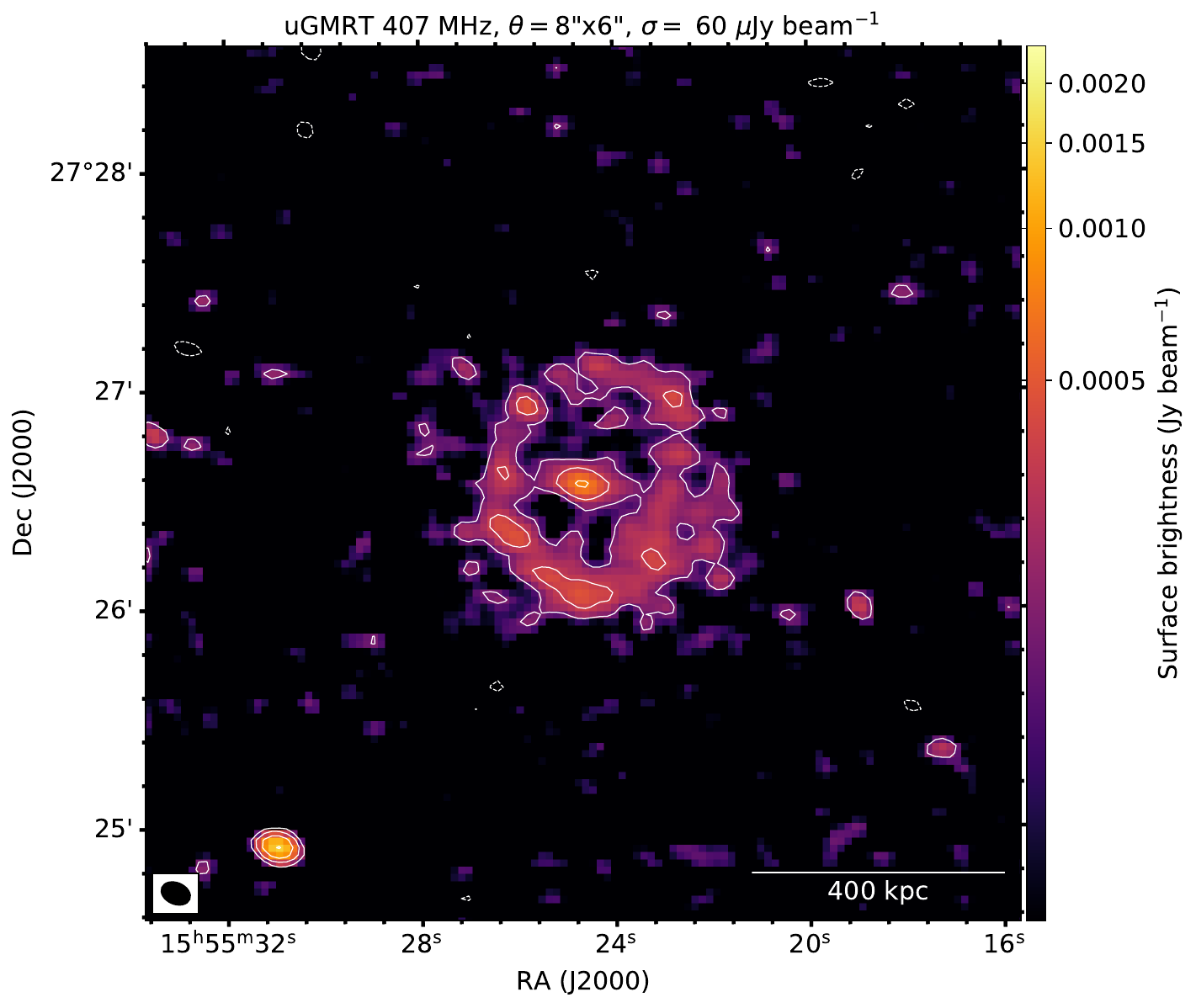}
        \includegraphics[width=0.40\textwidth]{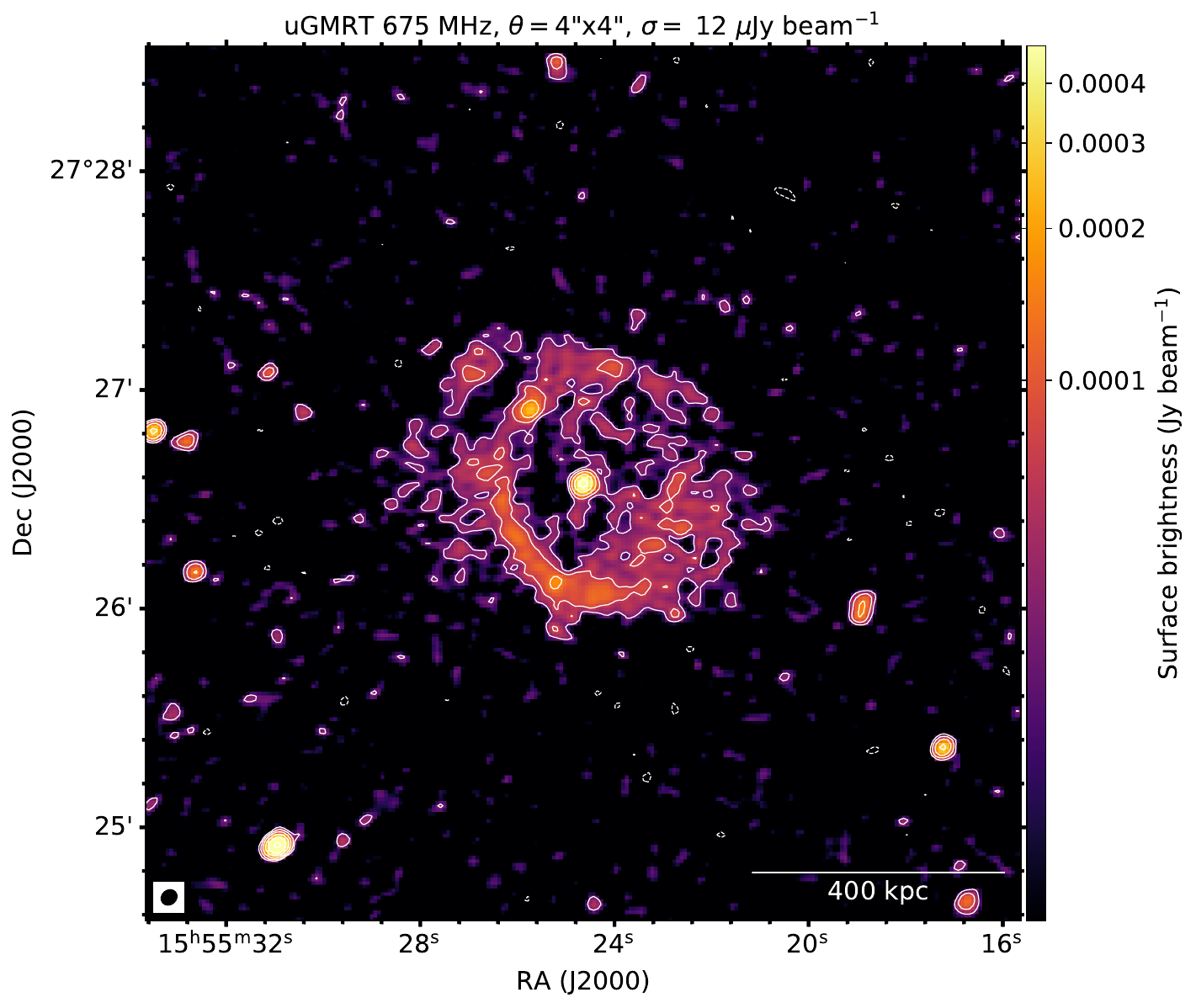}
        \includegraphics[width=0.40\textwidth]{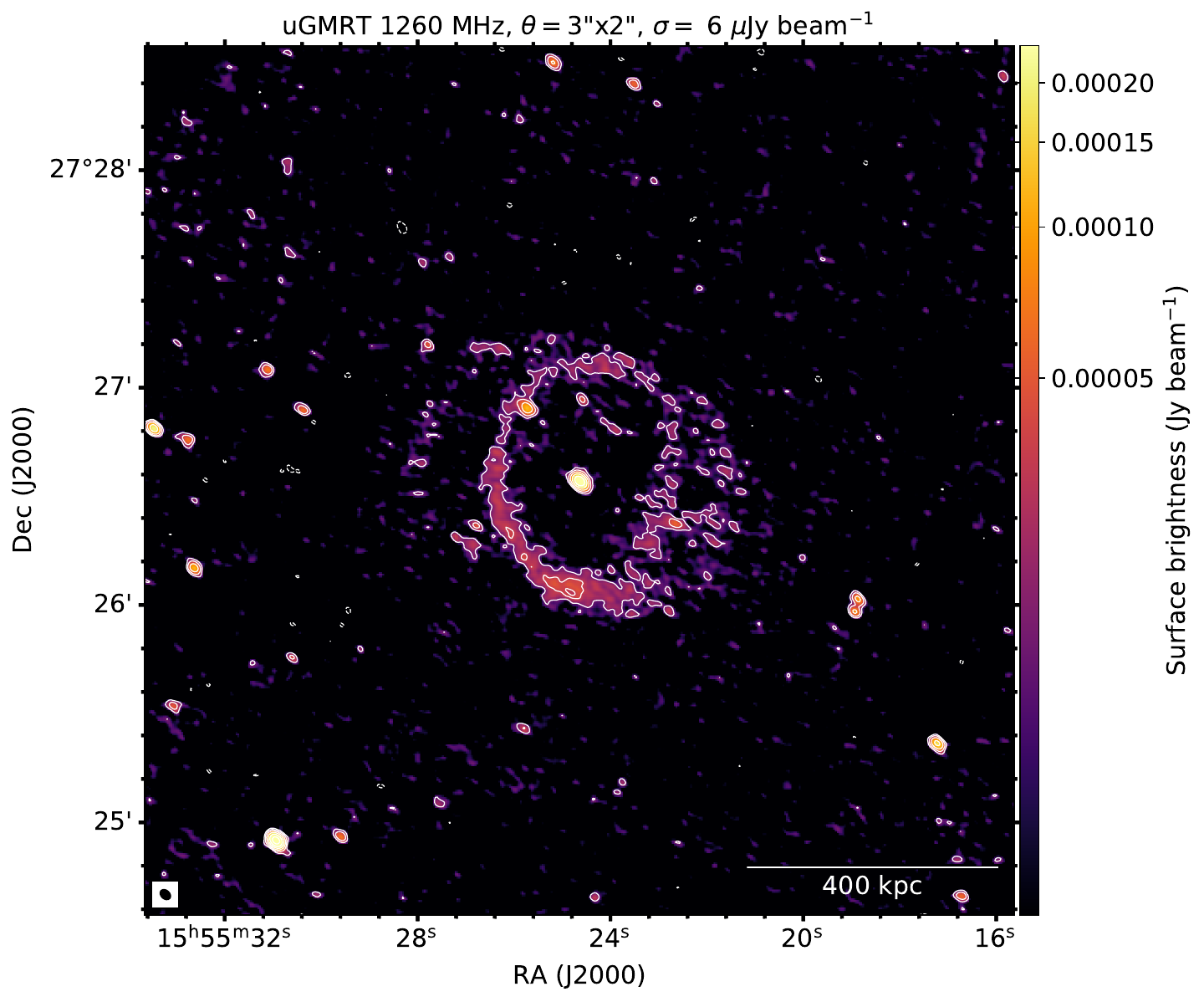}
        \includegraphics[width=0.40\textwidth]{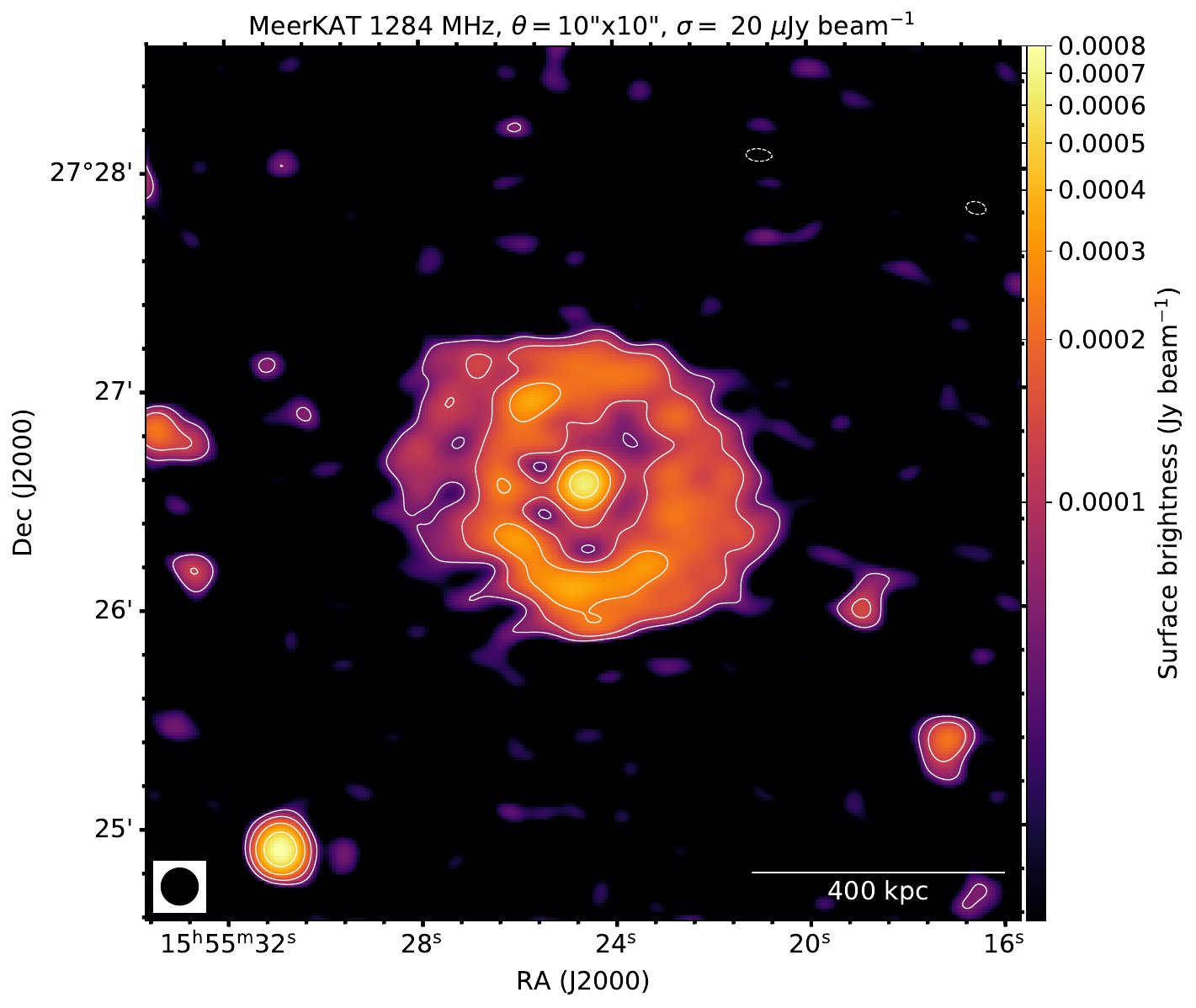}
        \smallskip
        
        \caption{Images of ORC4. From top left to bottom right: LOFAR at 50 MHz, LOFAR at 143 MHz, uGMRT at 407 MHz, uGMRT at 675 MHz, uGMRT at 1260 MHz, and MeerKAT at 1284 MHz. The resolution and noise are reported on top of each panel. The contour levels are $[\pm3, \;6, \;12, \; ...]\times \sigma$.  }
        \label{fig: imgfullres}
\end{figure*}

We present the radio images of ORC4 in Fig. \ref{fig: imgfullres}. The displayed images were obtained by adopting a \cite{briggs95} weighting scheme with robust -0.5 for LOFAR, 0 for uGMRT band 3 and band 4, and 0.5 for uGMRT band 5. Auxiliary radio images from our data and public surveys that we will use in our analysis are shown in Appendix \ref{app: img low res}. 

Source G is detected in all of our images, except at 50 MHz, where it is too faint compared to the surrounding emission. It is fairly compact above 675 MHz, while at lower frequencies it appears extended, but there is likely contamination from the ORC emission. Moreover, high-frequency images at  high resolution suggest the presence of extended components on small scales, which however remain unresolved by current data. We refer to  Appendix \ref{app: Central AGN spectrum} for a more detailed discussion on source G.

The 1260 MHz image reveals that the main ring is not completely symmetric, as its western region is patchier and more extended than the eastern region. Consistently, the LOFAR and MeerKAT images show a brighter component, identified with the main ring, surrounded by a fainter and slightly larger structure likely associated with the secondary ring. Interestingly, the inner region enclosed by the ring is filled with diffuse emission in the low-resolution images, although with a frequency-dependent appearance. At 50 and 143 MHz the emission displays no significant brightness gradients, while at 1284 MHz it is much fainter than the ring. The morphological behaviour observed with MeerKAT at high frequency is confirmed by uGMRT images at 675 and 1260 MHz at lower resolution ($\sim 10''$; see also Sect. \ref{sect: Surface brightness profiles} and Fig. \ref{fig: img lowres}), obtained by tapering the baselines to increase the sensitivity to the diffuse emission.

\subsection{Surface brightness profiles}
\label{sect: Surface brightness profiles}
 
\begin{figure*}
        \centering
        \includegraphics[width=0.4\textwidth]{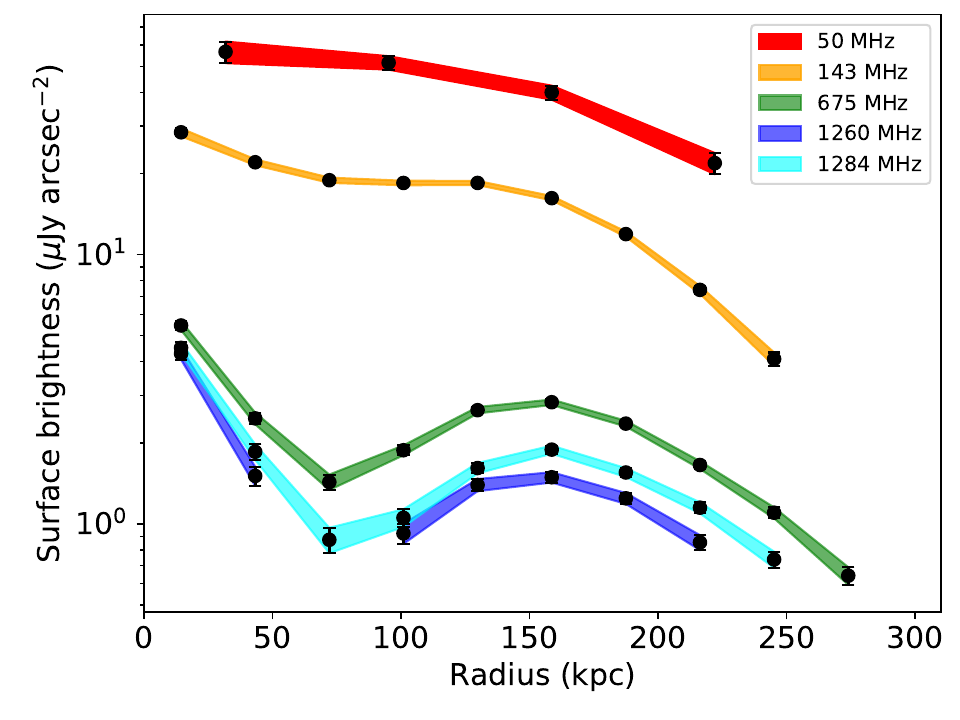}
        \includegraphics[width=0.33\textwidth]{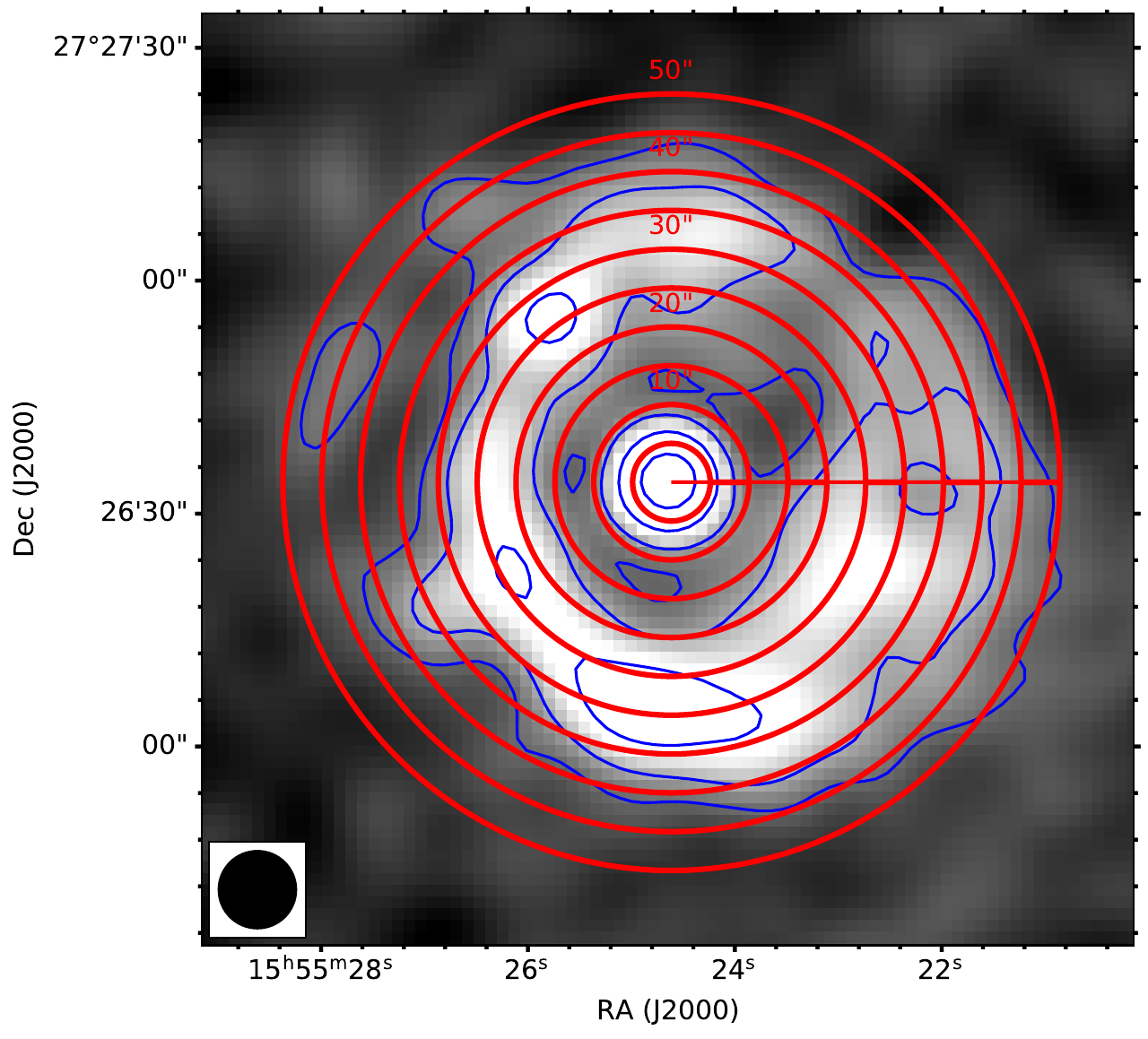} 
        \smallskip
        
        \caption{Azimuthally-averaged surface brightness profiles. The used images are shown in Fig. \ref{fig: imgfullres} (50 and 1284 MHz) and in Fig. \ref{fig: img lowres} (143, 675, and 1260 MHz). An example of the sampling regions adopted in our analysis is shown in the right panel, overlaid to the 1260 MHz image ($\theta=10''$).
        The sampling annuli have widths equal to half of the beam FWHM. Data points with surface brightness $<3\sigma$ were excluded. }
        \label{fig: sb}
\end{figure*}

In Sect. \ref{sect: Radio morphology} we mentioned that the morphology of ORC4 may be dependent on both the resolution and the observing frequency. To further investigate such dependency, we used the azimuthally-averaged surface brightness profiles reported in Fig.~\ref{fig: sb}. These were measured from the 50 and 1284 MHz images in Fig. \ref{fig: imgfullres} and from the 143, 675, and 1260 MHz images at $10''$ resolution shown in Fig. \ref{fig: img lowres}, sampled with annuli of width equal to half of the beam FWHM. 

The surface brightness profiles exhibit different shapes. At $21''$ resolution, the brightness at 50 MHz decreases monotonically following a smooth trend. All the profiles show a radial decline within the inner $\sim 75$ kpc, which is more abrupt with the increasing frequency. At radii of $\sim 75-175$ kpc, in the region of the main ring, the brightness rapidly increases at 675, 1260, and 1284 MHz, while it remains approximately constant at 143 MHz. Beyond radii of $\sim 175$ kpc, the brightness decreases at all frequencies. Although the shape of profiles at 1260 and 1284 MHz is well consistent, MeerKAT provides slightly higher values (by $\sim10-20\%$, see also Table \ref{tab: flussiperspettri}); this may be driven either by an higher sensitivity to the diffuse emission due to its denser \textit{uv}-coverage at short spacings, or to less accurate amplitude solutions and/or primary beam correction far from the pointing centre.

The analysis of the 143, 675, 1260, and 1284 MHz profiles at the same resolution confirms that the morphology of ORC4 is frequency-dependent. The main ring, which is the defining morphological feature of ORCs, is clear at high frequencies, but it becomes less defined at low frequencies; indeed, the emission of the ring and that embedded in it have a comparable brightness at 143 MHz. At 50 MHz it is not trivial to quantify the relative impact of the low observing frequency and poor resolution in smoothing brightness gradients; however, as shown by the RACS image at 888 MHz ($\theta=25''\times25''$) in Fig. \ref{fig: img lowres}, insufficient resolution can wash out the ring structure, making it indistinguishable from the diffuse emission even at high frequency.

\subsection{Integrated spectrum and radio power}
\label{sect: integrated spectrum}

\begin{table}
\centering
                \caption{Flux density measurements of ORC4.  }
    \label{tab: flussiperspettri} 
    \resizebox{0.48\textwidth}{!}{
    \begin{tabular}{ccccccc}
        \hline
        \noalign{\smallskip}
      $S_{50}$  & $S_{143}$ & $S_{407}$ & $S_{675}$ & $S_{888}$ & $S_{1260}$ & $S_{1284}$ \\  
       (mJy)  & (mJy) & (mJy) & (mJy) & (mJy) & (mJy) & (mJy) \\  
        \noalign{\smallskip}
        \hline
        \noalign{\smallskip}

$232  \pm  26$ & $74  \pm  7$ & $26  \pm  2$ & $13  \pm  1$ & $9  \pm  1$ & $7.6  \pm  0.4$ & $8.6  \pm  0.4$ \\

        \noalign{\smallskip}
        \hline
        \end{tabular}
        }
         \begin{tablenotes}
                \centering
       \item   {\small \textbf{Notes}. The reported values were measured within a circular annulus of radii $10''$ and $1'$ centred on source G.     } 
       \end{tablenotes}
   \end{table}

\begin{figure}
        \centering
        \includegraphics[width=0.4\textwidth]{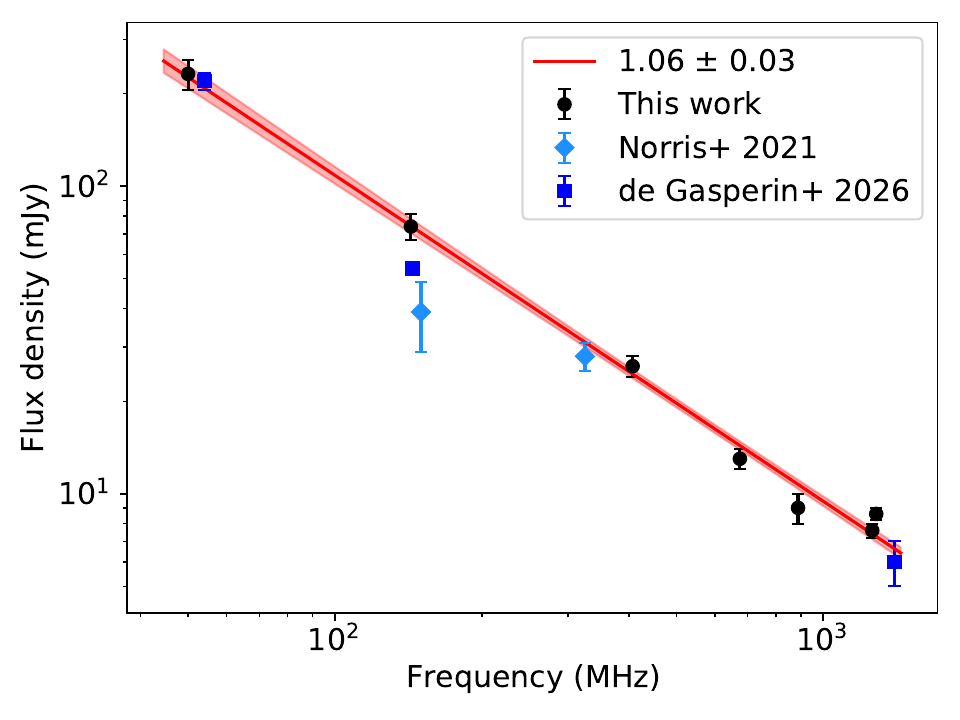}
        \smallskip
        
        \caption{Integrated spectrum of ORC4. The black data points (Table \ref{tab: flussiperspettri}) were fitted with a power law through bootstrap of the flux densities within their uncertainty. The red line and shaded region represent the mean and standard deviation of 1000 fit realisations. Blue and cyan data points are taken from the literature. }
        \label{fig: spettro}
\end{figure}

To derive the integrated spectrum of ORC4 we produced images with a similar resolution of $\sim 10''-15''$, by adopting adequate tapering and weighting schemes. We measured the flux densities within a circular annulus centred on source G of inner and outer radii of $10''$ and $1'$, thus filtering out contamination from the AGN and its surrounding in a circle of diameter $\sim 115$ kpc. The compact radio source projected on the ring at NE\footnote{This lies $25.2''$ ($\sim 145$ kpc at ORC4 redshift) NE to source~G (Fig. \ref{fig: radio-ottico}). It is hosted by the galaxy J155525.73+272654.2 at photometric redshift $z=0.57$ \citep{Alam15}.} was masked, although its contribution to the flux density of ORC4 is small. Errors on the flux densities were computed as 
\begin{equation}
\Delta S= \sqrt{ \left( \sigma \cdot \sqrt{N_{\rm b}} \right)^2 + \left(  \xi \cdot S \right) ^2}\; \; \;,
\label{eq: erroronflux}
\end{equation}
where $N_{\rm b}$ is the number of beams within the considered region and $\xi$ is the absolute flux scale uncertainty, for which we assumed typical values of $\xi=10\%$ for LOFAR \citep{shimwell22LOTSSDR2,degasperin23}, $\xi=5\%$ for uGMRT \citep{lal20a,lal22} and MeerKAT \citep{riseley24}, and $\xi=7\%$ for RACS \citep{Mcconnell20racs}. The obtained flux densities are listed in Table \ref{tab: flussiperspettri} and shown in Fig. \ref{fig: spettro}.

Following a bootstrap approach, the value of each flux density measurement was randomly extracted 1000 times from a Gaussian distribution within its uncertainty, and the data were then fitted with a power law. The integrated spectrum is shown in Fig. \ref{fig: spettro}. The obtained spectral index is $\alpha=1.06\pm0.03$, where the quoted value and error represent the mean and standard deviation, respectively, of the slope distribution of the 1000 realisations.

The inferred spectral index is flatter than previously reported by \cite{DeGasperin26}. This is driven by a discrepancy on flux density measurements at $\sim150$ MHz, which are over-plotted in Fig. \ref{fig: spettro}. Both $S_{144}=54\pm1$ mJy\footnote{The reported error does not include the flux scale uncertainty.} by \cite{DeGasperin26} and $S_{150}=39\pm10$ mJy by \cite{Norris21}, obtained from LOFAR \citep{Shimwell26LOTSS} and GMRT \citep{intema17} surveys, respectively, are lower than our measured $S_{143}$. We checked for the flux density scale of our LOFAR HBA data by deriving the spectra of point sources around ORC4, finding no clear offset. Furthermore, we retrieved the public LOFAR survey image, which was obtained without any post-processing (extraction and self-calibration), and we measured $S_{144}=97\pm10$ mJy. Therefore, our data provides a flux density that is the average of those obtained from images produced by \cite{Shimwell26LOTSS} and \cite{DeGasperin26}, reflecting different depth, quality, and post-processing calibration. Overall, we note that the spectrum of ORC4 itself is well constrained up to GHz-frequencies, supporting the fidelity of our measurements and yielding to a more accurate estimate of $\alpha$.

We derived the radio power as $P_{\rm \nu}= 4 \pi D_{\rm L}^{2}S_{\rm \nu}(1+z)^{{\alpha-1}}$. This yields $P_{\rm 150}= (5.4\pm 0.5)\times 10^{25} \; {\rm W \; Hz^{-1}}$ at 150 MHz and $P_{\rm 1400}= (5.1\pm 0.6)\times 10^{24} \; {\rm W \; Hz^{-1}}$ at 1400 MHz.

\subsection{Spatial spectral distribution}
\label{sect: Spatial spectral distribution}

\begin{figure*}
        \centering
        \includegraphics[width=0.33\textwidth]{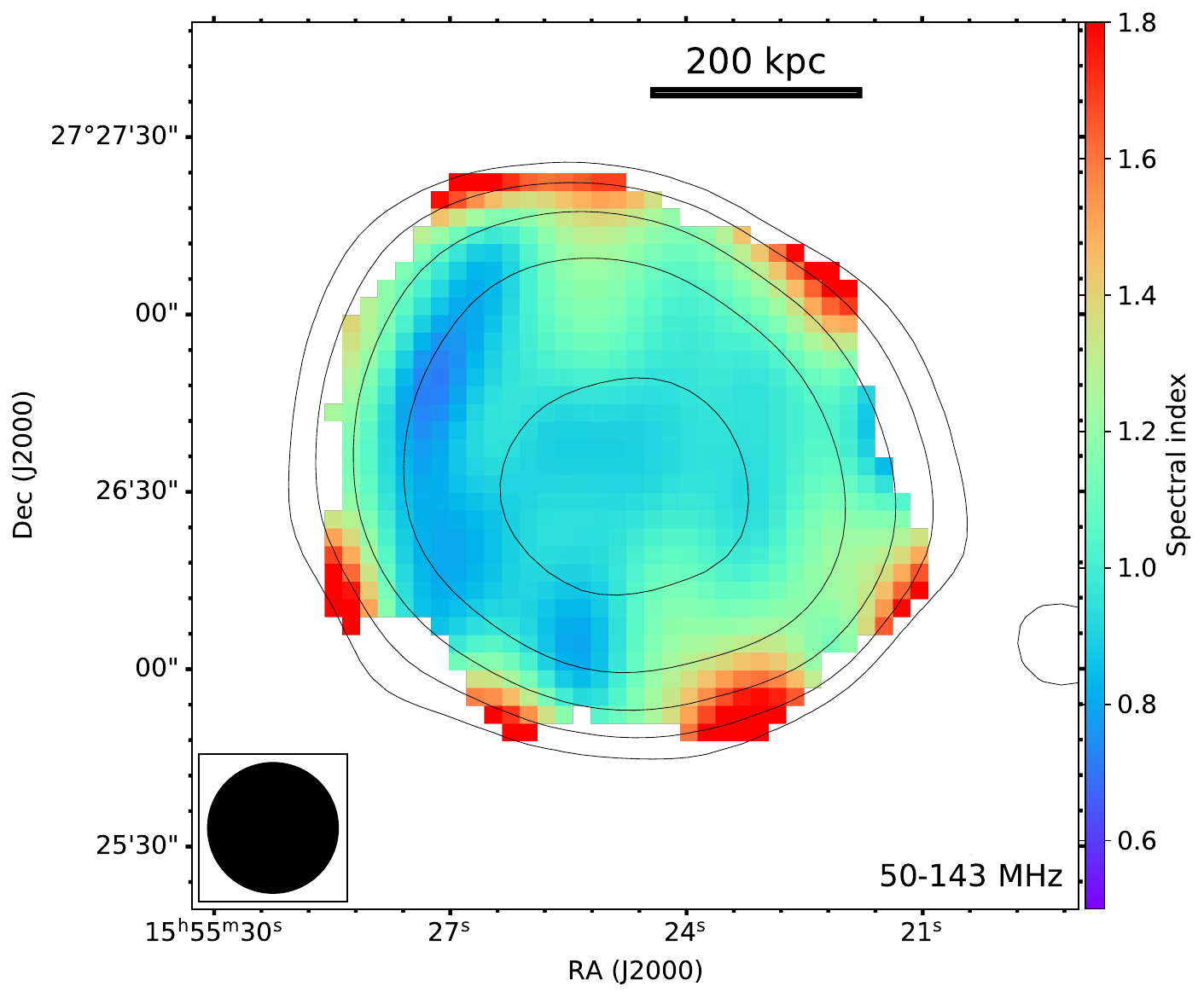}
        \includegraphics[width=0.33\textwidth]{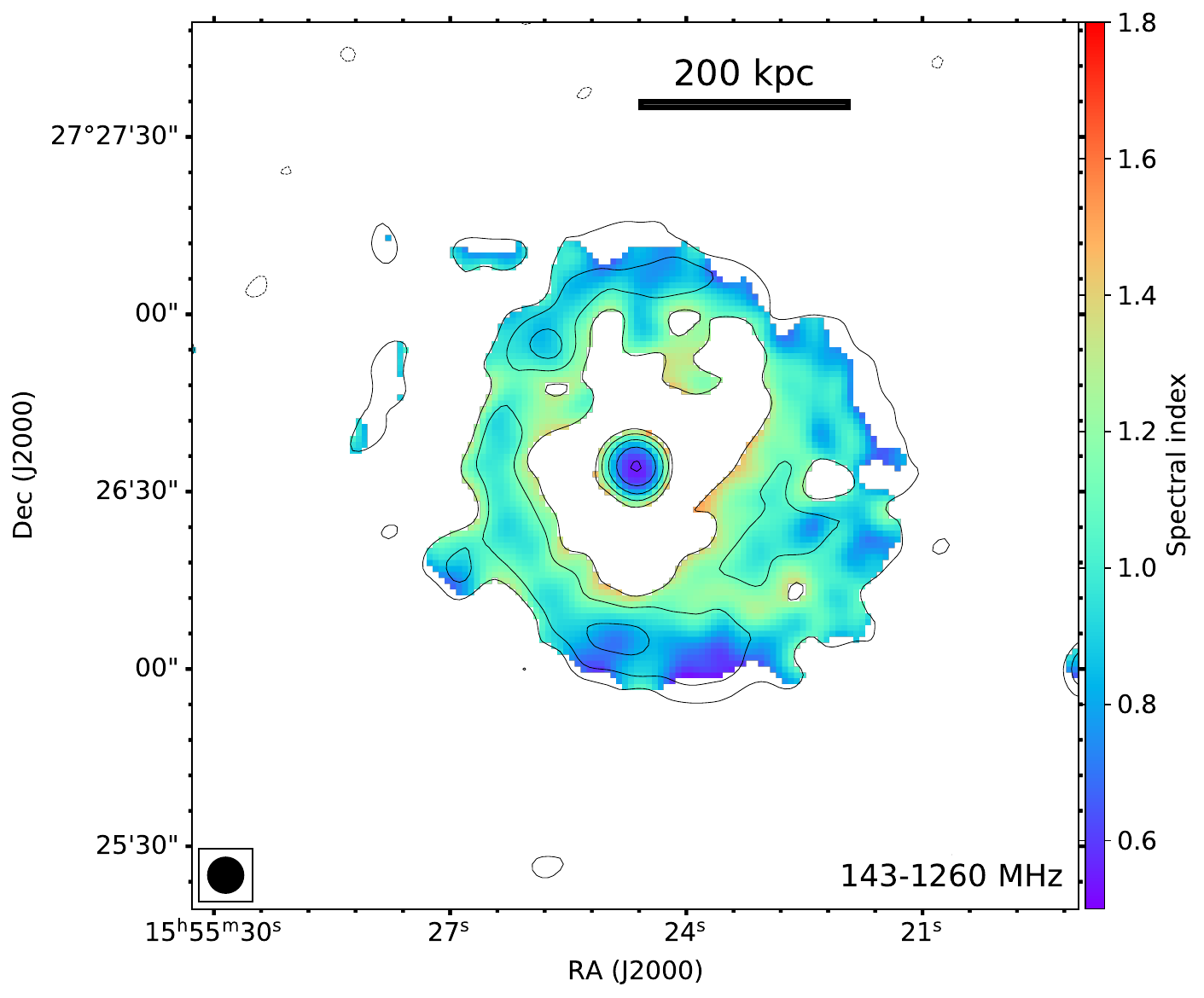}
        \includegraphics[width=0.33\textwidth]{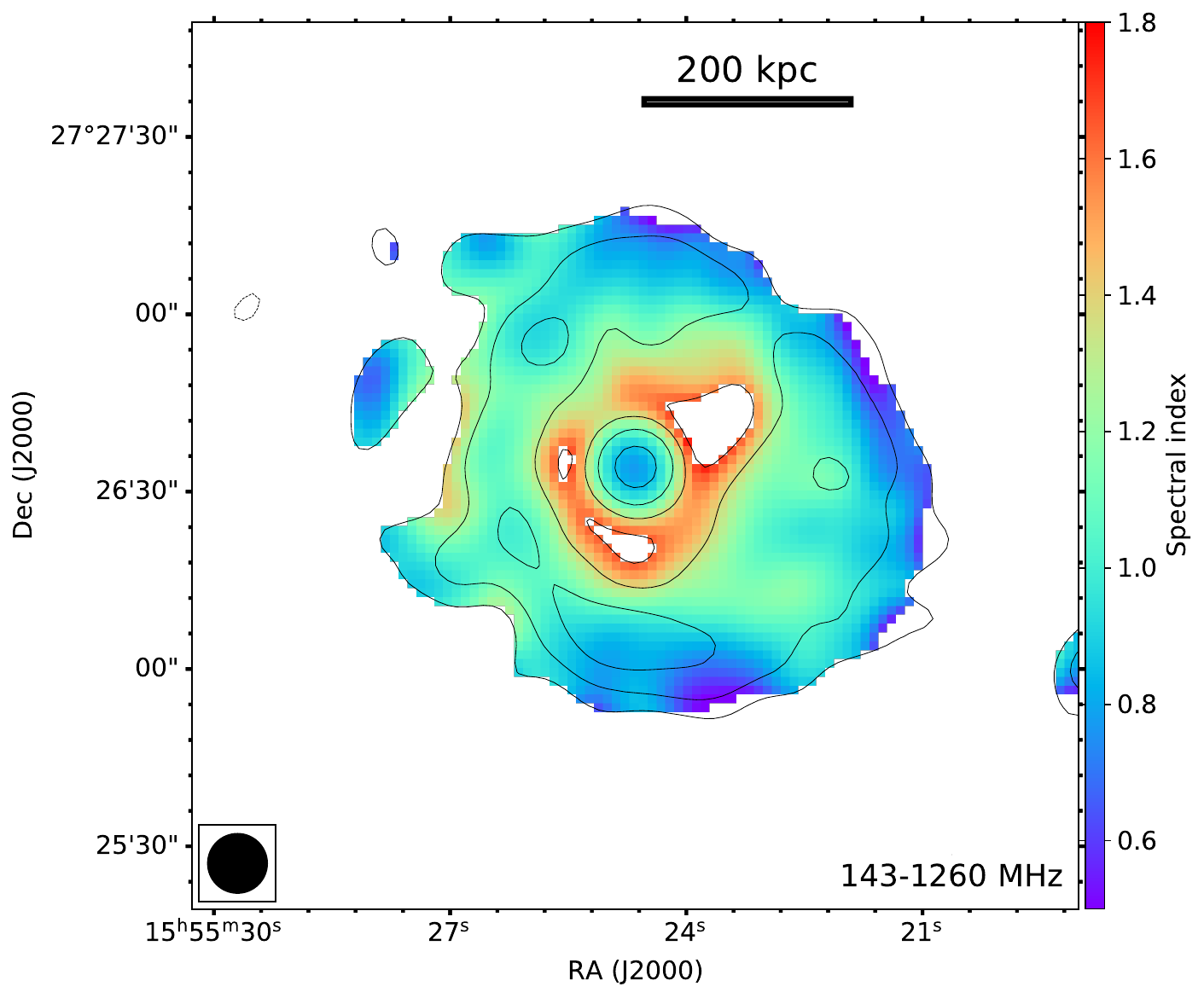}
         \includegraphics[width=0.33\textwidth]{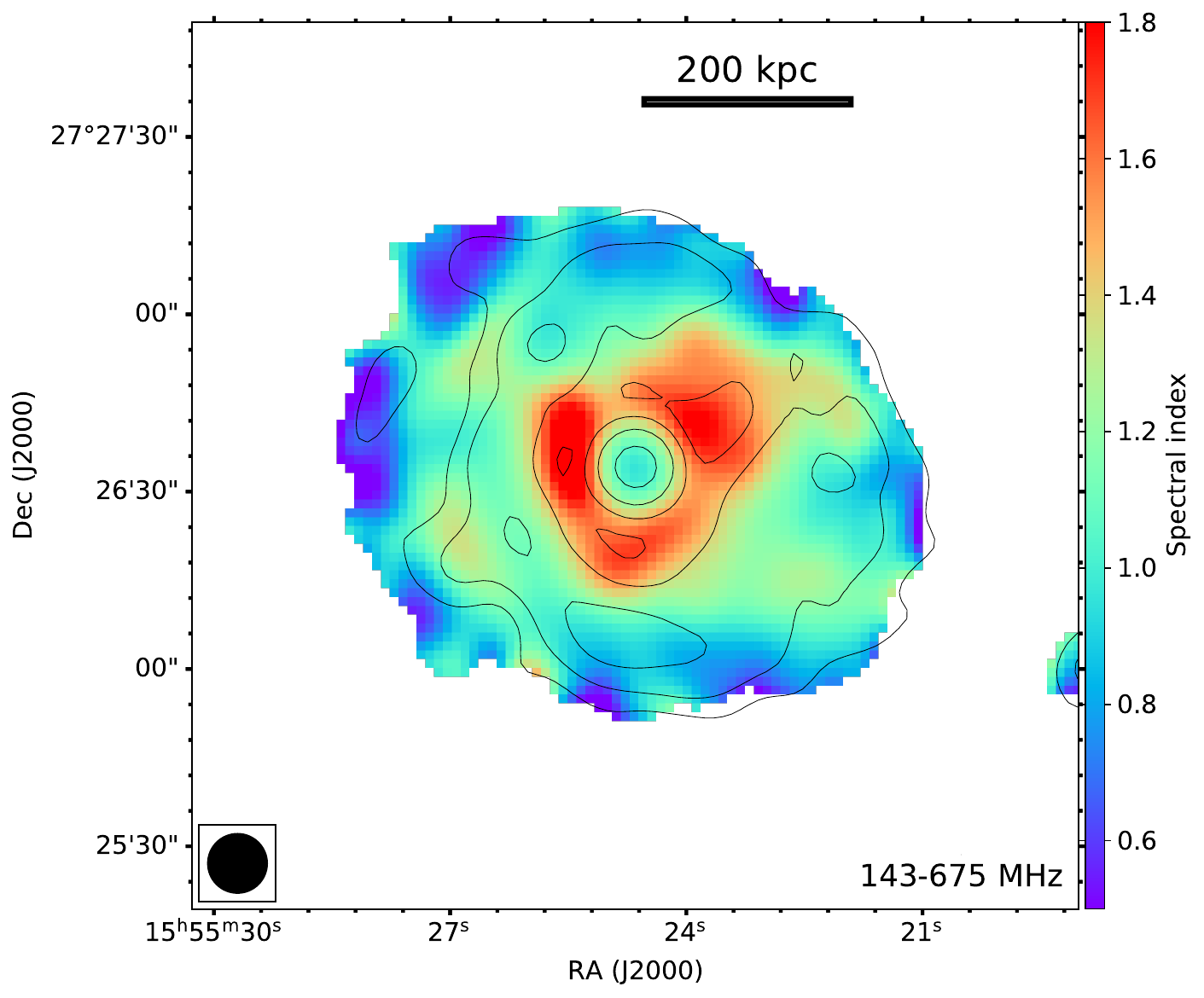}
        \includegraphics[width=0.33\textwidth]{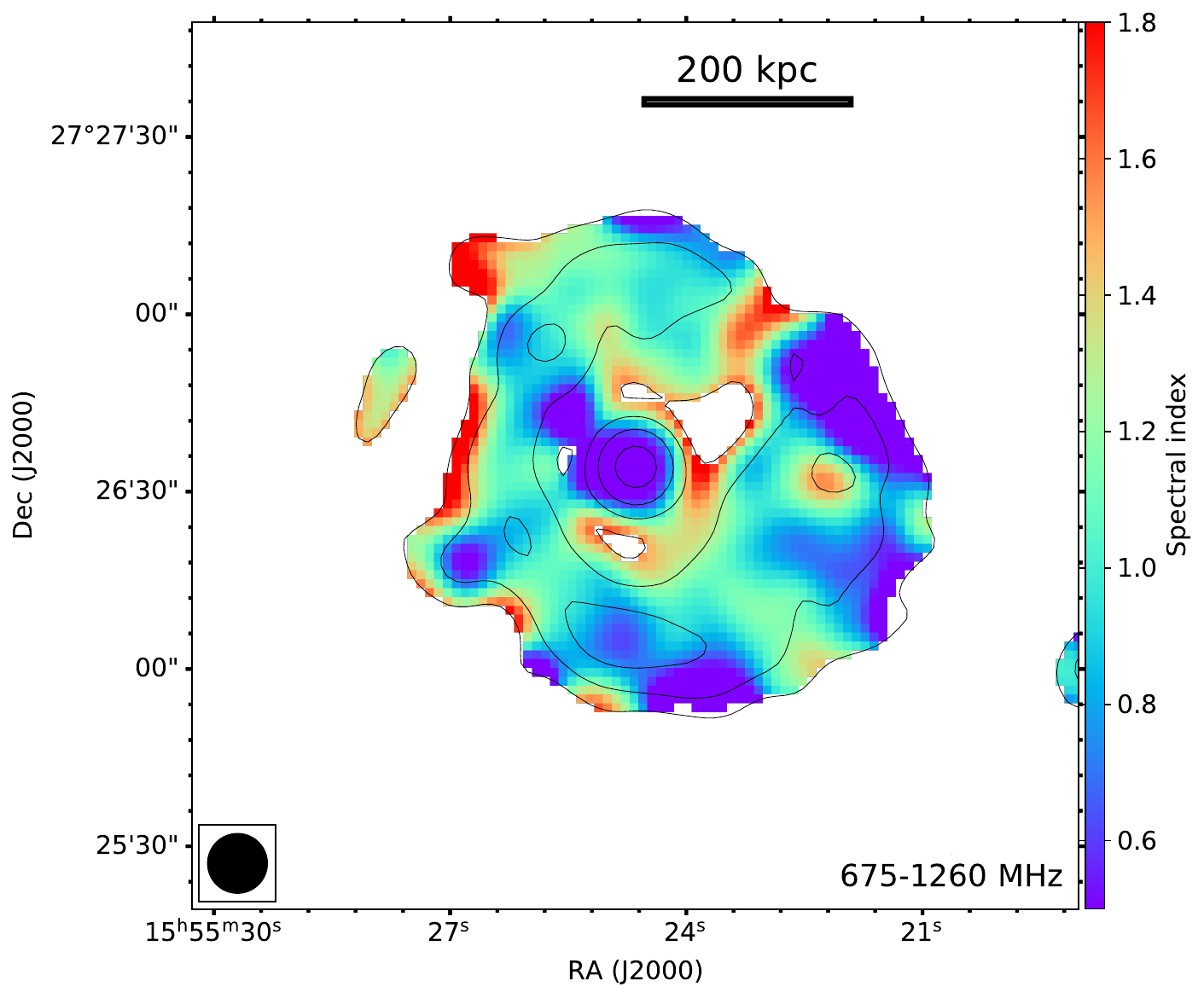}
        \includegraphics[width=0.33\textwidth]{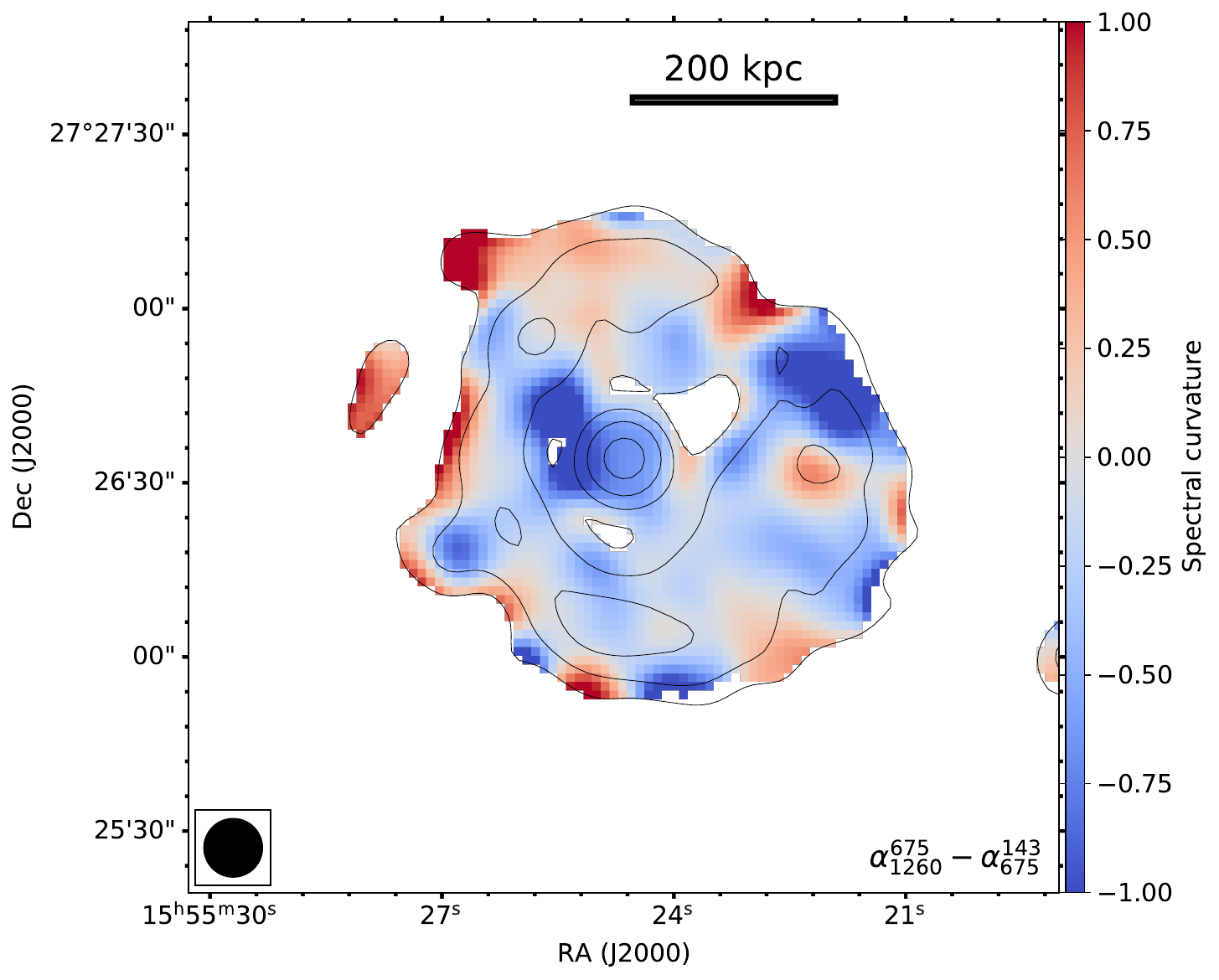}
        
        \smallskip
        
        \caption{Spectral index and curvature maps. From top left to bottom middle: spectral index maps at 50-143 MHz ($22''$), 143-1260 MHz ($6''$), 143-1260 MHz ($10''$), 143-675 MHz ($10''$), and 675-1260 MHz ($10''$). Bottom right: spectral curvature map obtained as $\alpha^{675}_{1260}-\alpha^{143}_{675}$ ($10''$).
        The radio contours represent $[\pm3, \; 6, \; 12,\; ...]\times \sigma$ from the 143 MHz image for the top left panel, and from the 1260 MHz image for all other panels. The corresponding error maps are shown in Fig. \ref{fig: errspixmap}.  }
        \label{fig: spixmap}
\end{figure*}

\begin{figure*}
        \centering
        \includegraphics[width=0.45\textwidth]{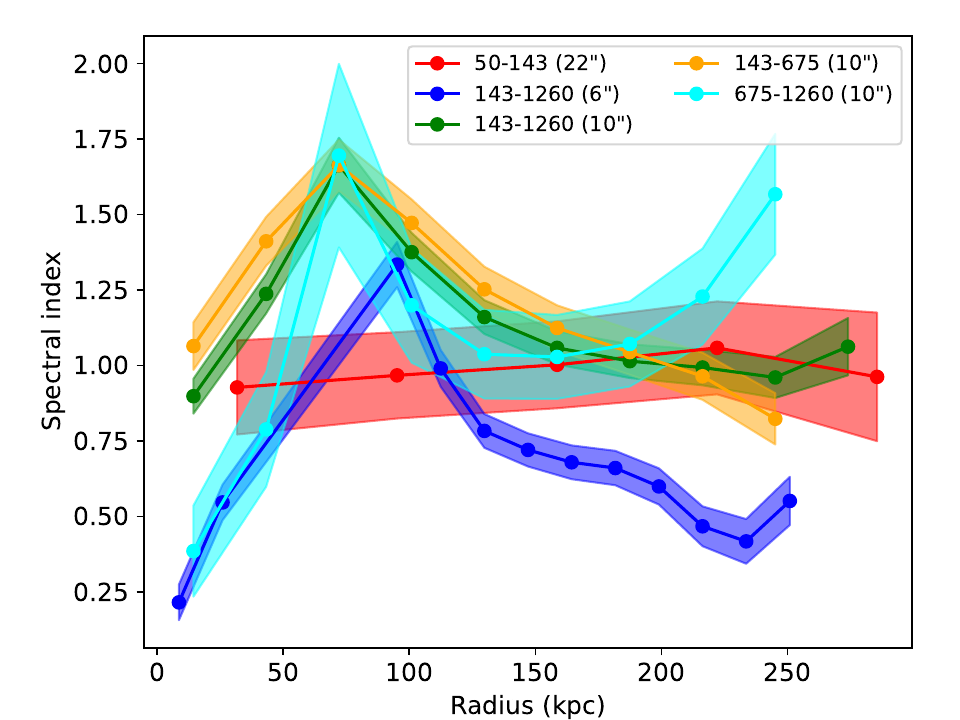}
        \includegraphics[width=0.45\textwidth]{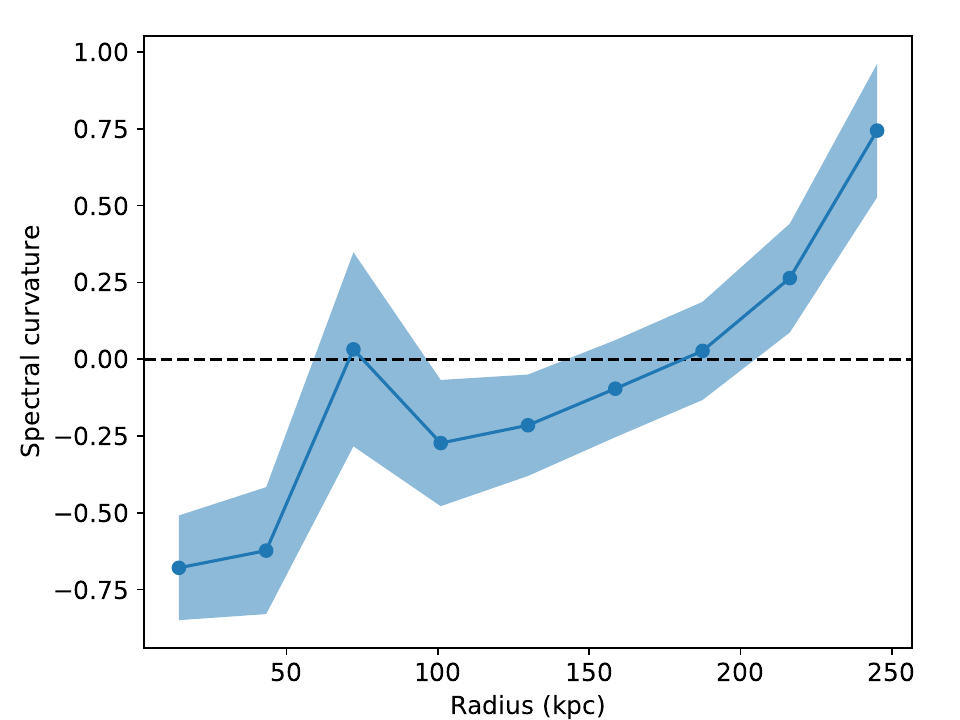}
        
        \smallskip
        
        \caption{Azimuthally-averaged spectral index and curvature profiles for the maps shown in Fig. \ref{fig: spixmap}. The profiles are measured within circular annuli of width equal to half of the beam FWHM (see Fig. \ref{fig: sb} for an example).  }
        \label{fig: spixprof}
\end{figure*}

\begin{figure}
        \centering
        \includegraphics[width=0.4\textwidth]{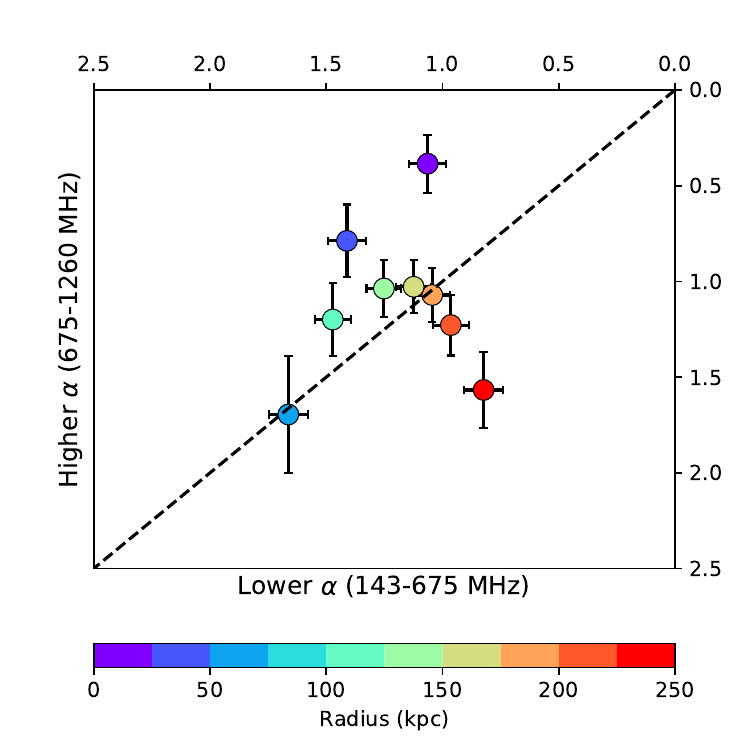}
        
        \smallskip
        
        \caption{Radio colour-colour diagram. The black dashed line (one-to-one line) represents the injection power law spectrum, while the region below this line is associated with ageing. The data points are obtained from azimuthally-averaged profiles  (Fig. \ref{fig: spixprof}) and are coloured based on their radial distance. }
        \label{fig: ccp}
\end{figure}

In this section we investigate the spatial distribution of the spectral properties. We produced spectral index maps using data at 50, 143, 675, and 1260 MHz. For each combination of frequencies $\nu_1$ and $\nu_2$, we matched the \textit{uv}-coverage during imaging, and then convolved the obtained images with the same circular beam. After masking regions below a $3\sigma$ threshold, the spectral index and its uncertainty were computed pixel-by-pixel as
\begin{equation}
 \alpha = - \frac{\ln{(S_1)}-\ln{(S_2)}}{\ln{(\nu_1)}-\ln{(\nu_2)}} \; \pm \; \left\lvert \frac{1}{\ln{ \left( \frac{\nu_{\rm 1}}{\nu_{\rm 2} } \right) }}\right\lvert \sqrt{ \left( \frac{\Delta S_{\rm 1}}{S_{\rm 1}}\right)^2 + \left( \frac{\Delta S_{\rm 2}}{S_{\rm 2}}\right)^2 } \; \; \; .
\label{eq: spectralindexformula}
\end{equation}

In Fig. \ref{fig: spixmap} we report a set of spectral index maps (error maps are shown in Fig. \ref{fig: errspixmap}). In Fig. \ref{fig: spixprof} we show the corresponding azimuthally-averaged spectral index profiles, which were computed from concentric circular annuli of width equal to half of the FWHM of the image beam. At 50-143 MHz, the spectral index is approximately uniform across the whole source, with $\alpha\sim1$ on average, but the poor resolution ($22''=127$ kpc) and large uncertainties at these frequencies prevents any conclusion on possible spatial gradients. Conversely, clear gradients are observed at higher resolution. The 143-1260 MHz map, which is the highest resolution ($6''=34$ kpc) spectral index map currently available for ORCs, reveals fine details. This map shows that the spectral index flattens outwards along the ring radius, with $\alpha\sim 0.5$ at the outermost regions. This gradient is confirmed by the 143-1260 MHz at slightly lower resolution ($10''=58$ kpc), which is more sensitive to the faint emission embedded in the ring. Within the inner $\sim 75$ kpc, the spectrum rapidly steepens up to $\alpha \sim 1.5$, and then smoothly flattens. Although the two 143-1260 MHz profiles have a similar shape, the $10''$-resolution profile appears shifted towards steeper spectral indices. This apparent inconsistency is plausibly driven by a worse resolution, which causes a mixing of adjacent regions with flatter and steeper spectra in the azimuthally-averaged profiles. A consistent shape, with a peak at $\sim75$ kpc followed by a smooth flattening, is found for the 143-675 MHz and 675-1260 MHz profiles. The 675-1260 MHz map also shows that the profile inverts the flattening trend at $ \sim 200$ kpc, but this abrupt steepening might be driven by projection effects associated with the intervening emission of the secondary ring.

To further characterise the spectral distribution and evolution of the target, we used the spectral index maps at 143-675 MHz and 675-1260 MHz at $10''$-resolution to derive a spectral curvature map and a radio colour-colour diagram (RCCD; \citealt{katz-stone93}). We computed the spectral curvature and its uncertainty pixel-by-pixel as 
\begin{equation}
 \mathcal{SC} =( \alpha_{675}^{1260}- \alpha_{143}^{675}) \pm \sqrt{ (\Delta \alpha_{675}^{1260})^2+(\Delta \alpha_{143}^{675})^2 } \; \; \; .
\label{eq: spectralcurvatureformula}
\end{equation}
With this definition and the assumed convention on the sign of the spectral index, $\mathcal{SC}>0$ indicates a standard concave spectrum (steepening at high frequencies), while $\mathcal{SC}<0$ indicates an unusual convex spectrum (flattening at high frequencies). Complementary to the spectral curvature map, RCCDs are powerful diagnostic plots to probe the local shape of radio spectra computed from two pairs of frequencies. In RCCDs, the one-to-one line represents the injection power-law spectrum ($\alpha=\alpha_{\rm inj}$), while data points below this line ($\alpha>\alpha_{\rm inj}$) are subject to ageing.

We show the obtained spectral curvature map and corresponding azimuthally-averaged profile in Fig. \ref{fig: spixmap} (see Fig. \ref{fig: errspixmap} for the error map) and Fig. \ref{fig: spixprof}, respectively, whereas the RCCD is reported in Fig. \ref{fig: ccp}. These complementary analysis tools reveal a coherent, but complex, picture for ORC4. The bulk of the emission is characterised by a moderately convex spectrum, with an average value of $\mathcal{SC}\sim -0.2$ between radii of $\sim 75$ and $\sim 160$ kpc. The curvature is not constant, but progressively increases with radius, becoming positive above $\sim 200$ kpc. Within the innermost 75 kpc, the convex spectrum appears to be more pronounced (down to $\mathcal{SC}\sim -0.7$). However, such central regions are more sensitive to small residual errors (a fraction of pixel) in the alignment of the radio images for spectral mapping, especially in proximity of source G, and they thus need to be interpreted with caution. If we ignore the inner $\sim50$ kpc, which are likely not reliable, the remaining data points are distributed along a clear parabolic path in the RCCD. Specifically, data points at radii $<160$ kpc lie above the line, in a prohibited region of the plot for any standard ageing model \citep{kardashev62,pacholczyk70,jaffe73,tribble93,komissarov&gubanov94}. Data points beyond 160 kpc are located in the allowed region of the plot, but the evolutionary path predicted by ageing models is directed towards the bottom left corner of the plot, while for ORC4 the observed path is directed towards the bottom right corner. In summary, the whole spectral distribution of the target cannot be explained in terms of standard ageing.

\section{Discussion}
\label{sect: Discussion}

The diverse ranges of frequency, resolution, and sensitivity covered by our data revealed insightful details on the properties of the target. In this section we discuss our results.

\subsection{3D morphology}
\label{sect: 3d morphology}

\begin{figure}
        \centering
        \includegraphics[width=0.49\textwidth]{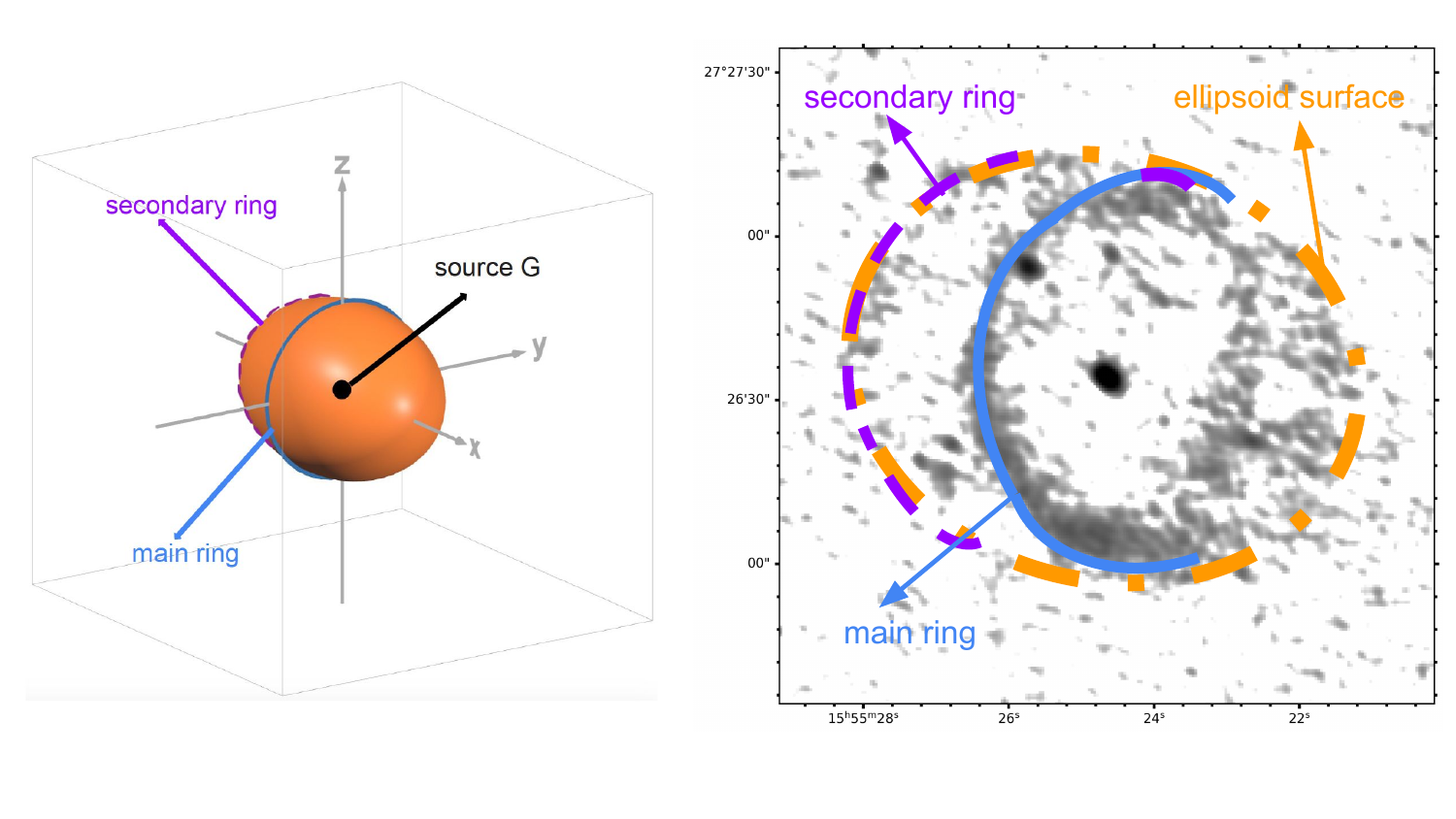}
        
        \smallskip
        
        \caption{Sketch reconstruction of ORC4 in 3D. The equatorial plane and polar axis coincide with the $y-z$ plane and $x$ axis, respectively. The axis ratio and inclination angle are indicative values that roughly reproduce the observed 2D morphology of ORC4, as highlighted in the right panel (this reports the 1260 MHz image shown in Fig. \ref{fig: imgfullres}).}
        \label{fig: 3dsketch}
\end{figure}  

The 3D geometry of ORCs remains unclear. For ORC1, various configurations have been proposed, including a thin and approximately spherical shell, in which the main and secondary rings arise from the expansion of a shock through an inhomogeneous medium (\citealt{norris22}; see the first panel in their Fig. 14). The high-resolution image of ORC4 at 1260 MHz (Fig. \ref{fig: imgfullres}) suggests some morphological similarities with ORC1. Based on this, we tentatively reconstructed the 3D geometry of ORC4, as sketched in Fig. \ref{fig: 3dsketch}. In our interpretation, which is slightly different from that proposed for ORC1, ORC4 has an overall ellipsoidal (egg-like) shape. The main ring traces the equatorial plane of the ellipsoid and is interpreted as the (projected) location of an expanding shock front (see discussion in Sects. \ref{sect: Origin}, \ref{sect: Shock Mach number}). The secondary ring and the western extension of the main ring trace emission from opposite regions along the polar axis, seen in projection. The shock front is associated with a thin shell, while faint emission (detected by our data especially at low frequencies) is diffuse across the entire ellipsoidal volume.

The proposed geometry of ORC4 should be considered as a phenomenological interpretation driven by the data and requires numerical simulations to be further tested. If confirmed, similar ellipsoidal shapes may be viable for other ORCs. In this case, the variety of ORC morphologies would be explained primarily in terms of the inclination angle and the combination of resolution and sensitivity required to detect and resolve both the primary and secondary ring. The observing frequency would also play a key role (Sect. \ref{sect: Surface brightness profiles}), as the main ring would be more easily distinguishable at high frequencies, where its surface brightness is much higher than the filling emission.

\subsection{Shock re-acceleration scenario}
\label{sect: Origin}

Pure radiative ageing cannot alone explain the observed spectral properties of ORC4, which instead suggest some kind of particle acceleration process. The most likely source of acceleration is a shock wave, whose passage is supported by the high concentration of [O II] detected on galactic scales ($\sim 40$ kpc; \citealt{Coil24}). The morphology, integrated spectrum ($\alpha\sim 1$), and radial outwards spectral flattening on scales $\gtrsim75$ kpc are consistent with acceleration from a shock expanding from the central regions, and the main ring would thus trace the location of the shock front. Particles on scales $\lesssim75$ kpc have reasonably undergone shock acceleration as well, but the observed radial spectral steepening suggests that at present radiative ageing is likely dominant; nevertheless, contamination from source G (Appendix \ref{app: Central AGN spectrum}) prevents further discussion for these inner regions. 

Recent simulations of galaxy mergers (\citealt{Dolag23,Ivleva26}) have tested the formation of ORCs via a diffusive shock acceleration (DSA; \citealt{Bell78,Blandford&Ostriker78}) mechanism. In these simulations, thermal particles are repeatedly scattered upstream and downstream the shock front, gaining kinetic energy at each encounter, and thus leading to the relativistic velocities required to emit via synchrotron in the radio band. DSA can reproduce the overall morphology, the steep integrated spectrum, and the radial spectral flattening, supporting the key role of shocks in the formation of ORCs. However, it underestimates the typical radio power of ORCs ($P_{\rm 150}\sim 10^{24}-10^{26} \; {\rm W \; Hz^{-1}}$) by $\sim 3$ orders of magnitude, and it predicts a null spectral curvature for the main ring, which progressively increases downstream due to radiative losses. 

A possible explanation for the high radio power of ORCs is that the emitting electrons are re-accelerated from a relativistic reservoir rather than from the thermal regime \citep[e.g.][]{Ivleva26}, as the former process is more efficient (e.g. \citealt{Kang&Ryu11,Caprioli18}). Our work provides observational indications for the presence of a pre-existing population of fossil relativistic electrons. The morphology of ORC4 is strongly dependent on the observing frequency (Figs. \ref{fig: imgfullres}, \ref{fig: sb}), as at low frequencies ($\lesssim 150$ MHz) the characteristic ring is less defined and the surface brightness is distributed more uniformly across the whole source. Moreover, the negative curvature (Fig. \ref{fig: spixmap}) and peculiar parabolic path in the RCCD (Fig. \ref{fig: ccp}) are not predicted by standard DSA and ageing models, as shown by a direct comparison with simulations reported in Appendix \ref{app: simulations}. Negative curvatures and RCCDs with data points located in the prohibited region or largely deviating from standard ageing curves are indicative of overlap of components with different spectra (e.g. \citealt{Young02,Rajpurohit21c,DeRubeis26}); this strongly supports the possibility that the observed spectral shape of ORC4 is the result of the combined contribution of re-accelerated electrons at different energies coexisting within the target volume. Specifically, re-accelerated low-energy fossil electrons are less subjected to radiative losses, thus smoothing brightness and spectral gradients, and their emission dominates at low frequencies; the emission of re-accelerated fossil electrons at higher energies is primarily important at high frequencies in proximity of the shock front, and becomes progressively fainter downstream due to radiative losses.

Possible events fuelling fossil reservoirs are intense galaxy starbursts and AGN outbursts (e.g. \citealt{vazza&botteon24}), which are both viable scenarios for ORC4. Specifically, the host galaxy experienced a strong starburst occurred $\sim 1$ Gyr ago \citep{Coil24} and source G may have injected fossil electrons during a previous AGN outburst phase. In the latter case, a tight connection would exist between ORCs and remnants from radio galaxies, as suggested by \cite{Shabala24}. The characteristic ring of ORCs may arise when a shock propagates through a reservoir of fossil relativistic electrons supplied by a previous active phase of the central radio galaxy. In this interpretation, rings may be potentially present in many fossil sources, but images at high resolution and frequency are necessary to better distinguish them.

\subsection{Shock Mach number}
\label{sect: Shock Mach number}
  
An important parameter for understanding the origin of ORCs is the strength of the shock. This can be constrained through either the integrated ($\alpha$) or injection ($\alpha_{\rm inj}$) spectrum; the latter is obtained via high resolution imaging, and assumed as the local spectral index at the shock front location, where particles are accelerated. Specifically, the DSA theory predicts that the Mach number is related to these spectral indices as (e.g. \citealt{drury83,Blandford&Eichler87,brunetti&jones14})
\begin{equation}
    \mathcal{M}^2=\frac{2\alpha_{\rm inj}+3}{2\alpha_{\rm inj}-1}=\frac{\alpha+1}{\alpha-1} \; \; \;. 
    \label{eq: mach}
\end{equation}
For ORC4, we measured $\alpha \sim 1$ (Fig. \ref{fig: spettro}) and $\alpha_{\rm inj}\sim 0.5$ (as obtained for the main ring from the $6''$-resolution spectral index map at 143-1260 MHz in Fig. \ref{fig: spixmap}). Both these values formally correspond to an extremely strong shock ($\mathcal{M}\rightarrow\infty $), but this interpretation requires caution.

Studies of radio relics, which are diffuse Mpc-scale sources tracing merger shocks in galaxy clusters \citep[e.g.][]{vanweeren19}, showed that $\mathcal{M}$ values derived from radio data in the DSA framework can be significantly overestimated if compared with those derived from X-ray discontinuities at the shock front (e.g. \citealt{botteon20A}). A likely explanation for this discrepancy is that radio and X-ray data trace different portions of a broad distribution of $\mathcal{M}$ along the shock surface (e.g. \citealt{wittor21}). Simulations of galaxy merger shocks for the formation of ORCs \citep{Dolag23,Ivleva26} also showed broad $\mathcal{M}$-distributions. The steep indices ($\alpha\sim1.2-1.5$) measured in downstream regions of ORC4 would be compatible with
particle populations injected via DSA by moderately strong shocks in the $\mathcal{M}\sim3-5$ range, as expected for merger shocks. Therefore, the radio data of ORC4 are consistent with a distribution of $\mathcal{M}$ values that includes a strong-shock regime. Unfortunately, current X-ray observations of the field of ORC4 are too shallow, preventing us to derive any independent constraint on $\mathcal{M}$.

We note that a subtle issue arises from the results of our analysis. If the proposed distribution of fossil re-accelerated electrons is present in ORC4, the 143-1260 MHz band might not be suitable to derive $\alpha_{\rm inj}$ because these frequencies would trace different re-acceleration and evolutionary stages, thus biasing the Mach number towards the dominant component. A more reliable measurement of $\alpha_{\rm inj}$ should be obtained from high-frequency data only. To this aim, we produced a 675-1260 MHz spectral index map at $6''$. While suggesting a slightly steeper injection index ($\alpha_{\rm inj}\sim0.7\pm 0.2$), this is still consistent with $\alpha_{\rm inj}\sim0.5$ within uncertainties, thus not providing alternative constraints on the shock strength.

\section{Summary and conclusions}
\label{sect: Summary and conclusions}

In this work, we presented new radio observations of ORC4 covering a remarkably wide frequency range (50-1284 MHz). We used these deep and high-resolution data to determine the morphological and spectral properties of the target, and investigate its origin. We summarise below our findings. 
In this work, we presented new radio observations of ORC4 covering a remarkably wide frequency range (50-1284 MHz). We used these deep and high-resolution data to determine the morphological and spectral properties of the target, and investigate its origin. We summarise below our findings. 
\begin{itemize}
    \item The morphology is dependent on the resolution (Figs. \ref{fig: imgfullres}, \ref{fig: img lowres}): poor resolution ($\sim 100$ kpc) can smooth brightness and spectral gradients and wash out the characteristic ring of ORCs. The morphology is also dependent on the observing frequency (Figs. \ref{fig: imgfullres}, \ref{fig: sb}): the ring is clearer at $\sim 1$ GHz, while it becomes progressively less dominant at lower frequencies, where the surface brightness contrast compared to the filling emission is smoother.
    \item We propose the possibility that the 3D geometry of ORC4 is ellipsoidal (Fig. \ref{fig: 3dsketch}). In this scenario, the main ring traces a thin expanding shock front, while faint emission is diffuse across the source volume.  
    \item The integrated spectrum at 50-1284 MHz (Fig. \ref{fig: spettro}) is well described by a power law with slope $\alpha=1.06\pm 0.03$. The radio power is $P_{\rm 150}= (5.4\pm 0.5)\times 10^{25} \; {\rm W \; Hz^{-1}}$ at 150 MHz. 
    \item The high-resolution ($\leq60$ kpc) spectral distribution (Figs. \ref{fig: spixmap}, \ref{fig: spixprof}) is characterised by a rapid steepening of the spectral index within the inner $\sim 75$ kpc (up to $\alpha \sim 1.5$), followed by a progressive flattening at larger radii. The spectral index of the ring is $\alpha \sim 0.5$ at 143-1260 MHz. The observed distributions provide on average a negative spectral curvature (Figs. \ref{fig: spixprof}, \ref{fig: ccp}) of $\mathcal{SC}\sim -0.2$, implying that the spectrum flattens with increasing frequency. 
    \item An expanding shock is the most likely explanation for the overall observed spectral trend. However, classical DSA followed by standard ageing models fail to predict the high $P_{\rm 150}$ and negative $\mathcal{SC}$ of ORC4 (Appendix \ref{app: simulations}). We thus propose a scenario in which fossil relativistic electrons have been re-accelerated by the shock passage (Sect. \ref{sect: Origin}). Higher-energy re-accelerated electrons dominate the emission of the ring (close to shock front) at higher frequencies, while lower-energy electrons are diffuse across the whole volume and dominate the emission at lower frequencies. The mixture of these re-accelerated fossil electrons with different spectra is broadly in agreement with the spectro-morphological features of the target. 
    \item Current radio observations of ORC4 are consistent with a broad distribution of Mach numbers, including a strong-shock regime (Sect. \ref{sect: Shock Mach number}).
\end{itemize}

Our analysis provides a fundamental piece of the puzzle to reconstruct the origin of ORC4. So far, simulations have focused on DSA of thermal electrons, while the fossil population is likely the key ingredient (see also \citealt{Shabala24,Wang&Heinz26}). Re-acceleration of fossil electrons has been suggested for other ORCs \citep{Koribalski24b,Koribalski24c}, and is invoked also for radio relics, as their radio power is underestimated by standard DSA theory, similar to that of ORCs, in spite of different environmental and physical conditions. Such fossil particles could have been supplied by the central radio source of ORC4 during a previous AGN outburst phase, similarly to the scenario suggested for ORC6 \citep{Koribalski26}, or by galactic outflows. Our data are poorly informative on the event triggering the shock, which may be related to an intense starburst \citep{Coil24}, a galaxy merger \citep{norris22,Dolag23,Ivleva26}, or large-scale accretion \citep{Yamasaki24}. Independent constraints on $\mathcal{M}$ are important to better test these scenarios.

The large amount of data collected for ORC4 revealed insightful features and some similarities with the archetypical ORC1. However, it is unclear if ORC4 represents an outlier compared to the population of ORCs, as their wide-band multi-frequency properties are generally poorly constrained. Follow-up observations for samples of ORCs are thus necessary for a proper comparison. Optical and X-ray data may constrain the environment and shock Mach number, whereas radio polarimetry can constrain the magnetic field distribution and 3D geometry. A key test for probing the role of re-accelerated fossil electrons will be the high frequency follow-up of the ORC sample observed by \cite{DeGasperin26} at low frequency.

\begin{acknowledgements}
We thank the referee for their comments and suggestions. LB thanks A. Pal for useful discussion. AI and KD acknowledge support by the COMPLEX project from the European Research Council (ERC) under
the European Union's Horizon 2020 research and innovation program grant agreement ERC-2019-AdG 882679.  CJR acknowledges financial support from the German Science Foundation DFG, via the Collaborative Research Center SFB1491 'Cosmic Interacting Matters – From Source to Signal'. JRC acknowledges funding from the European Union via the European Research Council (ERC) grant Epaphus (project number 101166008) and from the Dutch research council (NWO) under the talent programme (Vidi grant VI.Vidi.243.136). We thank the staff of the GMRT that made these observations possible. GMRT is run by the National Centre for Radio Astrophysics of the Tata Institute of Fundamental Research. LOFAR \citep{vanhaarlem13} is the Low Frequency Array designed and constructed by ASTRON. It has observing, data processing, and data storage facilities in several countries, which are owned by various parties (each with their own funding sources), and that are collectively operated by the ILT foundation under a joint scientific policy. The ILT resources have benefited from the following recent major funding sources: CNRS-INSU, Observatoire de Paris and Universit\'e d’Orl\'eans, France; BMBF, MIWF- NRW, MPG, Germany; Science Foundation Ireland (SFI), Department of Business, Enterprise and Innovation (DBEI), Ireland; NWO, The Netherlands; The Science and Technology Facilities Council, UK; Ministry of Science and Higher Education, Poland; The Istituto Nazionale di Astrofisica (INAF), Italy. This research made use of the Dutch national e-infrastructure with support of the SURF Cooperative (e-infra 180169) and the LOFAR e-infra group. The J\"ulich LOFAR Long Term Archive and the German LOFAR network are both coordinated and operated by the J\"ulich Supercomputing Centre (JSC), and computing resources on the supercomputer JUWELS at JSC were provided by the Gauss Centre for Supercomputing e.V. (grant CHTB00) through the John von Neumann Institute for Computing (NIC). This research made use of the University of Hertfordshire high-performance computing facility and the LOFAR-UK computing facility located at the University of Hertfordshire and supported by STFC [ST/P000096/1], and of the Italian LOFAR IT computing infrastructure supported and operated by INAF, and by the Physics Department of Turin University (under an agreement with Consorzio Interuniversitario per la Fisica Spaziale) at the C3S Supercomputing Centre, Italy. This research made use of the HOTCAT cluster \citep{bertocco20,taffoni20} at Osservatorio Astronomico di Trieste. The research leading to these results has received funding from the European Union’s Horizon 2020 research and innovation programme under grant agreement No 101004719 (ORP). The MeerKAT telescope is operated by the South African Radio Astronomy Observatory, which is a facility of the National Research Foundation, an agency of the Department of Science and Innovation. This paper includes archived data obtained through the CSIRO ASKAP Science Data Archive, CASDA (\url{http://data.csiro.au}). This publication makes use of data products from the Wide-field Infrared Survey Explorer, which is a joint project of the University of California, Los Angeles, and the Jet Propulsion Laboratory/California Institute of Technology, funded by the National Aeronautics and Space Administration. DESI construction and operations is managed by the Lawrence Berkeley National Laboratory. This research is supported by the U.S. Department of Energy, Office of Science, Office of High-Energy Physics, under Contract No. DE–AC02–05CH11231, and by the National Energy Research Scientific Computing Center, a DOE Office of Science User Facility under the same contract. Additional support for DESI is provided by the U.S. National Science Foundation, Division of Astronomical Sciences under Contract No. AST-0950945 to the NSF’s National Optical-Infrared Astronomy Research Laboratory; the Science and Technology Facilities Council of the United Kingdom; the Gordon and Betty Moore Foundation; the Heising-Simons Foundation; the French Alternative Energies and Atomic Energy Commission (CEA); the Secretariat of Science, Humanities, Technology and Innovation (SECIHTI) of Mexico; the Ministry of Science and Innovation of Spain, and by the DESI Member Institutions. The DESI collaboration is honored to be permitted to conduct astronomical research on Iolkam Du'ag (Kitt Peak), a mountain with particular significance to the Tohono O'odham Nation. This research has made use of i) SAOImageDS9, developed by Smithsonian Astrophysical Observatory \citep{ds9}, ii) the VizieR catalogue access tool, CDS, Strasbourg Astronomical Observatory, France (DOI: 10.26093/cds/vizier), iii) APLpy, an open-source plotting package for Python \citep{robitaille&bressert12APLPY},  iv) Astropy, a community-developed core Python package for Astronomy \citep{astropycollaboration13,astropycollaboration18}, v) Matplotlib \citep{hunter07MATPLOTLIB}, vi) Numpy \citep{harris20NUMPY}, and vii) SciPy \citep{scipy}.

\end{acknowledgements}

\bibliographystyle{aa}
\bibliography{bibliografia}

@ARTICLE{jaffe73,
	author = {{Jaffe}, W.~J. and {Perola}, G.~C.},
	journal = {\aap},
	year = "1973",
	volume = {26},
	pages = {423},
	       adsurl = {https://ui.adsabs.harvard.edu/abs/1973A&A....26..423J}
}

@ARTICLE{kardashev62,
	author = {{Kardashev}, N.~S.},
	journal = {\sovast},
	year = "1962",
	volume = {6},
	pages = {317},
	   adsurl = {https://ui.adsabs.harvard.edu/abs/1962SvA.....6..317K}
}

@BOOK{pacholczyk70,
	author = {{Pacholczyk}, A.~G.},
	title = "{Radio astrophysics. Nonthermal processes in galactic and extragalactic sources}",
	year = "1970",
	     adsurl = {https://ui.adsabs.harvard.edu/abs/1970ranp.book.....P}
}

@ARTICLE{tribble93,
	author = {{Tribble}, Peter C.},
	journal = {\mnras},
	year = "1993",
	volume = {261},
	pages = {57-62},
	     adsurl = {https://ui.adsabs.harvard.edu/abs/1993MNRAS.261...57T}
}

@ARTICLE{brunetti&jones14,
       author = {{Brunetti}, Gianfranco and {Jones}, Thomas W.},
      journal = {International Journal of Modern Physics D},
         year = 2014,
       volume = {23},
       number = {4},
        pages = {1430007-98},
 primaryClass = {astro-ph.CO},
       adsurl = {https://ui.adsabs.harvard.edu/abs/2014IJMPD..2330007B}
       }

@ARTICLE{vanweeren19,
       author = {{van Weeren}, R.~J. and {de Gasperin}, F. and {Akamatsu}, H. and
         {Br{\"u}ggen}, M. and {Feretti}, L. and {Kang}, H. and {Stroe}, A. and
         {Zandanel}, F.},
      journal = {\ssr},
         year = 2019,
       volume = {215},
       number = {1},
        pages = {16},
 primaryClass = {astro-ph.HE},
       adsurl = {https://ui.adsabs.harvard.edu/abs/2019SSRv..215...16V}
  }

@ARTICLE{offringa14,
       author = {{Offringa}, A.~R. and {McKinley}, B. and {Hurley-Walker}, N. and
         {Briggs}, F.~H. and {Wayth}, R.~B. and {Kaplan}, D.~L. and
         {Bell}, M.~E. and {Feng}, L. and {Neben}, A.~R. and {Hughes}, J.~D. and
         {Rhee}, J. and {Murphy}, T. and {Bhat}, N.~D.~R. and {Bernardi}, G. and
         {Bowman}, J.~D. and {Cappallo}, R.~J. and {Corey}, B.~E. and {Deshpand
        e}, A.~A. and {Emrich}, D. and {Ewall-Wice}, A. and {Gaensler}, B.~M. and
         {Goeke}, R. and {Greenhill}, L.~J. and {Hazelton}, B.~J. and
         {Hindson}, L. and {Johnston-Hollitt}, M. and {Jacobs}, D.~C. and
         {Kasper}, J.~C. and {Kratzenberg}, E. and {Lenc}, E. and
         {Lonsdale}, C.~J. and {Lynch}, M.~J. and {McWhirter}, S.~R. and
         {Mitchell}, D.~A. and {Morales}, M.~F. and {Morgan}, E. and
         {Kudryavtseva}, N. and {Oberoi}, D. and {Ord}, S.~M. and {Pindor}, B. and
         {Procopio}, P. and {Prabu}, T. and {Riding}, J. and {Roshi}, D.~A. and
         {Shankar}, N. Udaya and {Srivani}, K.~S. and {Subrahmanyan}, R. and
         {Tingay}, S.~J. and {Waterson}, M. and {Webster}, R.~L. and
         {Whitney}, A.~R. and {Williams}, A. and {Williams}, C.~L.},
      journal = {\mnras},
         year = 2014,
       volume = {444},
       number = {1},
        pages = {606-619},
       adsurl = {https://ui.adsabs.harvard.edu/abs/2014MNRAS.444..606O}
       }

@ARTICLE{offringa17,
       author = {{Offringa}, A.~R. and {Smirnov}, O.},
      journal = {\mnras},
         year = 2017,
       volume = {471},
       number = {1},
        pages = {301-316},
       adsurl = {https://ui.adsabs.harvard.edu/abs/2017MNRAS.471..301O}
       }

@ARTICLE{intema09,
       author = {{Intema}, H.~T. and {van der Tol}, S. and {Cotton}, W.~D. and
         {Cohen}, A.~S. and {van Bemmel}, I.~M. and {R{\"o}ttgering}, H.~J.~A.},
      journal = {\aap},
         year = 2009,
       volume = {501},
       number = {3},
        pages = {1185-1205},
       adsurl = {https://ui.adsabs.harvard.edu/abs/2009A&A...501.1185I}
       }

@ARTICLE{intema17,
       author = {{Intema}, H.~T. and {Jagannathan}, P. and {Mooley}, K.~P. and
         {Frail}, D.~A.},
        title = "{The GMRT 150 MHz all-sky radio survey. First alternative data release TGSS ADR1}",
      journal = {\aap},
         year = 2017,
       volume = {598},
        pages = {A78},
       adsurl = {https://ui.adsabs.harvard.edu/abs/2017A&A...598A..78I}
       }

@ARTICLE{botteon20A,
       author = {{Botteon}, A. and {Brunetti}, G. and {Ryu}, D. and {Roh}, S.},
        title = "{Shock acceleration efficiency in radio relics}",
      journal = {\aap},
         year = 2020,
        month = feb,
       volume = {634},
          eid = {A64},
       adsurl = {https://ui.adsabs.harvard.edu/abs/2020A&A...634A..64B}}

@INPROCEEDINGS{briggs95,
       author = {{Briggs}, D.~S.},
        title = "{High Fidelity Interferometric Imaging: Robust Weighting and NNLS Deconvolution}",
    booktitle = {American Astronomical Society Meeting Abstracts},
         year = 1995,
       series = {American Astronomical Society Meeting Abstracts},
       volume = {187},
        month = dec,
          eid = {112.02},
        pages = {112.02},
       adsurl = {https://ui.adsabs.harvard.edu/abs/1995AAS...18711202B}}

@ARTICLE{vanhaarlem13,
       author = {{van Haarlem}, M.~P. and {Wise}, M.~W. and {Gunst}, A.~W. and
         {Heald}, G. and {McKean}, J.~P. and {Hessels}, J.~W.~T. and
         {de Bruyn}, A.~G. and {Nijboer}, R. and {Swinbank}, J. and
         {Fallows}, R. and et al.},
        title = "{LOFAR: The LOw-Frequency ARray}",
      journal = {\aap},
         year = 2013,
        month = aug,
       volume = {556},
          eid = {A2},
        pages = {A2},
       adsurl = {https://ui.adsabs.harvard.edu/abs/2013A&A...556A...2V}}

@ARTICLE{vanweeren21,
       author = {{van Weeren}, R.~J. and {Shimwell}, T.~W. and {Botteon}, A. and {Brunetti}, G. and {Br{\"u}ggen}, M. and {Boxelaar}, J.~M. and {Cassano}, R. and {Di Gennaro}, G. and {Andrade-Santos}, F. and {Bonnassieux}, E. and {Bonafede}, A. and {Cuciti}, V. and {Dallacasa}, D. and {de Gasperin}, F. and {Gastaldello}, F. and {Hardcastle}, M.~J. and {Hoeft}, M. and {Kraft}, R.~P. and {Mandal}, S. and {Rossetti}, M. and {R{\"o}ttgering}, H.~J.~A. and {Tasse}, C. and {Wilber}, A.~G.},
        title = "{LOFAR observations of galaxy clusters in HETDEX. Extraction and self-calibration of individual LOFAR targets}",
      journal = {\aap},
         year = 2021,
        month = jul,
       volume = {651},
          eid = {A115},
        pages = {A115},
          doi = {10.1051/0004-6361/202039826},
archivePrefix = {arXiv},
       eprint = {2011.02387},
 primaryClass = {astro-ph.CO},
       adsurl = {https://ui.adsabs.harvard.edu/abs/2021A&A...651A.115V}
}

@ARTICLE{venturi17,
       author = {{Venturi}, T. and {Rossetti}, M. and {Brunetti}, G. and {Farnsworth}, D. and {Gastaldello}, F. and {Giacintucci}, S. and {Lal}, D.~V. and {Rudnick}, L. and {Shimwell}, T.~W. and {Eckert}, D. and {Molendi}, S. and {Owers}, M.},
        title = "{The two-component giant radio halo in the galaxy cluster Abell 2142}",
      journal = {\aap},
         year = 2017,
        month = jul,
       volume = {603},
          eid = {A125},
        pages = {A125},
          doi = {10.1051/0004-6361/201630014},
       adsurl = {https://ui.adsabs.harvard.edu/abs/2017A&A...603A.125V}}

@ARTICLE{shimwell22LOTSSDR2,
       author = {{Shimwell}, T.~W. and {Hardcastle}, M.~J. and {Tasse}, C. and {Best}, P.~N. and {R{\"o}ttgering}, H.~J.~A. and {Williams}, W.~L. and {Botteon}, A. and {Drabent}, A. and {Mechev}, A. and {Shulevski}, A. and {van Weeren}, R.~J. and {Bester}, L. and {Br{\"u}ggen}, M. and {Brunetti}, G. and {Callingham}, J.~R. and {Chy{\.z}y}, K.~T. and {Conway}, J.~E. and {Dijkema}, T.~J. and {Duncan}, K. and {de Gasperin}, F. and {Hale}, C.~L. and {Haverkorn}, M. and {Hugo}, B. and {Jackson}, N. and {Mevius}, M. and {Miley}, G.~K. and {Morabito}, L.~K. and {Morganti}, R. and {Offringa}, A. and {Oonk}, J.~B.~R. and {Rafferty}, D. and {Sabater}, J. and {Smith}, D.~J.~B. and {Schwarz}, D.~J. and {Smirnov}, O. and {O'Sullivan}, S.~P. and {Vedantham}, H. and {White}, G.~J. and {Albert}, J.~G. and {Alegre}, L. and {Asabere}, B. and {Bacon}, D.~J. and {Bonafede}, A. and {Bonnassieux}, E. and {Brienza}, M. and {Bilicki}, M. and {Bonato}, M. and {Calistro Rivera}, G. and {Cassano}, R. and {Cochrane}, R. and {Croston}, J.~H. and {Cuciti}, V. and {Dallacasa}, D. and {Danezi}, A. and {Dettmar}, R.~J. and {Di Gennaro}, G. and {Edler}, H.~W. and {En{\ss}lin}, T.~A. and {Emig}, K.~L. and {Franzen}, T.~M.~O. and {Garc{\'\i}a-Vergara}, C. and {Grange}, Y.~G. and {G{\"u}rkan}, G. and {Hajduk}, M. and {Heald}, G. and {Heesen}, V. and {Hoang}, D.~N. and {Hoeft}, M. and {Horellou}, C. and {Iacobelli}, M. and {Jamrozy}, M. and {Jeli{\'c}}, V. and {Kondapally}, R. and {Kukreti}, P. and {Kunert-Bajraszewska}, M. and {Magliocchetti}, M. and {Mahatma}, V. and {Ma{\l}ek}, K. and {Mandal}, S. and {Massaro}, F. and {Meyer-Zhao}, Z. and {Mingo}, B. and {Mostert}, R.~I.~J. and {Nair}, D.~G. and {Nakoneczny}, S.~J. and {Nikiel-Wroczy{\'n}ski}, B. and {Orr{\'u}}, E. and {Pajdosz-{\'S}mierciak}, U. and {Pasini}, T. and {Prandoni}, I. and {van Piggelen}, H.~E. and {Rajpurohit}, K. and {Retana-Montenegro}, E. and {Riseley}, C.~J. and {Rowlinson}, A. and {Saxena}, A. and {Schrijvers}, C. and {Sweijen}, F. and {Siewert}, T.~M. and {Timmerman}, R. and {Vaccari}, M. and {Vink}, J. and {West}, J.~L. and {Wo{\l}owska}, A. and {Zhang}, X. and {Zheng}, J.},
        title = "{The LOFAR Two-metre Sky Survey. V. Second data release}",
      journal = {\aap},
         year = 2022,
        month = mar,
       volume = {659},
          eid = {A1},
        pages = {A1},
          doi = {10.1051/0004-6361/202142484},
archivePrefix = {arXiv},
       eprint = {2202.11733},
 primaryClass = {astro-ph.GA},
       adsurl = {https://ui.adsabs.harvard.edu/abs/2022A&A...659A...1S}
}

@INPROCEEDINGS{taffoni20,
       author = {{Taffoni}, G. and {Becciani}, U. and {Garilli}, B. and {Maggio}, G. and {Pasian}, F. and {Umana}, G. and {Smareglia}, R. and {Vitello}, F.},
        title = "{CHIPP: INAF Pilot Project for HTC, HPC and HPDA}",
    booktitle = {Astronomical Data Analysis Software and Systems XXIX},
         year = 2020,
       editor = {{Pizzo}, R. and {Deul}, E.~R. and {Mol}, J.~D. and {de Plaa}, J. and {Verkouter}, H.},
       series = {Astronomical Society of the Pacific Conference Series},
       volume = {527},
        month = jan,
        pages = {307},
archivePrefix = {arXiv},
       eprint = {2002.01283},
 primaryClass = {astro-ph.IM},
       adsurl = {https://ui.adsabs.harvard.edu/abs/2020ASPC..527..307T}}

@INPROCEEDINGS{bertocco20,
       author = {{Bertocco}, S. and {Goz}, D. and {Tornatore}, L. and {Ragagnin}, A. and {Maggio}, G. and {Gasparo}, F. and {Vuerli}, C. and {Taffoni}, G. and {Molinaro}, M.},
        title = "{INAF Trieste Astronomical Observatory Information Technology Framework}",
    booktitle = {Astronomical Data Analysis Software and Systems XXIX},
         year = 2020,
       editor = {{Pizzo}, R. and {Deul}, E.~R. and {Mol}, J.~D. and {de Plaa}, J. and {Verkouter}, H.},
       series = {Astronomical Society of the Pacific Conference Series},
       volume = {527},
        month = jan,
        pages = {303},
archivePrefix = {arXiv},
       eprint = {1912.05340},
 primaryClass = {astro-ph.IM},
       adsurl = {https://ui.adsabs.harvard.edu/abs/2020ASPC..527..303B}}

@MISC{robitaille&bressert12APLPY,
       author = {{Robitaille}, Thomas and {Bressert}, Eli},
        title = "{APLpy: Astronomical Plotting Library in Python}",
 howpublished = {Astrophysics Source Code Library, record ascl:1208.017},
         year = 2012,
        month = aug,
          eid = {ascl:1208.017},
        pages = {ascl:1208.017},
archivePrefix = {ascl},
       eprint = {1208.017},
       adsurl = {https://ui.adsabs.harvard.edu/abs/2012ascl.soft08017R}
}

@ARTICLE{astropycollaboration13,
       author = {{Astropy Collaboration} and {Robitaille}, Thomas P. and {Tollerud}, Erik J. and {Greenfield}, Perry and {Droettboom}, Michael and {Bray}, Erik and {Aldcroft}, Tom and {Davis}, Matt and {Ginsburg}, Adam and {Price-Whelan}, Adrian M. and {Kerzendorf}, Wolfgang E. and {Conley}, Alexander and {Crighton}, Neil and {Barbary}, Kyle and {Muna}, Demitri and {Ferguson}, Henry and {Grollier}, Fr{\'e}d{\'e}ric and {Parikh}, Madhura M. and {Nair}, Prasanth H. and {Unther}, Hans M. and {Deil}, Christoph and {Woillez}, Julien and {Conseil}, Simon and {Kramer}, Roban and {Turner}, James E.~H. and {Singer}, Leo and {Fox}, Ryan and {Weaver}, Benjamin A. and {Zabalza}, Victor and {Edwards}, Zachary I. and {Azalee Bostroem}, K. and {Burke}, D.~J. and {Casey}, Andrew R. and {Crawford}, Steven M. and {Dencheva}, Nadia and {Ely}, Justin and {Jenness}, Tim and {Labrie}, Kathleen and {Lim}, Pey Lian and {Pierfederici}, Francesco and {Pontzen}, Andrew and {Ptak}, Andy and {Refsdal}, Brian and {Servillat}, Mathieu and {Streicher}, Ole},
        title = "{Astropy: A community Python package for astronomy}",
      journal = {\aap},
         year = 2013,
        month = oct,
       volume = {558},
          eid = {A33},
        pages = {A33},
          doi = {10.1051/0004-6361/201322068},
archivePrefix = {arXiv},
       eprint = {1307.6212},
 primaryClass = {astro-ph.IM},
       adsurl = {https://ui.adsabs.harvard.edu/abs/2013A&A...558A..33A}
}

@ARTICLE{astropycollaboration18,
       author = {{Astropy Collaboration} and {Price-Whelan}, A.~M. and {Sip{\H{o}}cz}, B.~M. and {G{\"u}nther}, H.~M. and {Lim}, P.~L. and {Crawford}, S.~M. and {Conseil}, S. and {Shupe}, D.~L. and {Craig}, M.~W. and {Dencheva}, N. and {Ginsburg}, A. and {VanderPlas}, J.~T. and {Bradley}, L.~D. and {P{\'e}rez-Su{\'a}rez}, D. and {de Val-Borro}, M. and {Aldcroft}, T.~L. and {Cruz}, K.~L. and {Robitaille}, T.~P. and {Tollerud}, E.~J. and {Ardelean}, C. and {Babej}, T. and {Bach}, Y.~P. and {Bachetti}, M. and {Bakanov}, A.~V. and {Bamford}, S.~P. and {Barentsen}, G. and {Barmby}, P. and {Baumbach}, A. and {Berry}, K.~L. and {Biscani}, F. and {Boquien}, M. and {Bostroem}, K.~A. and {Bouma}, L.~G. and {Brammer}, G.~B. and {Bray}, E.~M. and {Breytenbach}, H. and {Buddelmeijer}, H. and {Burke}, D.~J. and {Calderone}, G. and {Cano Rodr{\'\i}guez}, J.~L. and {Cara}, M. and {Cardoso}, J.~V.~M. and {Cheedella}, S. and {Copin}, Y. and {Corrales}, L. and {Crichton}, D. and {D'Avella}, D. and {Deil}, C. and {Depagne}, {\'E}. and {Dietrich}, J.~P. and {Donath}, A. and {Droettboom}, M. and {Earl}, N. and {Erben}, T. and {Fabbro}, S. and {Ferreira}, L.~A. and {Finethy}, T. and {Fox}, R.~T. and {Garrison}, L.~H. and {Gibbons}, S.~L.~J. and {Goldstein}, D.~A. and {Gommers}, R. and {Greco}, J.~P. and {Greenfield}, P. and {Groener}, A.~M. and {Grollier}, F. and {Hagen}, A. and {Hirst}, P. and {Homeier}, D. and {Horton}, A.~J. and {Hosseinzadeh}, G. and {Hu}, L. and {Hunkeler}, J.~S. and {Ivezi{\'c}}, {\v{Z}}. and {Jain}, A. and {Jenness}, T. and {Kanarek}, G. and {Kendrew}, S. and {Kern}, N.~S. and {Kerzendorf}, W.~E. and {Khvalko}, A. and {King}, J. and {Kirkby}, D. and {Kulkarni}, A.~M. and {Kumar}, A. and {Lee}, A. and {Lenz}, D. and {Littlefair}, S.~P. and {Ma}, Z. and {Macleod}, D.~M. and {Mastropietro}, M. and {McCully}, C. and {Montagnac}, S. and {Morris}, B.~M. and {Mueller}, M. and {Mumford}, S.~J. and {Muna}, D. and {Murphy}, N.~A. and {Nelson}, S. and {Nguyen}, G.~H. and {Ninan}, J.~P. and {N{\"o}the}, M. and {Ogaz}, S. and {Oh}, S. and {Parejko}, J.~K. and {Parley}, N. and {Pascual}, S. and {Patil}, R. and {Patil}, A.~A. and {Plunkett}, A.~L. and {Prochaska}, J.~X. and {Rastogi}, T. and {Reddy Janga}, V. and {Sabater}, J. and {Sakurikar}, P. and {Seifert}, M. and {Sherbert}, L.~E. and {Sherwood-Taylor}, H. and {Shih}, A.~Y. and {Sick}, J. and {Silbiger}, M.~T. and {Singanamalla}, S. and {Singer}, L.~P. and {Sladen}, P.~H. and {Sooley}, K.~A. and {Sornarajah}, S. and {Streicher}, O. and {Teuben}, P. and {Thomas}, S.~W. and {Tremblay}, G.~R. and {Turner}, J.~E.~H. and {Terr{\'o}n}, V. and {van Kerkwijk}, M.~H. and {de la Vega}, A. and {Watkins}, L.~L. and {Weaver}, B.~A. and {Whitmore}, J.~B. and {Woillez}, J. and {Zabalza}, V. and {Astropy Contributors}},
        title = "{The Astropy Project: Building an Open-science Project and Status of the v2.0 Core Package}",
      journal = {\aj},
         year = 2018,
        month = sep,
       volume = {156},
       number = {3},
          eid = {123},
        pages = {123},
          doi = {10.3847/1538-3881/aabc4f},
archivePrefix = {arXiv},
       eprint = {1801.02634},
 primaryClass = {astro-ph.IM},
       adsurl = {https://ui.adsabs.harvard.edu/abs/2018AJ....156..123A}
}

@ARTICLE{hunter07MATPLOTLIB,
       author = {{Hunter}, John D.},
        title = "{Matplotlib: A 2D Graphics Environment}",
      journal = {Computing in Science and Engineering},
         year = 2007,
        month = may,
       volume = {9},
       number = {3},
        pages = {90-95},
          doi = {10.1109/MCSE.2007.55},
       adsurl = {https://ui.adsabs.harvard.edu/abs/2007CSE.....9...90H}
}

@ARTICLE{harris20NUMPY,
       author = {{Harris}, Charles R. and {Millman}, K. Jarrod and {van der Walt}, St{\'e}fan J. and {Gommers}, Ralf and {Virtanen}, Pauli and {Cournapeau}, David and {Wieser}, Eric and {Taylor}, Julian and {Berg}, Sebastian and {Smith}, Nathaniel J. and {Kern}, Robert and {Picus}, Matti and {Hoyer}, Stephan and {van Kerkwijk}, Marten H. and {Brett}, Matthew and {Haldane}, Allan and {del R{\'\i}o}, Jaime Fern{\'a}ndez and {Wiebe}, Mark and {Peterson}, Pearu and {G{\'e}rard-Marchant}, Pierre and {Sheppard}, Kevin and {Reddy}, Tyler and {Weckesser}, Warren and {Abbasi}, Hameer and {Gohlke}, Christoph and {Oliphant}, Travis E.},
        title = "{Array programming with NumPy}",
      journal = {\nat},
         year = 2020,
        month = sep,
       volume = {585},
       number = {7825},
        pages = {357-362},
          doi = {10.1038/s41586-020-2649-2},
archivePrefix = {arXiv},
       eprint = {2006.10256},
 primaryClass = {cs.MS},
       adsurl = {https://ui.adsabs.harvard.edu/abs/2020Natur.585..357H}
}

@ARTICLE{Rajpurohit21c,
       author = {{Rajpurohit}, K. and {Wittor}, D. and {van Weeren}, R.~J. and {Vazza}, F. and {Hoeft}, M. and {Rudnick}, L. and {Locatelli}, N. and {Eilek}, J. and {Forman}, W.~R. and {Bonafede}, A. and {Bonnassieux}, E. and {Riseley}, C.~J. and {Brienza}, M. and {Brunetti}, G. and {Br{\"u}ggen}, M. and {Loi}, F. and {Rajpurohit}, A.~S. and {R{\"o}ttgering}, H.~J.~A. and {Botteon}, A. and {Clarke}, T.~E. and {Drabent}, A. and {Dom{\'\i}nguez-Fern{\'a}ndez}, P. and {Di Gennaro}, G. and {Gastaldello}, F.},
        title = "{Understanding the radio relic emission in the galaxy cluster MACS J0717.5+3745: Spectral analysis}",
      journal = {\aap},
         year = 2021,
        month = feb,
       volume = {646},
          eid = {A56},
        pages = {A56},
          doi = {10.1051/0004-6361/202039428},
archivePrefix = {arXiv},
       eprint = {2011.14436},
 primaryClass = {astro-ph.GA},
       adsurl = {https://ui.adsabs.harvard.edu/abs/2021A&A...646A..56R}
}

@ARTICLE{degasperin23,
       author = {{de Gasperin}, F. and {Edler}, H.~W. and {Williams}, W.~L. and {Callingham}, J.~R. and {Asabere}, B. and {Br{\"u}ggen}, M. and {Brunetti}, G. and {Dijkema}, T.~J. and {Hardcastle}, M.~J. and {Iacobelli}, M. and {Offringa}, A. and {Norden}, M.~J. and {R{\"o}ttgering}, H.~J.~A. and {Shimwell}, T. and {van Weeren}, R.~J. and {Tasse}, C. and {Bomans}, D.~J. and {Bonafede}, A. and {Botteon}, A. and {Cassano}, R. and {Chy{\.z}y}, K.~T. and {Cuciti}, V. and {Emig}, K.~L. and {Kadler}, M. and {Miley}, G. and {Mingo}, B. and {Oei}, M.~S.~S.~L. and {Prandoni}, I. and {Schwarz}, D.~J. and {Zarka}, P.},
        title = "{The LOFAR LBA Sky Survey. II. First data release}",
      journal = {\aap},
         year = 2023,
        month = may,
       volume = {673},
          eid = {A165},
        pages = {A165},
          doi = {10.1051/0004-6361/202245389},
archivePrefix = {arXiv},
       eprint = {2301.12724},
 primaryClass = {astro-ph.IM},
       adsurl = {https://ui.adsabs.harvard.edu/abs/2023A&A...673A.165D}
}

@ARTICLE{bruno23b,
       author = {{Bruno}, L. and {Botteon}, A. and {Shimwell}, T. and {Cuciti}, V. and {de Gasperin}, F. and {Brunetti}, G. and {Dallacasa}, D. and {Gastaldello}, F. and {Rossetti}, M. and {van Weeren}, R.~J. and {Venturi}, T. and {Russo}, S.~A. and {Taffoni}, G. and {Cassano}, R. and {Biava}, N. and {Lusetti}, G. and {Bonafede}, A. and {Ghizzardi}, S. and {De Grandi}, S.},
        title = "{A three-component giant radio halo: The puzzling case of the galaxy cluster Abell 2142}",
      journal = {\aap},
         year = 2023,
        month = oct,
       volume = {678},
          eid = {A133},
        pages = {A133},
          doi = {10.1051/0004-6361/202347245},
archivePrefix = {arXiv},
       eprint = {2308.07603},
 primaryClass = {astro-ph.CO},
       adsurl = {https://ui.adsabs.harvard.edu/abs/2023A&A...678A.133B}
}

@ARTICLE{katz-stone93,
       author = {{Katz-Stone}, Debora M. and {Rudnick}, Lawrence and {Anderson}, Martha C.},
        title = "{Determining the Shape of Spectra in Extended Radio Sources}",
      journal = {\apj},
         year = 1993,
        month = apr,
       volume = {407},
        pages = {549},
          doi = {10.1086/172536},
       adsurl = {https://ui.adsabs.harvard.edu/abs/1993ApJ...407..549K}
}

@ARTICLE{lacyVLASS20,
       author = {{Lacy}, M. and {Baum}, S.~A. and {Chandler}, C.~J. and {Chatterjee}, S. and {Clarke}, T.~E. and {Deustua}, S. and {English}, J. and {Farnes}, J. and {Gaensler}, B.~M. and {Gugliucci}, N. and {Hallinan}, G. and {Kent}, B.~R. and {Kimball}, A. and {Law}, C.~J. and {Lazio}, T.~J.~W. and {Marvil}, J. and {Mao}, S.~A. and {Medlin}, D. and {Mooley}, K. and {Murphy}, E.~J. and {Myers}, S. and {Osten}, R. and {Richards}, G.~T. and {Rosolowsky}, E. and {Rudnick}, L. and {Schinzel}, F. and {Sivakoff}, G.~R. and {Sjouwerman}, L.~O. and {Taylor}, R. and {White}, R.~L. and {Wrobel}, J. and {Andernach}, H. and {Beasley}, A.~J. and {Berger}, E. and {Bhatnager}, S. and {Birkinshaw}, M. and {Bower}, G.~C. and {Brandt}, W.~N. and {Brown}, S. and {Burke-Spolaor}, S. and {Butler}, B.~J. and {Comerford}, J. and {Demorest}, P.~B. and {Fu}, H. and {Giacintucci}, S. and {Golap}, K. and {G{\"u}th}, T. and {Hales}, C.~A. and {Hiriart}, R. and {Hodge}, J. and {Horesh}, A. and {Ivezi{\'c}}, {\v{Z}}. and {Jarvis}, M.~J. and {Kamble}, A. and {Kassim}, N. and {Liu}, X. and {Loinard}, L. and {Lyons}, D.~K. and {Masters}, J. and {Mezcua}, M. and {Moellenbrock}, G.~A. and {Mroczkowski}, T. and {Nyland}, K. and {O'Dea}, C.~P. and {O'Sullivan}, S.~P. and {Peters}, W.~M. and {Radford}, K. and {Rao}, U. and {Robnett}, J. and {Salcido}, J. and {Shen}, Y. and {Sobotka}, A. and {Witz}, S. and {Vaccari}, M. and {van Weeren}, R.~J. and {Vargas}, A. and {Williams}, P.~K.~G. and {Yoon}, I.},
        title = "{The Karl G. Jansky Very Large Array Sky Survey (VLASS). Science Case and Survey Design}",
      journal = {\pasp},
         year = 2020,
        month = mar,
       volume = {132},
       number = {1009},
          eid = {035001},
        pages = {035001},
          doi = {10.1088/1538-3873/ab63eb},
archivePrefix = {arXiv},
       eprint = {1907.01981},
 primaryClass = {astro-ph.IM},
       adsurl = {https://ui.adsabs.harvard.edu/abs/2020PASP..132c5001L}
}

@INPROCEEDINGS{ds9,
       author = {{Joye}, W.~A. and {Mandel}, E.},
        title = "{New Features of SAOImage DS9}",
    booktitle = {Astronomical Data Analysis Software and Systems XII},
         year = 2003,
       editor = {{Payne}, H.~E. and {Jedrzejewski}, R.~I. and {Hook}, R.~N.},
       series = {Astronomical Society of the Pacific Conference Series},
       volume = {295},
        month = jan,
        pages = {489},
       adsurl = {https://ui.adsabs.harvard.edu/abs/2003ASPC..295..489J}
}

@ARTICLE{riseley24,
       author = {{Riseley}, C.~J. and {Bonafede}, A. and {Bruno}, L. and {Botteon}, A. and {Rossetti}, M. and {Biava}, N. and {Bonnassieux}, E. and {Loi}, F. and {Vernstrom}, T. and {Balboni}, M.},
        title = "{A ``MeerKAT-meets-LOFAR'' study of the complex multi-component (mini-)halo in the extreme sloshing cluster Abell 2142}",
      journal = {\aap},
         year = 2024,
        month = jun,
       volume = {686},
          eid = {A44},
        pages = {A44},
          doi = {10.1051/0004-6361/202348944},
archivePrefix = {arXiv},
       eprint = {2403.00414},
 primaryClass = {astro-ph.CO},
       adsurl = {https://ui.adsabs.harvard.edu/abs/2024A&A...686A..44R}
}

@ARTICLE{vazza&botteon24,
       author = {{Vazza}, Franco and {Botteon}, Andrea},
        title = "{The Seeding of Cosmic Ray Electrons by Cluster Radio Galaxies: A Review}",
      journal = {Galaxies},
         year = 2024,
        month = apr,
       volume = {12},
       number = {2},
          eid = {19},
        pages = {19},
          doi = {10.3390/galaxies12020019},
archivePrefix = {arXiv},
       eprint = {2403.16068},
 primaryClass = {astro-ph.GA},
       adsurl = {https://ui.adsabs.harvard.edu/abs/2024Galax..12...19V}
}

@ARTICLE{komissarov&gubanov94,
       author = {{Komissarov}, S.~S. and {Gubanov}, A.~G.},
        title = "{Relic radio galaxies: evolution of synchrotron spectrum}",
      journal = {\aap},
         year = 1994,
        month = may,
       volume = {285},
        pages = {27-43},
       adsurl = {https://ui.adsabs.harvard.edu/abs/1994A&A...285...27K}
}

@ARTICLE{Norris21,
       author = {{Norris}, Ray P. and {Intema}, Huib T. and {Kapi{\'n}ska}, Anna D. and {Koribalski}, B{\"a}rbel S. and {Lenc}, Emil and {Rudnick}, L. and {Alsaberi}, Rami Z.~E. and {Anderson}, Craig and {Anderson}, G.~E. and {Crawford}, E. and {Crocker}, Roland and {English}, Jayanne and {Filipovi{\'c}}, Miroslav D. and {Galvin}, Tim J. and {Hopkins}, Andrew M. and {Hurley-Walker}, Natasha and {Inoue}, Susumu and {Luken}, Kieran and {Macgregor}, Peter J. and {Manojlovi{\'c}}, Pero and {Marvil}, Josh and {O'Brien}, Andrew N. and {Park}, Laurence and {Raja}, Wasim and {Shobhana}, Devika and {Venturi}, Tiziana and {Collier}, Jordan D. and {Hale}, Catherine and {Hotan}, Aidan and {Moss}, Vanessa and {Whiting}, Matthew},
        title = "{Unexpected circular radio objects at high Galactic latitude}",
      journal = {\pasa},
         year = 2021,
        month = jan,
       volume = {38},
          eid = {e003},
        pages = {e003},
          doi = {10.1017/pasa.2020.52},
archivePrefix = {arXiv},
       eprint = {2006.14805},
 primaryClass = {astro-ph.GA},
       adsurl = {https://ui.adsabs.harvard.edu/abs/2021PASA...38....3N}
}

@ARTICLE{DeGasperin26,
       author = {{De Gasperin}, F. and {Br{\"u}ggen}, M. and {Pasini}, T. and {Andernach}, H. and {Ivleva}, A. and {Spingola}, C. and {B{\"o}ss}, L.~M. and {Callingham}, J. and {Hardcastle}, M. and {Di Gennaro}, G.},
        title = "{Single- and double-headed odd radio circles in the LOFAR surveys}",
      journal = {\aap},
         year = 2026,
        month = may,
       volume = {709},
          eid = {A154},
        pages = {A154},
          doi = {10.1051/0004-6361/202658988},
archivePrefix = {arXiv},
       eprint = {2602.18294},
 primaryClass = {astro-ph.GA},
       adsurl = {https://ui.adsabs.harvard.edu/abs/2026A&A...709A.154D}
}

@ARTICLE{Shimwell26LOTSS,
       author = {{Shimwell}, T.~W. and {Hardcastle}, M.~J. and {Tasse}, C. and {Drabent}, A. and {Botteon}, A. and {Williams}, W.~L. and {Best}, P.~N. and {R{\"o}ttgering}, H.~J.~A. and {Br{\"u}ggen}, M. and {Brunetti}, G. and {Callingham}, J.~R. and {Chy{\.z}y}, K.~T. and {Conway}, J.~E. and {De Gasperin}, F. and {Haverkorn}, M. and {Horellou}, C. and {Jackson}, N. and {Miley}, G.~K. and {Morabito}, L.~K. and {Morganti}, R. and {O'Sullivan}, S.~P. and {Schwarz}, D.~J. and {Smith}, D.~J.~B. and {van Weeren}, R.~J. and {Vedantham}, H.~K. and {White}, G.~J. and {Ahmadi}, A. and {Alegre}, L. and {Arias}, M. and {Asabere}, B. and {Bahr-Kalus}, B. and {Barkus}, B. and {Bilicki}, M. and {B{\"o}hme}, L. and {Brentjens}, M. and {Brienza}, M. and {Bomans}, D.~J. and {Bonafede}, A. and {Bonato}, M. and {Bonnassieux}, E. and {Boxelaar}, J.~M. and {Camera}, S. and {Cassano}, R. and {Chilufya}, J. and {Cianfaglione}, M. and {Croston}, J.~H. and {Cuciti}, V. and {Dabhade}, P. and {De Rubeis}, E. and {de Jong}, J.~M.~G.~H.~J. and {Dallacasa}, D. and {Dettmar}, R.~J. and {Duncan}, K.~J. and {Di Gennaro}, G. and {Edler}, H.~W. and {Groeneveld}, C. and {G{\"u}rkan}, G. and {Hajduk}, M. and {Hale}, C.~L. and {Heesen}, V. and {Hoang}, D.~N. and {Hoeft}, M. and {Holties}, H. and {Horton}, M.~A. and {Iacobelli}, M. and {Jamrozy}, M. and {Jarvis}, M.~J. and {Jelic}, V. and {Kadler}, M. and {Kondapally}, R. and {Kunert-Bajraszewska}, M. and {Loose}, M. and {Magliocchetti}, M. and {Ma{\l}ek}, K. and {Manzano}, C. and {McKean}, J.~P. and {Mevius}, M. and {Mingo}, B. and {Miskolczi}, A. and {Misra}, A. and {Mold{\'o}n}, J. and {Nair}, D.~G. and {Nakoneczny}, S.~J. and {Orru}, E. and {Pashapour-Ahmadabadi}, M. and {Pasini}, T. and {Petley}, J. and {Pierce}, J.~C.~S. and {Prandoni}, I. and {Rafferty}, D. and {Rajpurohit}, K. and {Riseley}, C.~J. and {Roberts}, I.~D. and {Sethi}, S. and {Shulevski}, A. and {Stein}, M. and {Stuardi}, C. and {Sweijen}, F. and {ter Veen}, S. and {Timmerman}, R. and {Vaccari}, M. and {Wijnholds}, S.},
        title = "{The LOFAR Two-metre Sky Survey: VII. Third Data Release}",
      journal = {\aap},
         year = 2026,
        month = mar,
       volume = {707},
          eid = {A198},
        pages = {A198},
          doi = {10.1051/0004-6361/202557749},
archivePrefix = {arXiv},
       eprint = {2602.15949},
 primaryClass = {astro-ph.GA},
       adsurl = {https://ui.adsabs.harvard.edu/abs/2026A&A...707A.198S}
}

@ARTICLE{Coil24,
       author = {{Coil}, Alison L. and {Perrotta}, Serena and {Rupke}, David S.~N. and {Lochhaas}, Cassandra and {Tremonti}, Christy A. and {Diamond-Stanic}, Aleks and {Fielding}, Drummond and {Geach}, James E. and {Hickox}, Ryan C. and {Moustakas}, John and {Rudnick}, Gregory H. and {Sell}, Paul and {Whalen}, Kelly E.},
        title = "{Ionized gas extends over 40 kpc in an odd radio circle host galaxy}",
      journal = {\nat},
         year = 2024,
        month = jan,
       volume = {625},
       number = {7995},
        pages = {459-462},
          doi = {10.1038/s41586-023-06752-8},
archivePrefix = {arXiv},
       eprint = {2310.15162},
 primaryClass = {astro-ph.GA},
       adsurl = {https://ui.adsabs.harvard.edu/abs/2024Natur.625..459C}
}

@ARTICLE{Norris21b,
       author = {{Norris}, Ray P. and {Crawford}, Evan and {Macgregor}, Peter},
        title = "{Odd Radio Circles and Their Environment}",
      journal = {Galaxies},
         year = 2021,
        month = oct,
       volume = {9},
       number = {4},
          eid = {83},
        pages = {83},
          doi = {10.3390/galaxies9040083},
archivePrefix = {arXiv},
       eprint = {2111.01269},
 primaryClass = {astro-ph.GA},
       adsurl = {https://ui.adsabs.harvard.edu/abs/2021Galax...9...83N}
}

@ARTICLE{Koribalski21,
       author = {{Koribalski}, B{\"a}rbel S. and {Norris}, Ray P. and {Andernach}, Heinz and {Rudnick}, Lawrence and {Shabala}, Stanislav and {Filipovi{\'c}}, Miroslav and {Lenc}, Emil},
        title = "{Discovery of a new extragalactic circular radio source with ASKAP: ORC J0102-2450}",
      journal = {\mnras},
         year = 2021,
        month = jul,
       volume = {505},
       number = {1},
        pages = {L11-L15},
          doi = {10.1093/mnrasl/slab041},
archivePrefix = {arXiv},
       eprint = {2104.13055},
 primaryClass = {astro-ph.GA},
       adsurl = {https://ui.adsabs.harvard.edu/abs/2021MNRAS.505L..11K}
}

@ARTICLE{Koribalski24b,
       author = {{Koribalski}, B{\"a}rbel S. and {Veronica}, Angie and {Dolag}, Klaus and {Reiprich}, Thomas H. and {Br{\"u}ggen}, Marcus and {Heywood}, Ian and {Andernach}, Heinz and {Dettmar}, Ralf-J{\"u}rgen and {Hoeft}, Matthias and {Zhang}, Xiaoyuan and {Bulbul}, Esra and {Garrel}, Christian and {J{\'o}zsa}, Gyula I.~G. and {English}, Jayanne},
        title = "{MeerKAT discovery of a double radio relic and odd radio circle: connecting cluster and galaxy merger shocks}",
      journal = {\mnras},
         year = 2024,
        month = jul,
       volume = {531},
       number = {3},
        pages = {3357-3372},
          doi = {10.1093/mnras/stae1254},
archivePrefix = {arXiv},
       eprint = {2304.11784},
 primaryClass = {astro-ph.GA},
       adsurl = {https://ui.adsabs.harvard.edu/abs/2024MNRAS.531.3357K}
}

@ARTICLE{Koribalski24c,
       author = {{Koribalski}, B{\"a}rbel S. and {Khabibullin}, Ildar and {Dolag}, Klaus and {Churazov}, Eugene and {Norris}, Ray P. and {Carretti}, Ettore and {Hopkins}, Andrew M. and {Vernstrom}, Tessa and {Shabala}, Stanislav S. and {Gupta}, Nikhel},
        title = "{The Physalis system: discovery of ORC-like radio shells around a massive pair of interacting early-type galaxies with offset X-ray emission}",
      journal = {\mnras},
         year = 2024,
        month = aug,
       volume = {532},
       number = {4},
        pages = {3682-3693},
          doi = {10.1093/mnras/stae1669},
archivePrefix = {arXiv},
       eprint = {2404.09522},
 primaryClass = {astro-ph.GA},
       adsurl = {https://ui.adsabs.harvard.edu/abs/2024MNRAS.532.3682K}
}

@ARTICLE{Norris25,
       author = {{Norris}, Ray P. and {Koribalski}, B{\"a}rbel S. and {Hale}, Catherine L. and {Jarvis}, Matt J. and {Macgregor}, Peter J. and {Taylor}, A. Russell},
        title = "{MeerKAT discovery of a MIGHTEE Odd Radio Circle}",
      journal = {\mnras},
         year = 2025,
        month = feb,
       volume = {537},
       number = {1},
        pages = {L42-L48},
          doi = {10.1093/mnrasl/slae114},
archivePrefix = {arXiv},
       eprint = {2411.17311},
 primaryClass = {astro-ph.GA},
       adsurl = {https://ui.adsabs.harvard.edu/abs/2025MNRAS.537L..42N}
}

@ARTICLE{Gupta25,
       author = {{Gupta}, Nikhel and {Norris}, Ray P. and {Hayder}, Zeeshan and {Huynh}, Minh and {Andernach}, Heinz and {Hopkins}, Andrew and {Shabala}, Stanislav and {Rudnick}, Lawrence and {Filipovi{\'c}}, Miroslav D. and {Koribalski}, B{\"a}rbel Silvia and {Petersson}, Lars and {Wang}, Rosalind},
        title = "{Discovery of odd radio circles and other peculiars in the first year of the EMU survey using object detection}",
      journal = {\pasa},
         year = 2025,
        month = jun,
       volume = {42},
          eid = {e097},
        pages = {e097},
          doi = {10.1017/pasa.2025.10061},
archivePrefix = {arXiv},
       eprint = {2506.08439},
 primaryClass = {astro-ph.GA},
       adsurl = {https://ui.adsabs.harvard.edu/abs/2025PASA...42...97G}
}

@ARTICLE{Taziaux25,
       author = {{Taziaux}, Sam and {Bomans}, Dominik J. and {Riseley}, Christopher J. and {Thomson}, Alec J.~M. and {Norris}, Ray P. and {Basu}, Aritra and {Heald}, George H. and {Galvin}, Timothy J. and {Adebahr}, Bj{\"o}rn and {Filipovi{\'c}}, Miroslav D. and {Gupta}, Nikhel and {Shabala}, Stas and {Zafar}, Tayyaba},
        title = "{Deep polarimetry study reveals double ring odd radio circle-like structures}",
      journal = {\aap},
         year = 2025,
        month = oct,
       volume = {702},
          eid = {A219},
        pages = {A219},
          doi = {10.1051/0004-6361/202556065},
archivePrefix = {arXiv},
       eprint = {2509.04981},
 primaryClass = {astro-ph.GA},
       adsurl = {https://ui.adsabs.harvard.edu/abs/2025A&A...702A.219T}
}

@ARTICLE{Hota25,
       author = {{Hota}, Ananda and {Dabhade}, Pratik and {Machado}, Prasun and {Das}, Joydeep and {Muley}, Aarti and {Purohit}, Arundhati},
        title = "{RAD@home discovery of extragalactic radio rings and odd radio circles: clues to their origins}",
      journal = {\mnras},
         year = 2025,
        month = oct,
       volume = {543},
       number = {2},
        pages = {1048-1057},
          doi = {10.1093/mnras/staf1531},
archivePrefix = {arXiv},
       eprint = {2510.01999},
 primaryClass = {astro-ph.GA},
       adsurl = {https://ui.adsabs.harvard.edu/abs/2025MNRAS.543.1048H}
}

@ARTICLE{Norris22,
       author = {{Norris}, Ray P. and {Collier}, J.~D. and {Crocker}, Roland M. and {Heywood}, Ian and {Macgregor}, Peter and {Rudnick}, L. and {Shabala}, Stas and {Andernach}, Heinz and {da Cunha}, Elisabete and {English}, Jayanne and {Filipovi{\'c}}, Miroslav and {Koribalski}, B{\"a}rbel S. and {Luken}, Kieran and {Robotham}, Aaron and {Sekhar}, Srikrishna and {Thorne}, Jessica E. and {Vernstrom}, Tessa},
        title = "{MeerKAT uncovers the physics of an odd radio circle}",
      journal = {\mnras},
         year = 2022,
        month = jun,
       volume = {513},
       number = {1},
        pages = {1300-1316},
          doi = {10.1093/mnras/stac701},
archivePrefix = {arXiv},
       eprint = {2203.10669},
 primaryClass = {astro-ph.GA},
       adsurl = {https://ui.adsabs.harvard.edu/abs/2022MNRAS.513.1300N}
}

@ARTICLE{Rupke24,
       author = {{Rupke}, David S.~N. and {Coil}, Alison L. and {Whalen}, Kelly E. and {Moustakas}, John and {Tremonti}, Christy A. and {Perrotta}, Serena},
        title = "{The Intrinsic Sizes of Odd Radio Circles}",
      journal = {\apj},
         year = 2024,
        month = may,
       volume = {967},
       number = {1},
          eid = {51},
        pages = {51},
          doi = {10.3847/1538-4357/ad3934},
archivePrefix = {arXiv},
       eprint = {2312.06387},
 primaryClass = {astro-ph.GA},
       adsurl = {https://ui.adsabs.harvard.edu/abs/2024ApJ...967...51R}
}

@ARTICLE{Lochner23,
       author = {{Lochner}, M. and {Rudnick}, L. and {Heywood}, I. and {Knowles}, K. and {Shabala}, S.~S.},
        title = "{A unique, ring-like radio source with quadrilateral structure detected with machine learning}",
      journal = {\mnras},
         year = 2023,
        month = mar,
       volume = {520},
       number = {1},
        pages = {1439-1446},
          doi = {10.1093/mnras/stad074},
archivePrefix = {arXiv},
       eprint = {2211.02062},
 primaryClass = {astro-ph.GA},
       adsurl = {https://ui.adsabs.harvard.edu/abs/2023MNRAS.520.1439L}
}

@ARTICLE{Bulbul24,
       author = {{Bulbul}, E. and {Zhang}, X. and {Kluge}, M. and {Br{\"u}ggen}, M. and {Koribalski}, B. and {Liu}, A. and {Artis}, E. and {Bahar}, Y.~E. and {Balzer}, F. and {Garrel}, C. and et al.},
        title = "{The galaxy group merger origin of the Cloverleaf odd radio circle system}",
      journal = {\aap},
         year = 2024,
        month = may,
       volume = {685},
          eid = {L2},
        pages = {L2},
          doi = {10.1051/0004-6361/202449900},
archivePrefix = {arXiv},
       eprint = {2403.09808},
 primaryClass = {astro-ph.GA},
       adsurl = {https://ui.adsabs.harvard.edu/abs/2024A&A...685L...2B}
}

@ARTICLE{Shabala24,
       author = {{Shabala}, S.~S. and {Yates-Jones}, P.~M. and {Jerrim}, L.~A. and {Turner}, R.~J. and {Krause}, M.~G.~H. and {Norris}, R.~P. and {Koribalski}, B.~S. and {Filipovi{\'c}}, M. and {Rudnick}, L. and {Power}, C. and et al.},
        title = "{Are Odd Radio Circles phoenixes of powerful radio galaxies?}",
      journal = {\pasa},
         year = 2024,
        month = jan,
       volume = {41},
          eid = {e024},
        pages = {e024},
          doi = {10.1017/pasa.2024.11},
archivePrefix = {arXiv},
       eprint = {2402.09708},
 primaryClass = {astro-ph.GA},
       adsurl = {https://ui.adsabs.harvard.edu/abs/2024PASA...41...24S}
}

@ARTICLE{Dolag23,
       author = {{Dolag}, Klaus and {B{\"o}ss}, Ludwig M. and {Koribalski}, B{\"a}rbel S. and {Steinwandel}, Ulrich P. and {Valentini}, Milena},
        title = "{Insights on the Origin of Odd Radio Circles from Cosmological Simulations}",
      journal = {\apj},
         year = 2023,
        month = mar,
       volume = {945},
       number = {1},
          eid = {74},
        pages = {74},
          doi = {10.3847/1538-4357/acb5f5},
archivePrefix = {arXiv},
       eprint = {2208.15003},
 primaryClass = {astro-ph.GA},
       adsurl = {https://ui.adsabs.harvard.edu/abs/2023ApJ...945...74D}
}

@ARTICLE{Omar22,
       author = {{Omar}, Amitesh},
        title = "{On energetics and progenitors of odd radio circles: a causal connection with tidal disruption of stars?}",
      journal = {\mnras},
         year = 2022,
        month = oct,
       volume = {516},
       number = {1},
        pages = {L43-L47},
          doi = {10.1093/mnrasl/slac081},
archivePrefix = {arXiv},
       eprint = {2204.08427},
 primaryClass = {astro-ph.GA},
       adsurl = {https://ui.adsabs.harvard.edu/abs/2022MNRAS.516L..43O}
}

@ARTICLE{Yamasaki24,
       author = {{Yamasaki}, Shotaro and {Sarkar}, Kartick C. and {Li}, Zhaozhou},
        title = "{Are odd radio circles virial shocks around massive galaxies? Implications for cosmic-ray diffusion in the circumgalactic medium}",
      journal = {\mnras},
         year = 2024,
        month = feb,
       volume = {528},
       number = {2},
        pages = {3854-3863},
          doi = {10.1093/mnras/stae281},
archivePrefix = {arXiv},
       eprint = {2309.17451},
 primaryClass = {astro-ph.HE},
       adsurl = {https://ui.adsabs.harvard.edu/abs/2024MNRAS.528.3854Y}
}

@ARTICLE{Ivleva26,
       author = {{Ivleva}, Anna and {B{\"o}ss}, Ludwig M. and {Dolag}, Klaus and {Koribalski}, B{\"a}rbel S. and {Khabibullin}, Ildar},
        title = "{Cosmological zoom-in simulation of odd radio circles as merger-driven shocks in galaxy groups}",
      journal = {\aap},
         year = 2026,
        month = feb,
       volume = {706},
          eid = {A80},
        pages = {A80},
          doi = {10.1051/0004-6361/202556716},
archivePrefix = {arXiv},
       eprint = {2508.00981},
 primaryClass = {astro-ph.GA},
       adsurl = {https://ui.adsabs.harvard.edu/abs/2026A&A...706A..80I}
}

@ARTICLE{Alam15,
       author = {{Alam}, Shadab and {Albareti}, Franco D. and {Allende Prieto}, Carlos and {Anders}, F. and {Anderson}, Scott F. and {Anderton}, Timothy and {Andrews}, Brett H. and {Armengaud}, Eric and {Aubourg}, {\'E}ric and {Bailey}, Stephen and {Basu}, Sarbani and {Bautista}, Julian E. and {Beaton}, Rachael L. and {Beers}, Timothy C. and {Bender}, Chad F. and {Berlind}, Andreas A. and {Beutler}, Florian and {Bhardwaj}, Vaishali and {Bird}, Jonathan C. and {Bizyaev}, Dmitry and {Blake}, Cullen H. and {Blanton}, Michael R. and {Blomqvist}, Michael and {Bochanski}, John J. and {Bolton}, Adam S. and {Bovy}, Jo and {Shelden Bradley}, A. and {Brandt}, W.~N. and {Brauer}, D.~E. and {Brinkmann}, J. and {Brown}, Peter J. and {Brownstein}, Joel R. and {Burden}, Angela and {Burtin}, Etienne and {Busca}, Nicol{\'a}s G. and {Cai}, Zheng and {Capozzi}, Diego and {Carnero Rosell}, Aurelio and {Carr}, Michael A. and {Carrera}, Ricardo and {Chambers}, K.~C. and {Chaplin}, William James and {Chen}, Yen-Chi and {Chiappini}, Cristina and {Chojnowski}, S. Drew and {Chuang}, Chia-Hsun and {Clerc}, Nicolas and {Comparat}, Johan and {Covey}, Kevin and {Croft}, Rupert A.~C. and {Cuesta}, Antonio J. and {Cunha}, Katia and {da Costa}, Luiz N. and {Da Rio}, Nicola and {Davenport}, James R.~A. and {Dawson}, Kyle S. and {De Lee}, Nathan and {Delubac}, Timoth{\'e}e and {Deshpande}, Rohit and {Dhital}, Saurav and {Dutra-Ferreira}, Let{\'\i}cia and {Dwelly}, Tom and {Ealet}, Anne and {Ebelke}, Garrett L. and {Edmondson}, Edward M. and {Eisenstein}, Daniel J. and {Ellsworth}, Tristan and {Elsworth}, Yvonne and {Epstein}, Courtney R. and {Eracleous}, Michael and {Escoffier}, Stephanie and {Esposito}, Massimiliano and {Evans}, Michael L. and {Fan}, Xiaohui and {Fern{\'a}ndez-Alvar}, Emma and {Feuillet}, Diane and {Filiz Ak}, Nurten and {Finley}, Hayley and {Finoguenov}, Alexis and {Flaherty}, Kevin and {Fleming}, Scott W. and {Font-Ribera}, Andreu and {Foster}, Jonathan and {Frinchaboy}, Peter M. and {Galbraith-Frew}, J.~G. and {Garc{\'\i}a}, Rafael A. and {Garc{\'\i}a-Hern{\'a}ndez}, D.~A. and {Garc{\'\i}a P{\'e}rez}, Ana E. and {Gaulme}, Patrick and {Ge}, Jian and {G{\'e}nova-Santos}, R. and {Georgakakis}, A. and {Ghezzi}, Luan and {Gillespie}, Bruce A. and {Girardi}, L{\'e}o and {Goddard}, Daniel and {Gontcho}, Satya Gontcho A. and {Gonz{\'a}lez Hern{\'a}ndez}, Jonay I. and {Grebel}, Eva K. and {Green}, Paul J. and {Grieb}, Jan Niklas and {Grieves}, Nolan and {Gunn}, James E. and {Guo}, Hong and {Harding}, Paul and {Hasselquist}, Sten and {Hawley}, Suzanne L. and {Hayden}, Michael and {Hearty}, Fred R. and {Hekker}, Saskia and {Ho}, Shirley and {Hogg}, David W. and {Holley-Bockelmann}, Kelly and {Holtzman}, Jon A. and {Honscheid}, Klaus and {Huber}, Daniel and {Huehnerhoff}, Joseph and {Ivans}, Inese I. and {Jiang}, Linhua and {Johnson}, Jennifer A. and {Kinemuchi}, Karen and {Kirkby}, David and {Kitaura}, Francisco and {Klaene}, Mark A. and {Knapp}, Gillian R. and {Kneib}, Jean-Paul and {Koenig}, Xavier P. and {Lam}, Charles R. and {Lan}, Ting-Wen and {Lang}, Dustin and {Laurent}, Pierre and {Le Goff}, Jean-Marc and {Leauthaud}, Alexie and {Lee}, Khee-Gan and {Lee}, Young Sun and {Licquia}, Timothy C. and {Liu}, Jian and {Long}, Daniel C. and {L{\'o}pez-Corredoira}, Mart{\'\i}n and {Lorenzo-Oliveira}, Diego and {Lucatello}, Sara and {Lundgren}, Britt and {Lupton}, Robert H. and {Mack}, III, Claude E. and {Mahadevan}, Suvrath and {Maia}, Marcio A.~G. and {Majewski}, Steven R. and {Malanushenko}, Elena and {Malanushenko}, Viktor and {Manchado}, A. and {Manera}, Marc and {Mao}, Qingqing and {Maraston}, Claudia and {Marchwinski}, Robert C. and {Margala}, Daniel and {Martell}, Sarah L. and {Martig}, Marie and {Masters}, Karen L. and {Mathur}, Savita and {McBride}, Cameron K. and {McGehee}, Peregrine M. and {McGreer}, Ian D. and {McMahon}, Richard G. and {M{\'e}nard}, Brice and {Menzel}, Marie-Luise and {Merloni}, Andrea and {M{\'e}sz{\'a}ros}, Szabolcs and {Miller}, Adam A. and {Miralda-Escud{\'e}}, Jordi and {Miyatake}, Hironao and {Montero-Dorta}, Antonio D. and {More}, Surhud and {Morganson}, Eric and {Morice-Atkinson}, Xan and {Morrison}, Heather L. and {Mosser}, Ben{\^o}it and {Muna}, Demitri and {Myers}, Adam D. and {Nandra}, Kirpal and {Newman}, Jeffrey A. and {Neyrinck}, Mark and {Nguyen}, Duy Cuong and {Nichol}, Robert C. and {Nidever}, David L. and {Noterdaeme}, Pasquier and {Nuza}, Sebasti{\'a}n E. and {O'Connell}, Julia E. and {O'Connell}, Robert W. and {O'Connell}, Ross and {Ogando}, Ricardo L.~C. and {Olmstead}, Matthew D. and {Oravetz}, Audrey E. and {Oravetz}, Daniel J. and {Osumi}, Keisuke and {Owen}, Russell and {Padgett}, Deborah L. and {Padmanabhan}, Nikhil and {Paegert}, Martin and {Palanque-Delabrouille}, Nathalie and {Pan}, Kaike},
        title = "{The Eleventh and Twelfth Data Releases of the Sloan Digital Sky Survey: Final Data from SDSS-III}",
      journal = {\apjs},
         year = 2015,
        month = jul,
       volume = {219},
       number = {1},
          eid = {12},
        pages = {12},
          doi = {10.1088/0067-0049/219/1/12},
archivePrefix = {arXiv},
       eprint = {1501.00963},
 primaryClass = {astro-ph.IM},
       adsurl = {https://ui.adsabs.harvard.edu/abs/2015ApJS..219...12A}
}

@software{Mohan&Rafferty15,
       author = {{Mohan}, Niruj and {Rafferty}, David},
        title = "{PyBDSF: Python Blob Detection and Source Finder}",
 howpublished = {Astrophysics Source Code Library, record ascl:1502.007},
         year = 2015,
        month = feb,
          eid = {ascl:1502.007},
archivePrefix = {ascl},
       eprint = {1502.007},
       adsurl = {https://ui.adsabs.harvard.edu/abs/2015ascl.soft02007M}
}

@ARTICLE{CASAteam22,
       author = {{CASA Team} and {Bean}, Ben and {Bhatnagar}, Sanjay and {Castro}, Sandra and {Donovan Meyer}, Jennifer and {Emonts}, Bjorn and {Garcia}, Enrique and {Garwood}, Robert and {Golap}, Kumar and {Gonzalez Villalba}, Justo and {Harris}, Pamela and {Hayashi}, Yohei and {Hoskins}, Josh and {Hsieh}, Mingyu and {Jagannathan}, Preshanth and {Kawasaki}, Wataru and {Keimpema}, Aard and {Kettenis}, Mark and {Lopez}, Jorge and {Marvil}, Joshua and {Masters}, Joseph and {McNichols}, Andrew and {Mehringer}, David and {Miel}, Renaud and {Moellenbrock}, George and {Montesino}, Federico and {Nakazato}, Takeshi and {Ott}, Juergen and {Petry}, Dirk and {Pokorny}, Martin and {Raba}, Ryan and {Rau}, Urvashi and {Schiebel}, Darrell and {Schweighart}, Neal and {Sekhar}, Srikrishna and {Shimada}, Kazuhiko and {Small}, Des and {Steeb}, Jan-Willem and {Sugimoto}, Kanako and {Suoranta}, Ville and {Tsutsumi}, Takahiro and {van Bemmel}, Ilse M. and {Verkouter}, Marjolein and {Wells}, Akeem and {Xiong}, Wei and {Szomoru}, Arpad and {Griffith}, Morgan and {Glendenning}, Brian and {Kern}, Jeff},
        title = "{CASA, the Common Astronomy Software Applications for Radio Astronomy}",
      journal = {\pasp},
         year = 2022,
        month = nov,
       volume = {134},
       number = {1041},
          eid = {114501},
        pages = {114501},
          doi = {10.1088/1538-3873/ac9642},
archivePrefix = {arXiv},
       eprint = {2210.02276},
 primaryClass = {astro-ph.IM},
       adsurl = {https://ui.adsabs.harvard.edu/abs/2022PASP..134k4501C}
}

@ARTICLE{Bell78,
       author = {{Bell}, A.~R.},
        title = "{The acceleration of cosmic rays in shock fronts - I.}",
      journal = {\mnras},
         year = 1978,
        month = jan,
       volume = {182},
        pages = {147-156},
          doi = {10.1093/mnras/182.2.147},
       adsurl = {https://ui.adsabs.harvard.edu/abs/1978MNRAS.182..147B}
}

@ARTICLE{Blandford&Ostriker78,
       author = {{Blandford}, R.~D. and {Ostriker}, J.~P.},
        title = "{Particle acceleration by astrophysical shocks.}",
      journal = {\apjl},
         year = 1978,
        month = apr,
       volume = {221},
        pages = {L29-L32},
          doi = {10.1086/182658},
       adsurl = {https://ui.adsabs.harvard.edu/abs/1978ApJ...221L..29B}
}

@ARTICLE{Wang&Heinz26,
       author = {{Wang}, Yiting and {Heinz}, Sebastian},
        title = "{Odd Radio Circles Modeled by Shock-bubble Interactions}",
      journal = {\apj},
         year = 2026,
        month = apr,
       volume = {1001},
       number = {2},
          eid = {232},
        pages = {232},
          doi = {10.3847/1538-4357/ae563c},
       adsurl = {https://ui.adsabs.harvard.edu/abs/2026ApJ..1001..232W}
}

@ARTICLE{Lal20a,
       author = {{Lal}, Dharam V.},
        title = "{Upgraded GMRT Observations of the Coma Cluster of Galaxies: The Observations}",
      journal = {\apjs},
         year = 2020,
        month = sep,
       volume = {250},
       number = {1},
          eid = {22},
        pages = {22},
          doi = {10.3847/1538-4365/abacd2},
archivePrefix = {arXiv},
       eprint = {2009.07145},
 primaryClass = {astro-ph.GA},
       adsurl = {https://ui.adsabs.harvard.edu/abs/2020ApJS..250...22L}
}

@ARTICLE{Lal22,
       author = {{Lal}, D.~V. and {Lyskova}, N. and {Zhang}, C. and {Venturi}, T. and {Forman}, W.~R. and {Jones}, C. and {Churazov}, E.~M. and {van Weeren}, R.~J. and {Bonafede}, A. and {Miller}, N.~A. and {Roberts}, I.~D. and {Bykov}, A.~M. and {Di Mascolo}, L. and {Br{\"u}ggen}, M. and {Brunetti}, G.},
        title = "{High-resolution, High-sensitivity, Low-frequency uGMRT View of Coma Cluster of Galaxies}",
      journal = {\apj},
         year = 2022,
        month = aug,
       volume = {934},
       number = {2},
          eid = {170},
        pages = {170},
          doi = {10.3847/1538-4357/ac7a9b},
archivePrefix = {arXiv},
       eprint = {2207.06624},
 primaryClass = {astro-ph.GA},
       adsurl = {https://ui.adsabs.harvard.edu/abs/2022ApJ...934..170L}
}

@ARTICLE{Wittor21,
       author = {{Wittor}, D. and {Ettori}, S. and {Vazza}, F. and {Rajpurohit}, K. and {Hoeft}, M. and {Dom{\'\i}nguez-Fern{\'a}ndez}, P.},
        title = "{Exploring the spectral properties of radio relics - I: integrated spectral index and Mach number}",
      journal = {\mnras},
         year = 2021,
        month = sep,
       volume = {506},
       number = {1},
        pages = {396-414},
          doi = {10.1093/mnras/stab1735},
archivePrefix = {arXiv},
       eprint = {2106.08351},
 primaryClass = {astro-ph.CO},
       adsurl = {https://ui.adsabs.harvard.edu/abs/2021MNRAS.506..396W}
}

@ARTICLE{Dey19_DESI,
       author = {{Dey}, Arjun and {Schlegel}, David J. and {Lang}, Dustin and {Blum}, Robert and {Burleigh}, Kaylan and {Fan}, Xiaohui and {Findlay}, Joseph R. and {Finkbeiner}, Doug and {Herrera}, David and {Juneau}, St{\'e}phanie and {Landriau}, Martin and {Levi}, Michael and {McGreer}, Ian and {Meisner}, Aaron and {Myers}, Adam D. and {Moustakas}, John and {Nugent}, Peter and {Patej}, Anna and {Schlafly}, Edward F. and {Walker}, Alistair R. and {Valdes}, Francisco and {Weaver}, Benjamin A. and {Y{\`e}che}, Christophe and {Zou}, Hu and {Zhou}, Xu and {Abareshi}, Behzad and {Abbott}, T.~M.~C. and {Abolfathi}, Bela and {Aguilera}, C. and {Alam}, Shadab and {Allen}, Lori and {Alvarez}, A. and {Annis}, James and {Ansarinejad}, Behzad and {Aubert}, Marie and {Beechert}, Jacqueline and {Bell}, Eric F. and {BenZvi}, Segev Y. and {Beutler}, Florian and {Bielby}, Richard M. and {Bolton}, Adam S. and {Brice{\~n}o}, C{\'e}sar and {Buckley-Geer}, Elizabeth J. and {Butler}, Karen and {Calamida}, Annalisa and {Carlberg}, Raymond G. and {Carter}, Paul and {Casas}, Ricard and {Castander}, Francisco J. and {Choi}, Yumi and {Comparat}, Johan and {Cukanovaite}, Elena and {Delubac}, Timoth{\'e}e and {DeVries}, Kaitlin and {Dey}, Sharmila and {Dhungana}, Govinda and {Dickinson}, Mark and {Ding}, Zhejie and {Donaldson}, John B. and {Duan}, Yutong and {Duckworth}, Christopher J. and {Eftekharzadeh}, Sarah and {Eisenstein}, Daniel J. and {Etourneau}, Thomas and {Fagrelius}, Parker A. and {Farihi}, Jay and {Fitzpatrick}, Mike and {Font-Ribera}, Andreu and {Fulmer}, Leah and {G{\"a}nsicke}, Boris T. and {Gaztanaga}, Enrique and {George}, Koshy and {Gerdes}, David W. and {Gontcho}, Satya Gontcho A. and {Gorgoni}, Claudio and {Green}, Gregory and {Guy}, Julien and {Harmer}, Diane and {Hernandez}, M. and {Honscheid}, Klaus and {Huang}, Lijuan Wendy and {James}, David J. and {Jannuzi}, Buell T. and {Jiang}, Linhua and {Joyce}, Richard and {Karcher}, Armin and {Karkar}, Sonia and {Kehoe}, Robert and {Kneib}, Jean-Paul and {Kueter-Young}, Andrea and {Lan}, Ting-Wen and {Lauer}, Tod R. and {Le Guillou}, Laurent and {Le Van Suu}, Auguste and {Lee}, Jae Hyeon and {Lesser}, Michael and {Perreault Levasseur}, Laurence and {Li}, Ting S. and {Mann}, Justin L. and {Marshall}, Robert and {Mart{\'\i}nez-V{\'a}zquez}, C.~E. and {Martini}, Paul and {du Mas des Bourboux}, H{\'e}lion and {McManus}, Sean and {Meier}, Tobias Gabriel and {M{\'e}nard}, Brice and {Metcalfe}, Nigel and {Mu{\~n}oz-Guti{\'e}rrez}, Andrea and {Najita}, Joan and {Napier}, Kevin and {Narayan}, Gautham and {Newman}, Jeffrey A. and {Nie}, Jundan and {Nord}, Brian and {Norman}, Dara J. and {Olsen}, Knut A.~G. and {Paat}, Anthony and {Palanque-Delabrouille}, Nathalie and {Peng}, Xiyan and {Poppett}, Claire L. and {Poremba}, Megan R. and {Prakash}, Abhishek and {Rabinowitz}, David and {Raichoor}, Anand and {Rezaie}, Mehdi and {Robertson}, A.~N. and {Roe}, Natalie A. and {Ross}, Ashley J. and {Ross}, Nicholas P. and {Rudnick}, Gregory and {Safonova}, Sasha and {Saha}, Abhijit and {S{\'a}nchez}, F. Javier and {Savary}, Elodie and {Schweiker}, Heidi and {Scott}, Adam and {Seo}, Hee-Jong and {Shan}, Huanyuan and {Silva}, David R. and {Slepian}, Zachary and {Soto}, Christian and {Sprayberry}, David and {Staten}, Ryan and {Stillman}, Coley M. and {Stupak}, Robert J. and {Summers}, David L. and {Sien Tie}, Suk and {Tirado}, H. and {Vargas-Maga{\~n}a}, Mariana and {Vivas}, A. Katherina and {Wechsler}, Risa H. and {Williams}, Doug and {Yang}, Jinyi and {Yang}, Qian and {Yapici}, Tolga and {Zaritsky}, Dennis and {Zenteno}, A. and {Zhang}, Kai and {Zhang}, Tianmeng and {Zhou}, Rongpu and {Zhou}, Zhimin},
        title = "{Overview of the DESI Legacy Imaging Surveys}",
      journal = {\aj},
         year = 2019,
        month = may,
       volume = {157},
       number = {5},
          eid = {168},
        pages = {168},
          doi = {10.3847/1538-3881/ab089d},
archivePrefix = {arXiv},
       eprint = {1804.08657},
 primaryClass = {astro-ph.IM},
       adsurl = {https://ui.adsabs.harvard.edu/abs/2019AJ....157..168D}
}

@ARTICLE{scipy,
  author  = {Virtanen, Pauli and Gommers, Ralf and Oliphant, Travis E. and
            Haberland, Matt and Reddy, Tyler and Cournapeau, David and
            Burovski, Evgeni and Peterson, Pearu and Weckesser, Warren and
            Bright, Jonathan and {van der Walt}, St{\'e}fan J. and
            Brett, Matthew and Wilson, Joshua and Millman, K. Jarrod and
            Mayorov, Nikolay and Nelson, Andrew R. J. and Jones, Eric and
            Kern, Robert and Larson, Eric and Carey, C J and
            Polat, {\.I}lhan and Feng, Yu and Moore, Eric W. and
            {VanderPlas}, Jake and Laxalde, Denis and Perktold, Josef and
            Cimrman, Robert and Henriksen, Ian and Quintero, E. A. and
            Harris, Charles R. and Archibald, Anne M. and
            Ribeiro, Ant{\^o}nio H. and Pedregosa, Fabian and
            {van Mulbregt}, Paul and {SciPy 1.0 Contributors}},
  title   = {{{SciPy} 1.0: Fundamental Algorithms for Scientific
            Computing in Python}},
  journal = {Nature Methods},
  year    = {2020},
  volume  = {17},
  pages   = {261--272},
  adsurl  = {https://rdcu.be/b08Wh},
  doi     = {10.1038/s41592-019-0686-2},
}

@ARTICLE{Mcconnell20racs,
       author = {{McConnell}, D. and {Hale}, C.~L. and {Lenc}, E. and {Banfield}, J.~K. and {Heald}, George and {Hotan}, A.~W. and {Leung}, James K. and {Moss}, Vanessa A. and {Murphy}, Tara and {O'Brien}, Andrew and {Pritchard}, Joshua and {Raja}, Wasim and {Sadler}, Elaine M. and {Stewart}, Adam and {Thomson}, Alec J.~M. and {Whiting}, M. and {Allison}, James R. and {Amy}, S.~W. and {Anderson}, C. and {Ball}, Lewis and {Bannister}, Keith W. and {Bell}, Martin and {Bock}, Douglas C.-J. and {Bolton}, Russ and {Bunton}, J.~D. and {Chippendale}, A.~P. and {Collier}, J.~D. and {Cooray}, F.~R. and {Cornwell}, T.~J. and {Diamond}, P.~J. and {Edwards}, P.~G. and {Gupta}, N. and {Hayman}, Douglas B. and {Heywood}, Ian and {Jackson}, C.~A. and {Koribalski}, B{\"a}rbel S. and {Lee-Waddell}, Karen and {McClure-Griffiths}, N.~M. and {Ng}, Alan and {Norris}, Ray P. and {Phillips}, Chris and {Reynolds}, John E. and {Roxby}, Daniel N. and {Schinckel}, Antony E.~T. and {Shields}, Matt and {Tremblay}, Chenoa and {Tzioumis}, A. and {Voronkov}, M.~A. and {Westmeier}, Tobias},
        title = "{The Rapid ASKAP Continuum Survey I: Design and first results}",
      journal = {\pasa},
         year = 2020,
        month = nov,
       volume = {37},
          eid = {e048},
        pages = {e048},
          doi = {10.1017/pasa.2020.41},
archivePrefix = {arXiv},
       eprint = {2012.00747},
 primaryClass = {astro-ph.IM},
       adsurl = {https://ui.adsabs.harvard.edu/abs/2020PASA...37...48M}
}

@ARTICLE{hale21racs,
       author = {{Hale}, Catherine L. and {McConnell}, D. and {Thomson}, A.~J.~M. and {Lenc}, E. and {Heald}, G.~H. and {Hotan}, A.~W. and {Leung}, J.~K. and {Moss}, V.~A. and {Murphy}, T. and {Pritchard}, J. and {Sadler}, E.~M. and {Stewart}, A.~J. and {Whiting}, M.~T.},
        title = "{The Rapid ASKAP Continuum Survey Paper II: First Stokes I Source Catalogue Data Release}",
      journal = {\pasa},
         year = 2021,
        month = nov,
       volume = {38},
          eid = {e058},
        pages = {e058},
          doi = {10.1017/pasa.2021.47},
archivePrefix = {arXiv},
       eprint = {2109.00956},
 primaryClass = {astro-ph.GA},
       adsurl = {https://ui.adsabs.harvard.edu/abs/2021PASA...38...58H}
}

@ARTICLE{Duchesne25racs,
       author = {{Duchesne}, S. and {Ross}, K. and {Thomson}, A.~J.~M. and {Lenc}, E. and {Murphy}, T. and {Galvin}, T.~J. and {Hotan}, A.~W. and {Moss}, V.~A. and {Whiting}, M.~T.},
        title = "{The Rapid ASKAP Continuum Survey (RACS) VI: The RACS-high 1655.5 MHz images and catalogue.}",
      journal = {\pasa},
         year = 2025,
        month = jan,
       volume = {42},
          eid = {38},
        pages = {38},
          doi = {10.1017/pasa.2025.2},
archivePrefix = {arXiv},
       eprint = {2501.04978},
 primaryClass = {astro-ph.GA},
       adsurl = {https://ui.adsabs.harvard.edu/abs/2025PASA...42...38D}
}

@ARTICLE{Koribalski26,
       author = {{Koribalski}, B{\"a}rbel Silvia and {Dolag}, Klaus and {Khabibullin}, Ildar and {Boess}, Ludwig and {Ivleva}, Anna and {Norris}, Ray},
        title = "{Two intersecting radio shells: Relics of galaxy merger shocks?}",
      journal = {\pasa},
         year = 2026,
        month = apr,
       volume = {43},
          eid = {e061},
        pages = {e061},
          doi = {10.1017/pasa.2026.10193},
archivePrefix = {arXiv},
       eprint = {2507.11781},
 primaryClass = {astro-ph.GA},
       adsurl = {https://ui.adsabs.harvard.edu/abs/2026PASA...43...61K}
}

@ARTICLE{Drury83,
       author = {{Drury}, L. Oc.},
        title = "{REVIEW ARTICLE: An introduction to the theory of diffusive shock acceleration of energetic particles in tenuous plasmas}",
      journal = {Reports on Progress in Physics},
         year = 1983,
        month = aug,
       volume = {46},
       number = {8},
        pages = {973-1027},
          doi = {10.1088/0034-4885/46/8/002},
       adsurl = {https://ui.adsabs.harvard.edu/abs/1983RPPh...46..973D}
}

@ARTICLE{Blandford&Eichler87,
       author = {{Blandford}, Roger and {Eichler}, David},
        title = "{Particle acceleration at astrophysical shocks: A theory of cosmic ray origin}",
      journal = {\physrep},
         year = 1987,
        month = oct,
       volume = {154},
       number = {1},
        pages = {1-75},
          doi = {10.1016/0370-1573(87)90134-7},
       adsurl = {https://ui.adsabs.harvard.edu/abs/1987PhR...154....1B}
}

@ARTICLE{Caprioli18,
       author = {{Caprioli}, Damiano and {Zhang}, Horace and {Spitkovsky}, Anatoly},
        title = "{Diffusive shock re-acceleration}",
      journal = {Journal of Plasma Physics},
         year = 2018,
        month = jun,
       volume = {84},
       number = {3},
          eid = {715840301},
        pages = {715840301},
          doi = {10.1017/S0022377818000478},
archivePrefix = {arXiv},
       eprint = {1801.01510},
 primaryClass = {astro-ph.HE},
       adsurl = {https://ui.adsabs.harvard.edu/abs/2018JPlPh..84c7101C}
}

@ARTICLE{DeRubeis26,
       author = {{De Rubeis}, E. and {Bondi}, M. and {Botteon}, A. and {van Weeren}, R.~J. and {de Jong}, J.~M.~G.~H.~J. and {Brunetti}, G. and {Rudnick}, L. and {Br{\"u}ggen}, M. and {Bruno}, L. and {Escott}, E.~L. and {Gheller}, C. and {Morabito}, L.~K. and {Rajpurohit}, K. and {R{\"o}ttgering}, H.~J.~A.},
        title = "{Revealing the intricacies of radio galaxies and filaments in the merging galaxy cluster Abell 2255: II. Properties of filaments using multi-frequency radio data}",
      journal = {\aap},
         year = 2026,
        month = feb,
       volume = {707},
          eid = {A49},
        pages = {A49},
          doi = {10.1051/0004-6361/202557651},
archivePrefix = {arXiv},
       eprint = {2512.19471},
 primaryClass = {astro-ph.CO},
       adsurl = {https://ui.adsabs.harvard.edu/abs/2026A&A...707A..49D}
}

@ARTICLE{Young02,
       author = {{Young}, A. and {Rudnick}, L. and {Katz-Stone}, D.~M. and {O'Donoghue}, A.~A.},
        title = "{Electron population aging models for wide-angle tails}",
      journal = {\nar},
         year = 2002,
        month = may,
       volume = {46},
       number = {2-7},
        pages = {105-107},
          doi = {10.1016/S1387-6473(01)00164-6},
archivePrefix = {arXiv},
       eprint = {astro-ph/9909182},
 primaryClass = {astro-ph},
       adsurl = {https://ui.adsabs.harvard.edu/abs/2002NewAR..46..105Y}
}

@ARTICLE{Kang&Ryu11,
       author = {{Kang}, Hyesung and {Ryu}, Dongsu},
        title = "{Re-acceleration of Non-thermal Particles at Weak Cosmological Shock Waves}",
      journal = {\apj},
         year = 2011,
        month = jun,
       volume = {734},
       number = {1},
          eid = {18},
        pages = {18},
          doi = {10.1088/0004-637X/734/1/18},
archivePrefix = {arXiv},
       eprint = {1102.2561},
 primaryClass = {astro-ph.CO},
       adsurl = {https://ui.adsabs.harvard.edu/abs/2011ApJ...734...18K}
}

\begin{appendix}
\onecolumn

\section{Auxiliary images}
\label{app: img low res}

\begin{figure*}[!htbp]
        \centering
        \includegraphics[width=0.33\textwidth]{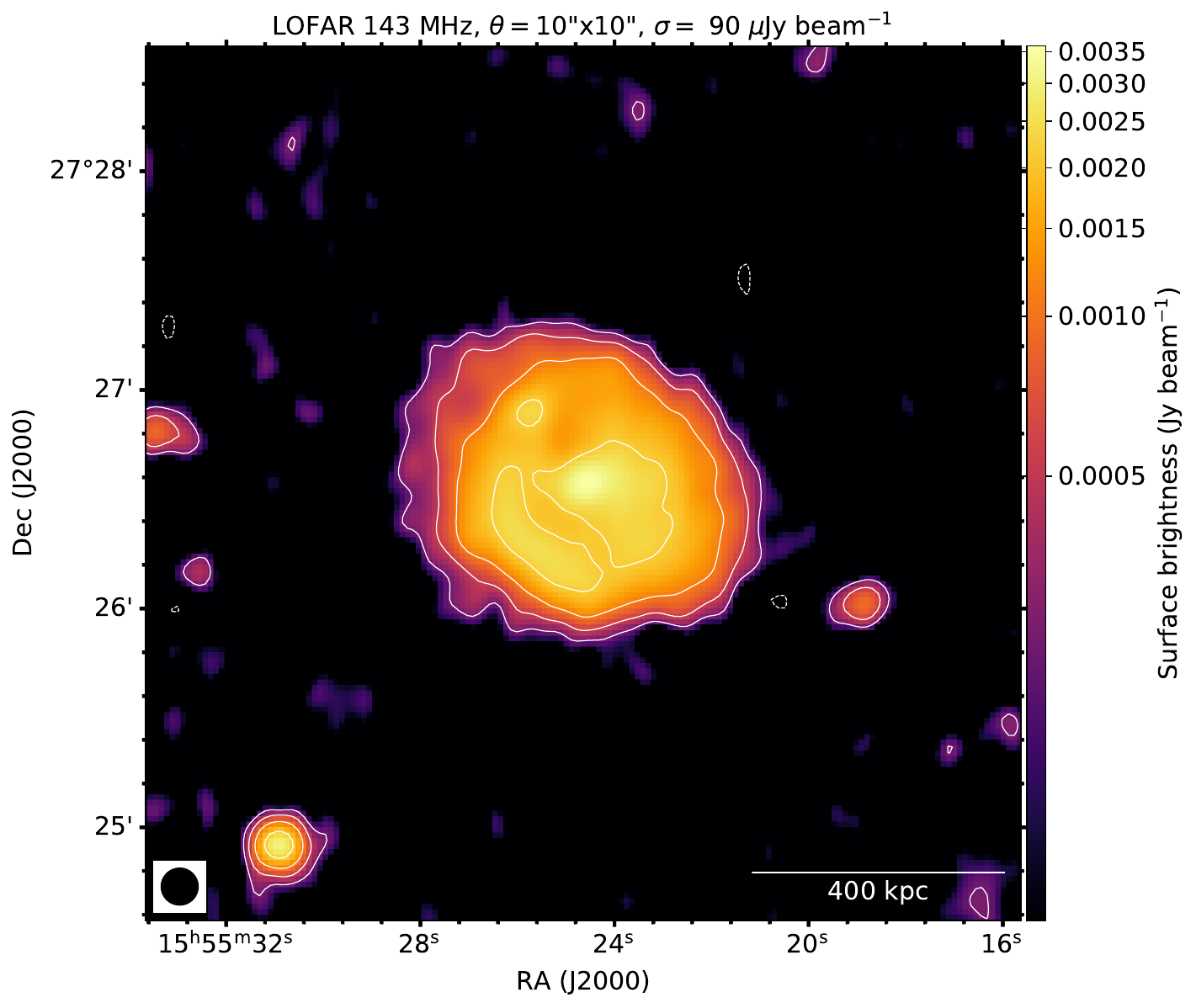}
        \includegraphics[width=0.33\textwidth]{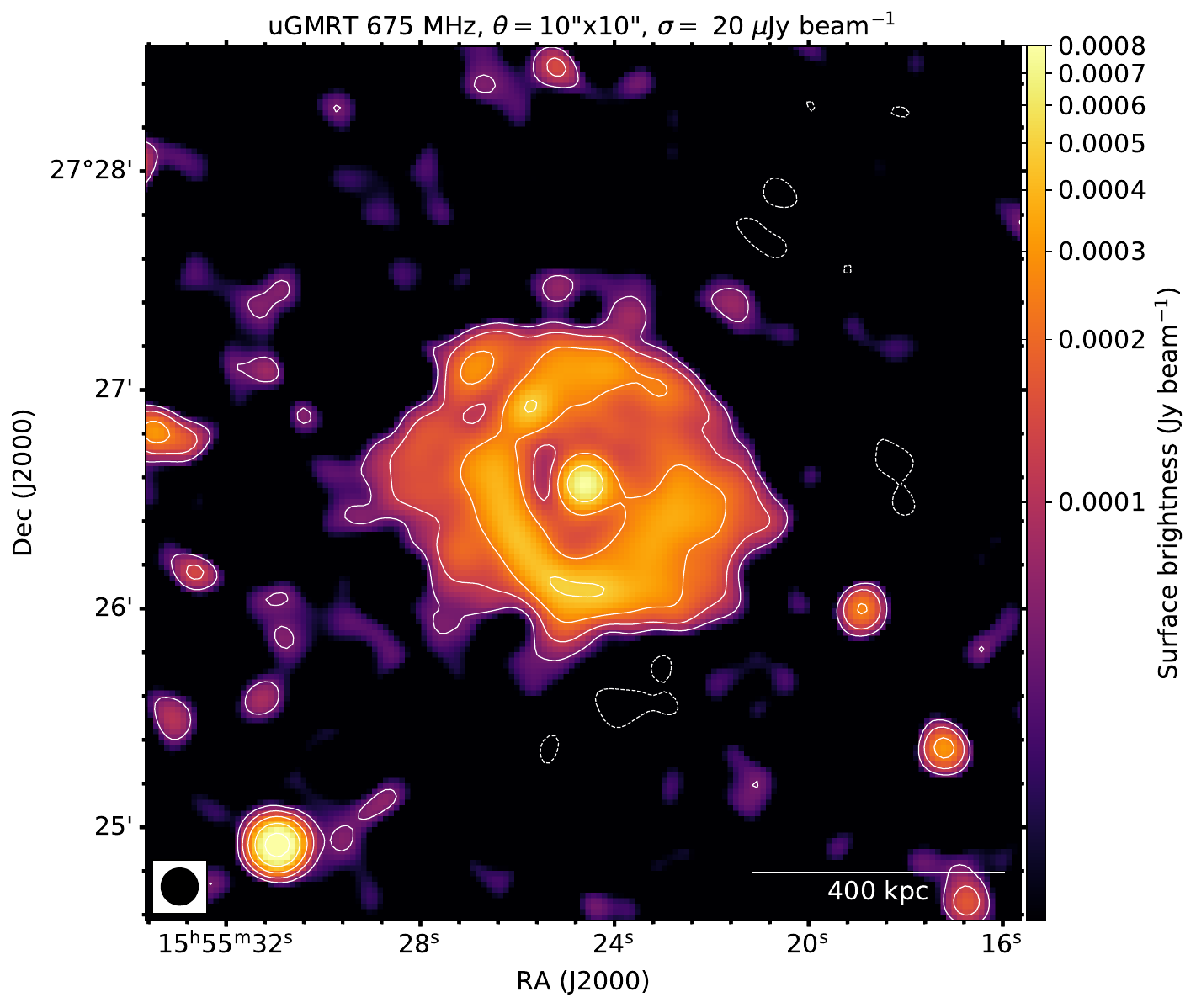}
         \includegraphics[width=0.33\textwidth]{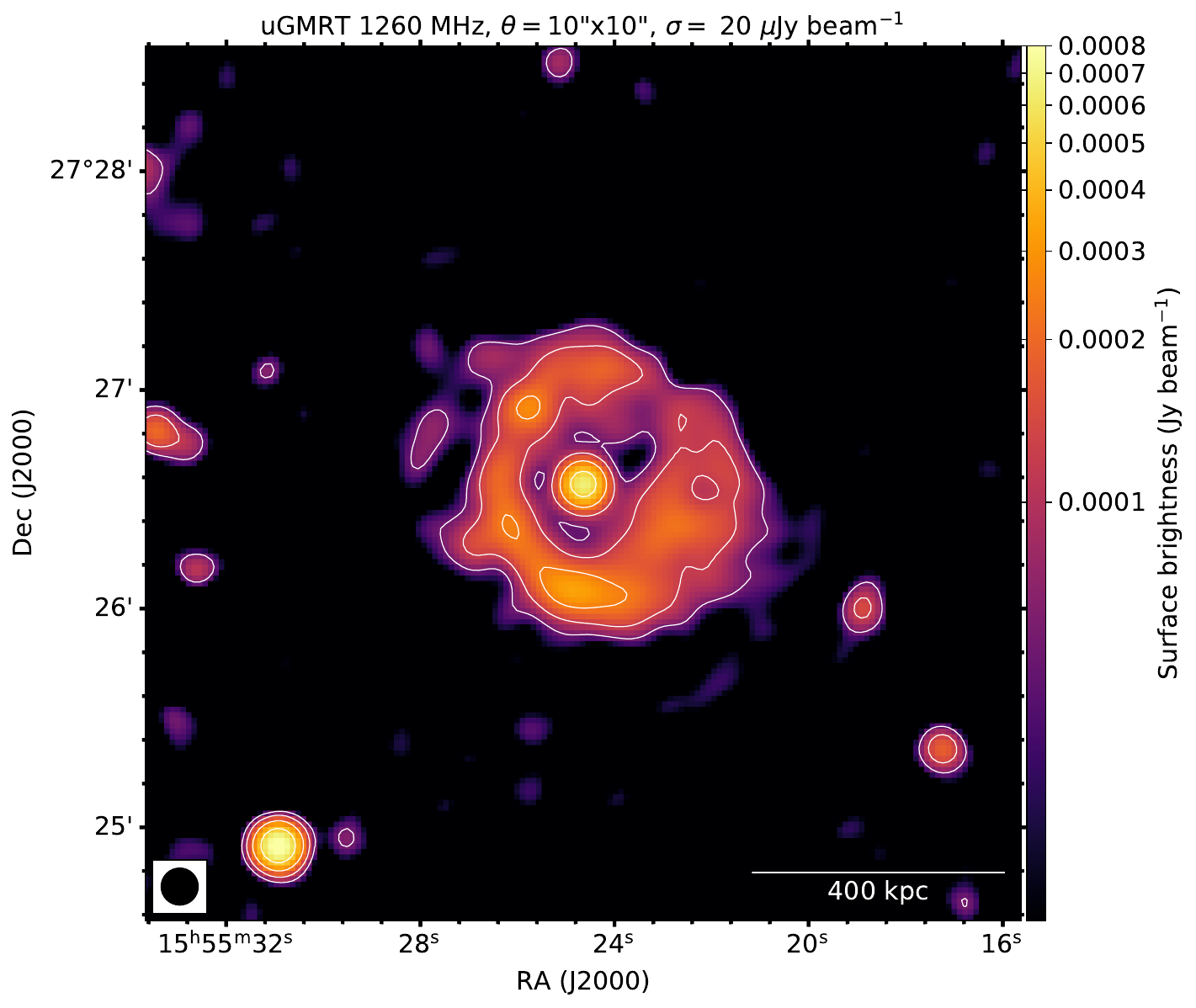}
         \includegraphics[width=0.33\textwidth]{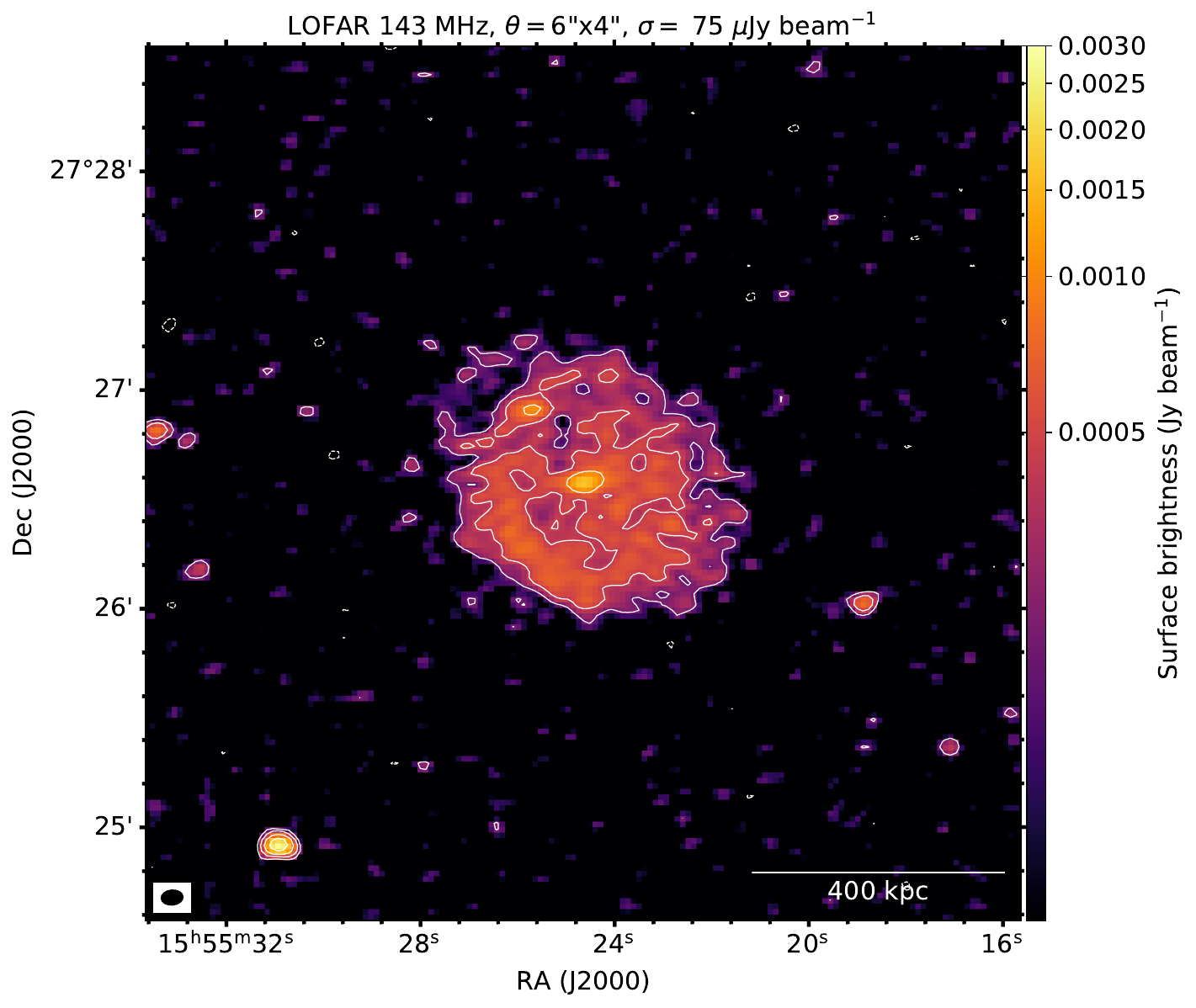}
         \includegraphics[width=0.33\textwidth]{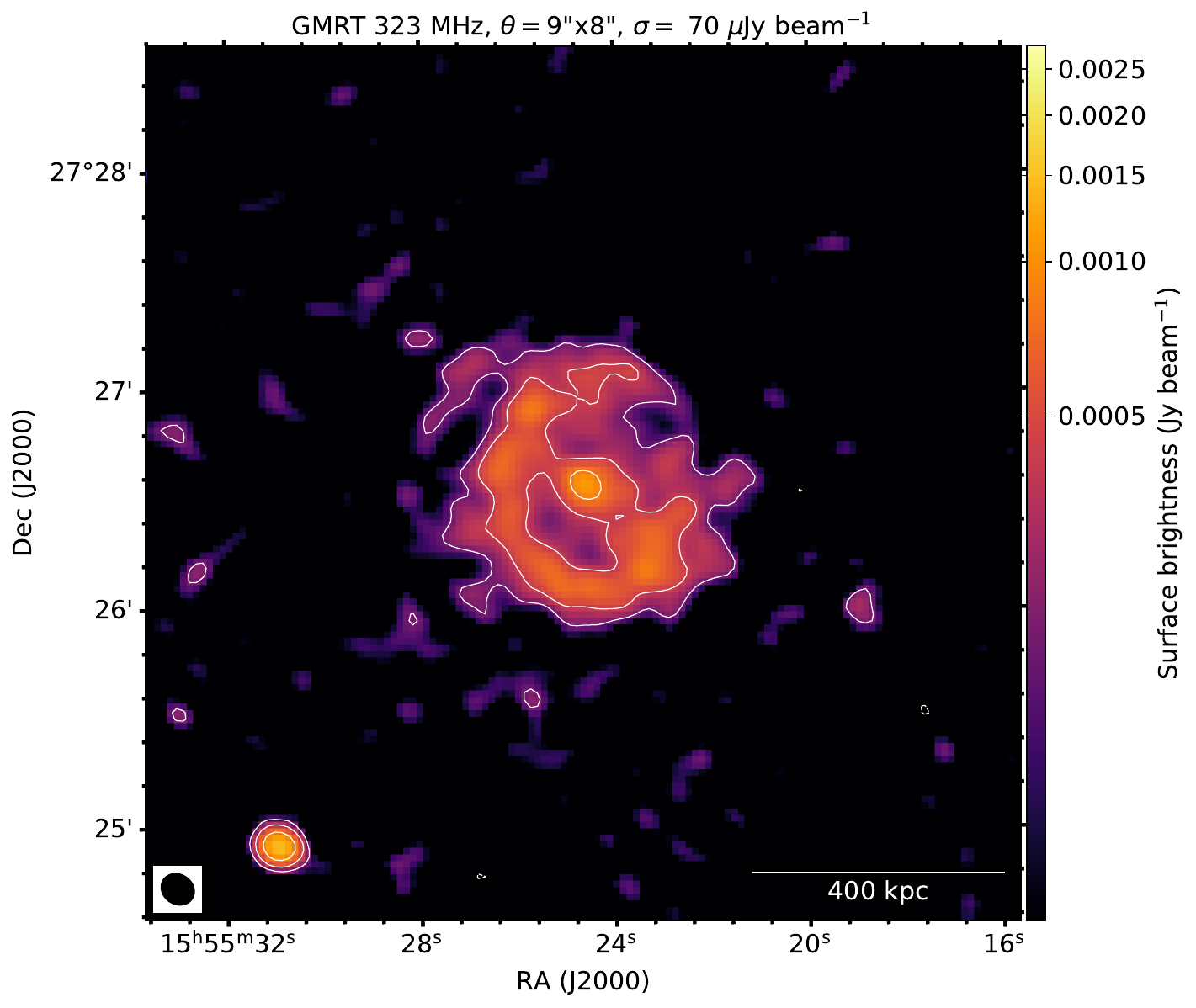}
         \includegraphics[width=0.33\textwidth]{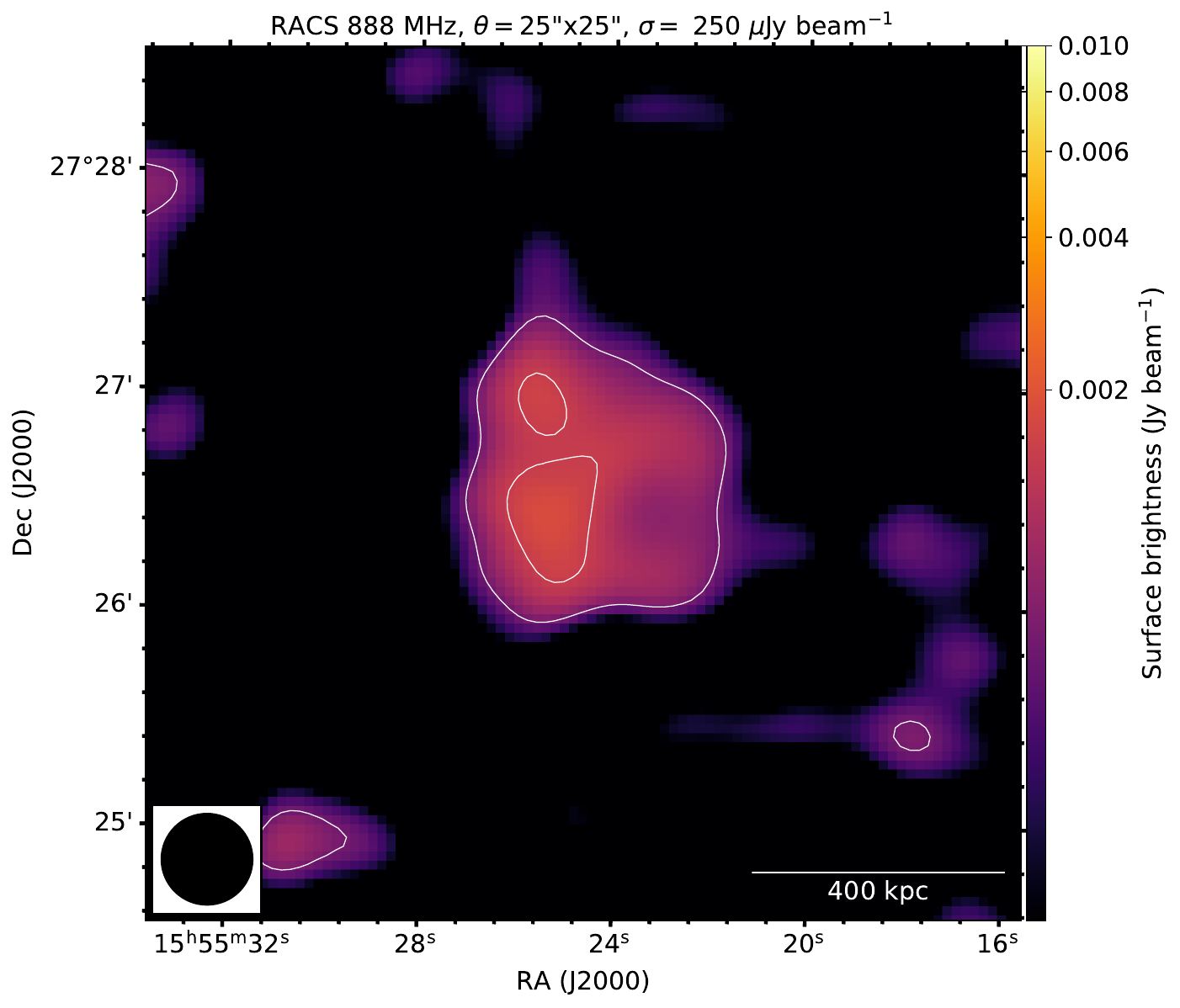}
         \includegraphics[width=0.33\textwidth]{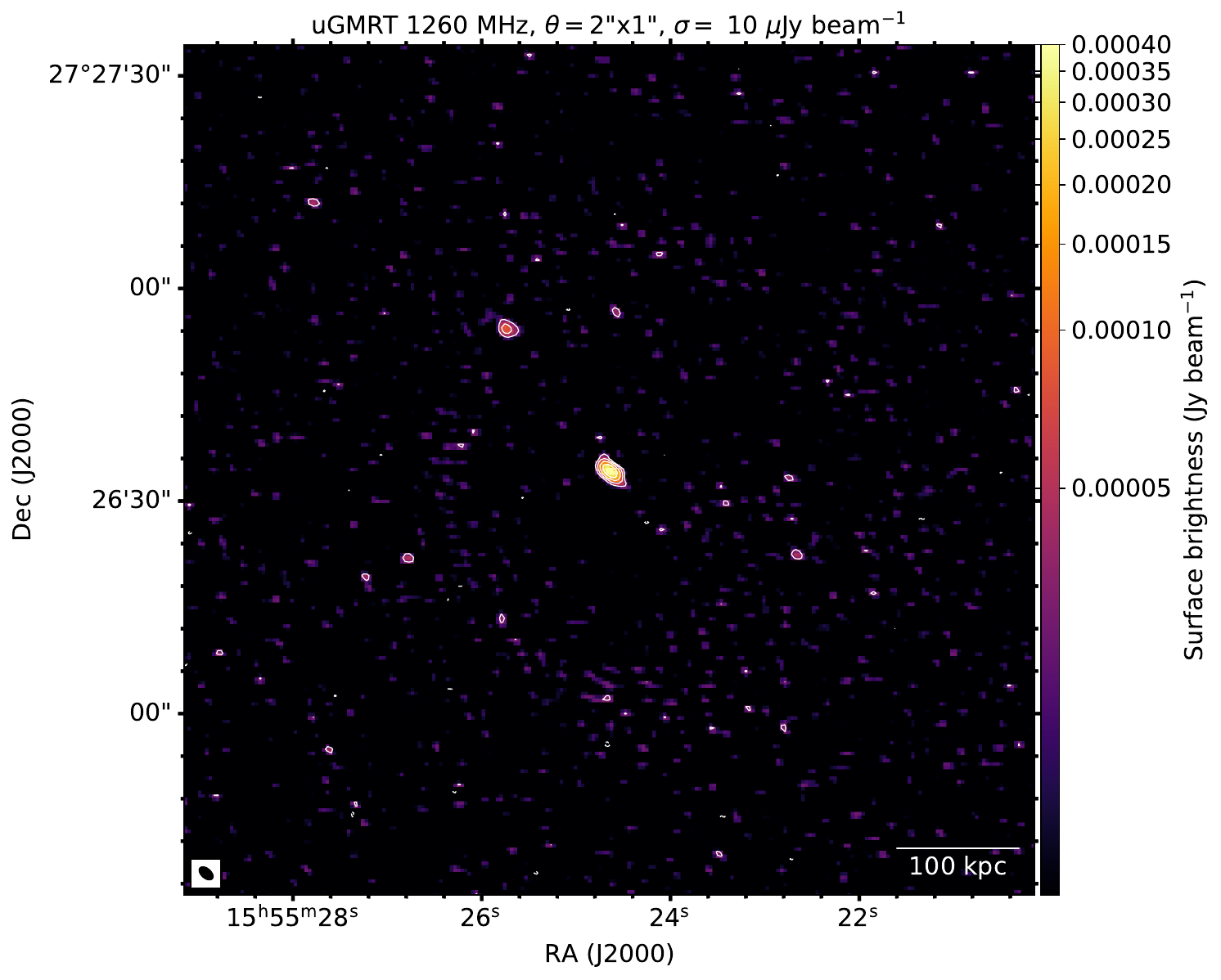}
         \includegraphics[width=0.33\textwidth]{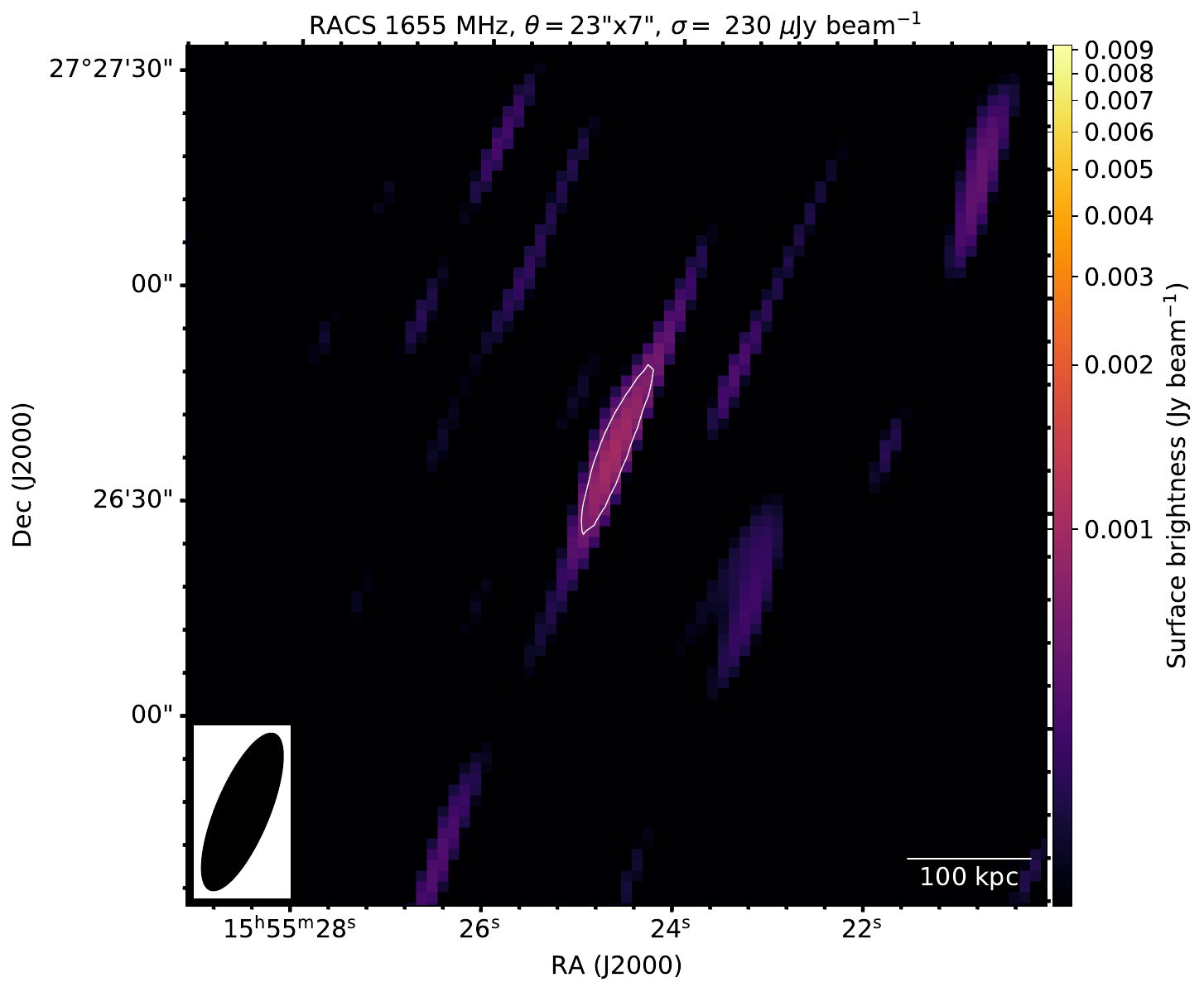}         
         \includegraphics[width=0.33\textwidth]{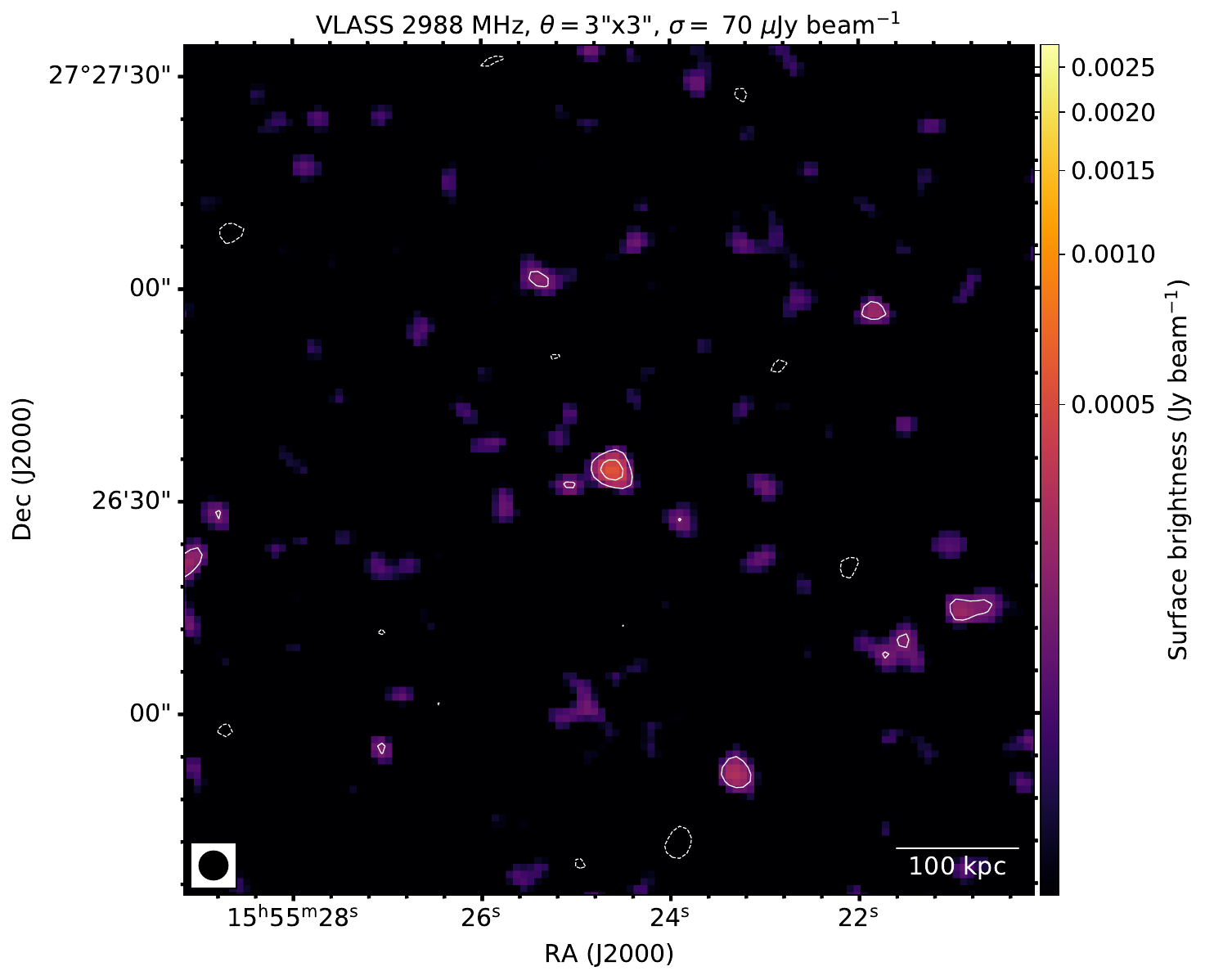}
        \smallskip
        
        \caption{Images of ORC4. From top left to bottom right: LOFAR at 143 MHz, uGMRT at 675 MHz, uGMRT at 1260 MHz, LOFAR at 143 MHz, GMRT at 323 MHz, RACS at 888 MHz, uGMRT at 1260 MHz, RACS at 1655 MHz, and VLASS at 2988 MHz. The resolution and noise are reported on top of each panel. The contour levels are $[\pm3, \;6, \;12, \; ...]\times \sigma$. }
        \label{fig: img lowres}
\end{figure*}

In Fig. \ref{fig: img lowres} we present additional radio images used in our analysis. The top panels report images at 143, 675, and 1260 MHz at the resolution of $10''$. These were obtained by matching the \textit{uv}-coverage ($0.5-32 \; {\rm k\lambda}$) and then convolving with a circular beam. We used these images to analyse the surface brightness (Fig. \ref{fig: sb}), spectral index and curvature (Fig. \ref{fig: spixmap}), and RCCD (Fig.~ \ref{fig: ccp}). Examples of the sampling regions used for these analyses are shown in Fig. \ref{fig: sb}.

In Fig. \ref{fig: img lowres} we also report images used for flux density measurements (Sect. \ref{sect: integrated spectrum}, Appendix \ref{app: Central AGN spectrum}). Among these, we included images from RACS at 888 MHz and 1655 MHz. ORC4 is detected at 888 MHz ($\theta= 25''\times25''$, $\sigma = 250 \; {\rm \mu Jy \; beam^{-1}}$), while source G is detected only at 1655 MHz ($\theta= 23''\times7''$, $\sigma = 230 \; {\rm \mu Jy \; beam^{-1}}$) at $\sim3\sigma$. Finally, we considered images from four epochs of VLASS at S band ($2-4$~GHz), each with a resolution of $\theta\sim 2.5''$ and a noise of $\sigma \sim 120 \; {\rm \mu Jy \; beam^{-1}}$. We convolved the four images with a circular ($3''$) beam, and then stacked them by computing the median. The stacked image is reported in Fig. \ref{fig: img lowres}, where source G is detected at $\sim 8\sigma$, while ORC4 is undetected. 

\begin{figure*}[!htbp]
        \centering
        \includegraphics[width=0.33\textwidth]{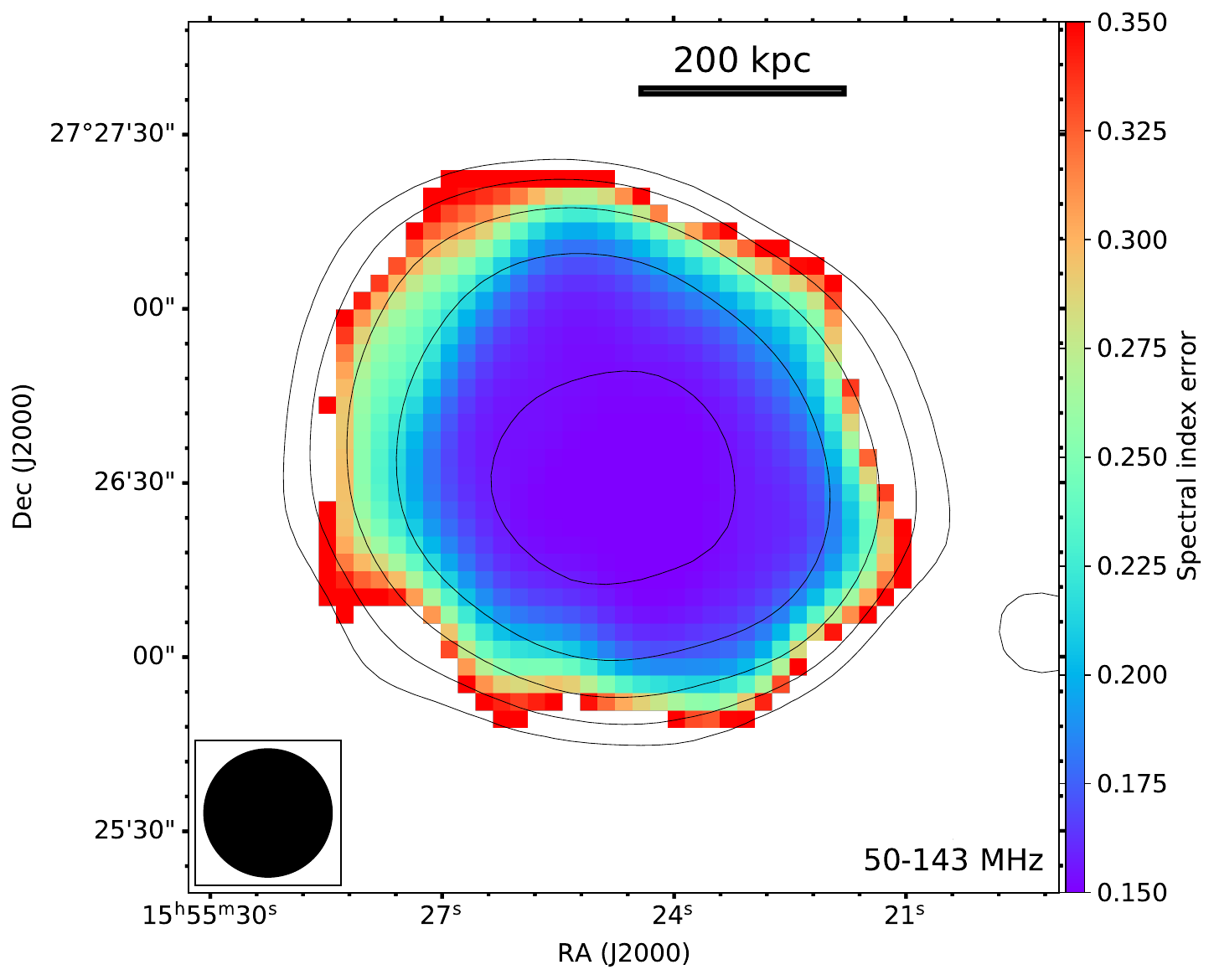}
        \includegraphics[width=0.33\textwidth]{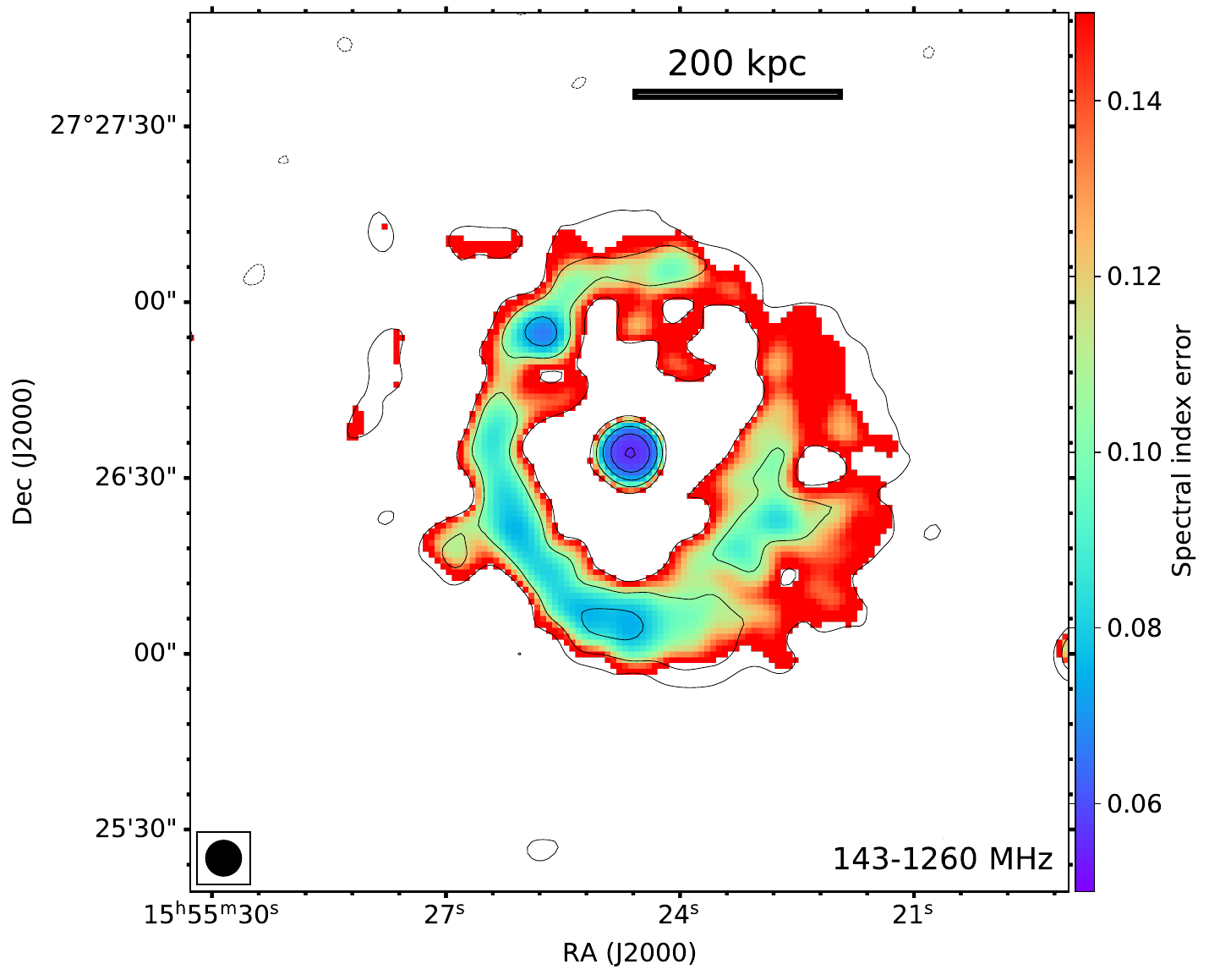}
         \includegraphics[width=0.33\textwidth]{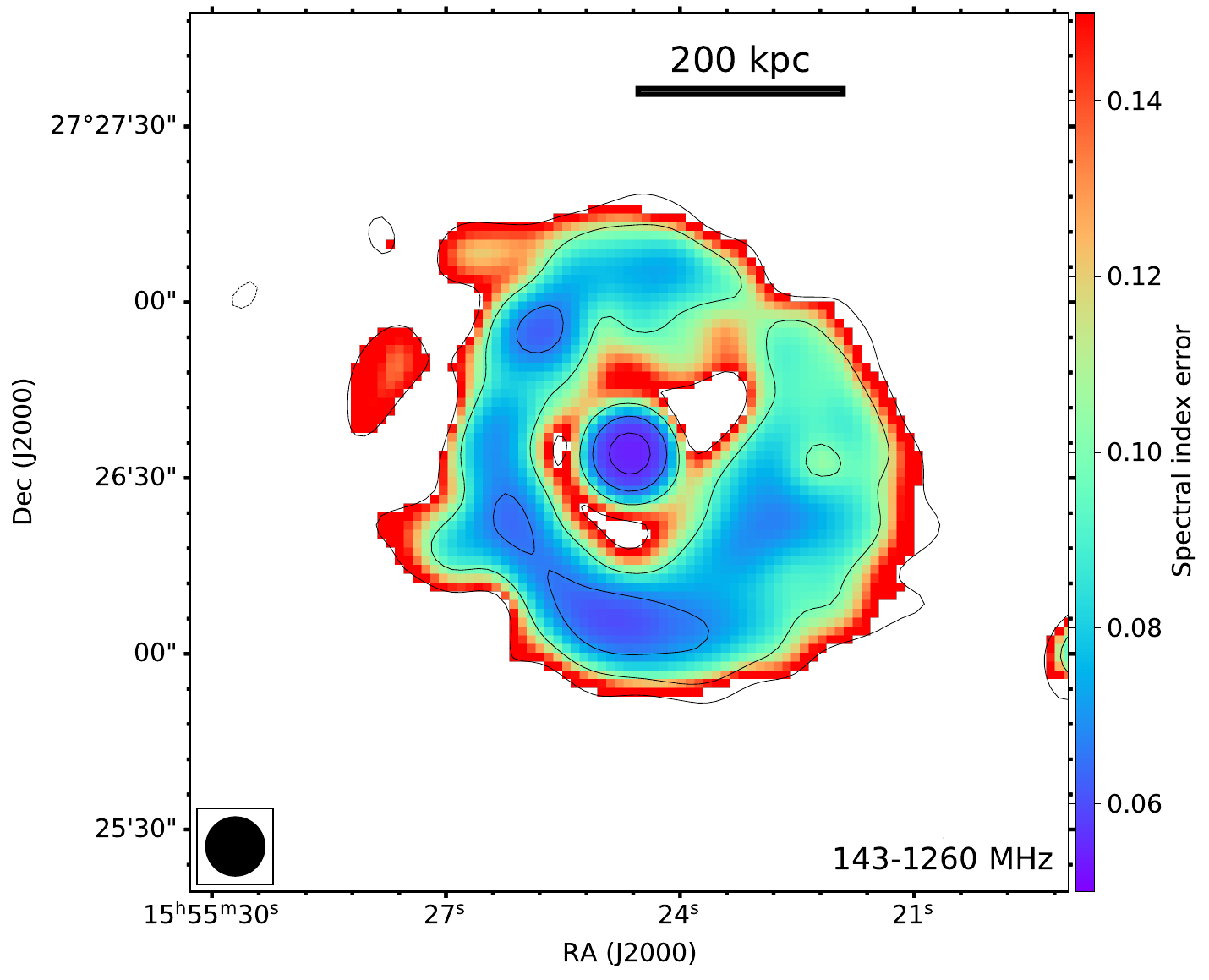}
         \includegraphics[width=0.33\textwidth]{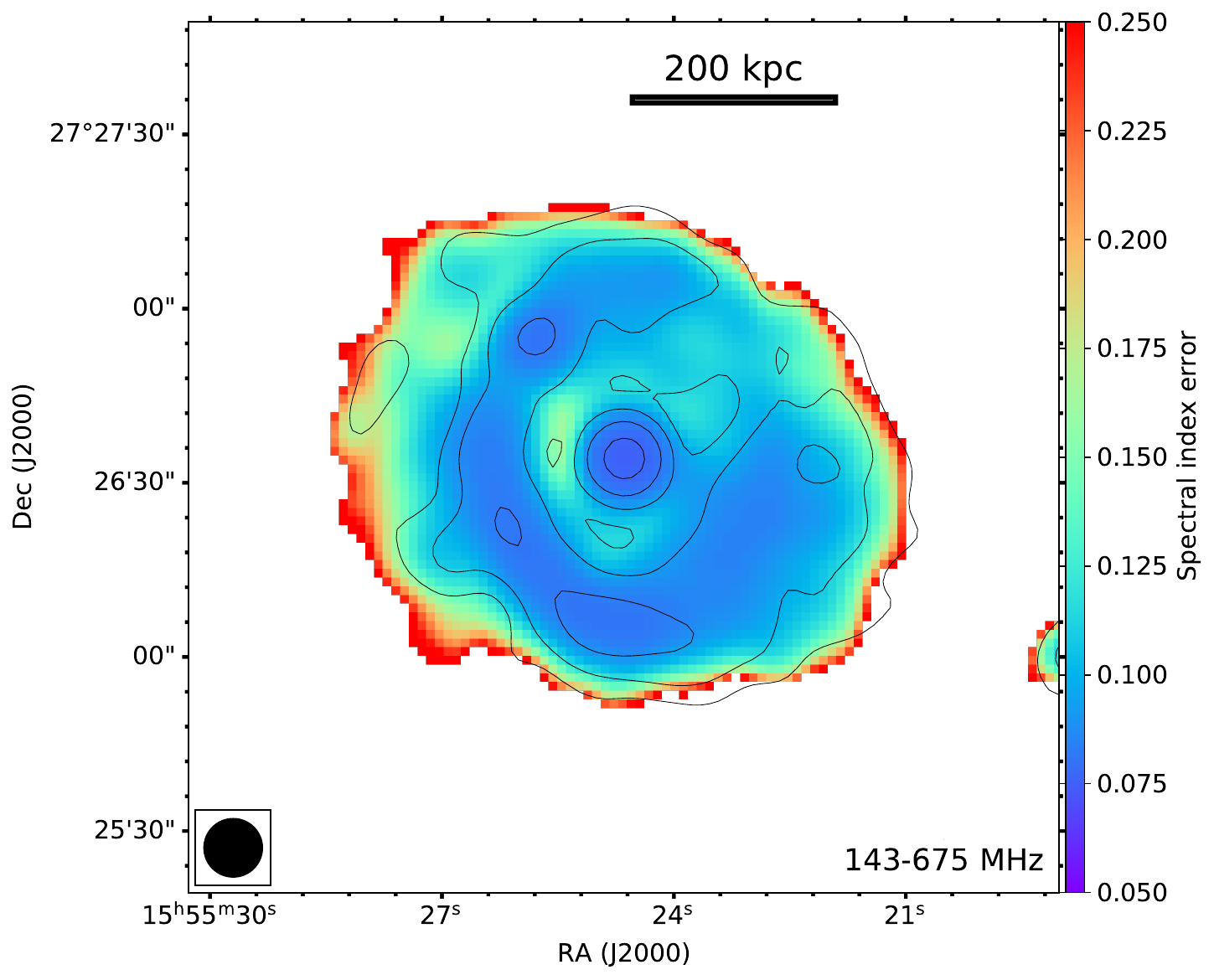}
        \includegraphics[width=0.33\textwidth]{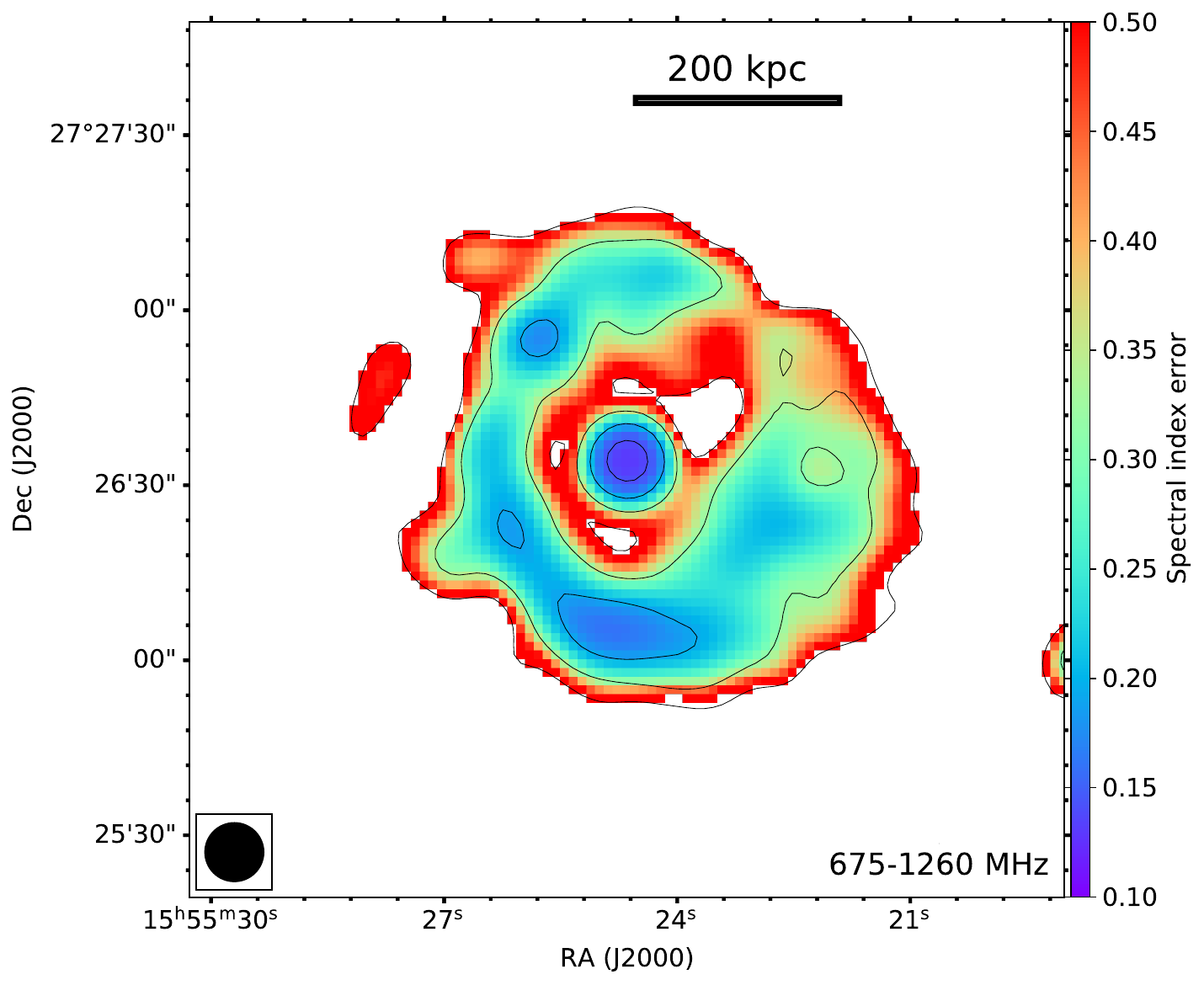}
        \includegraphics[width=0.33\textwidth]{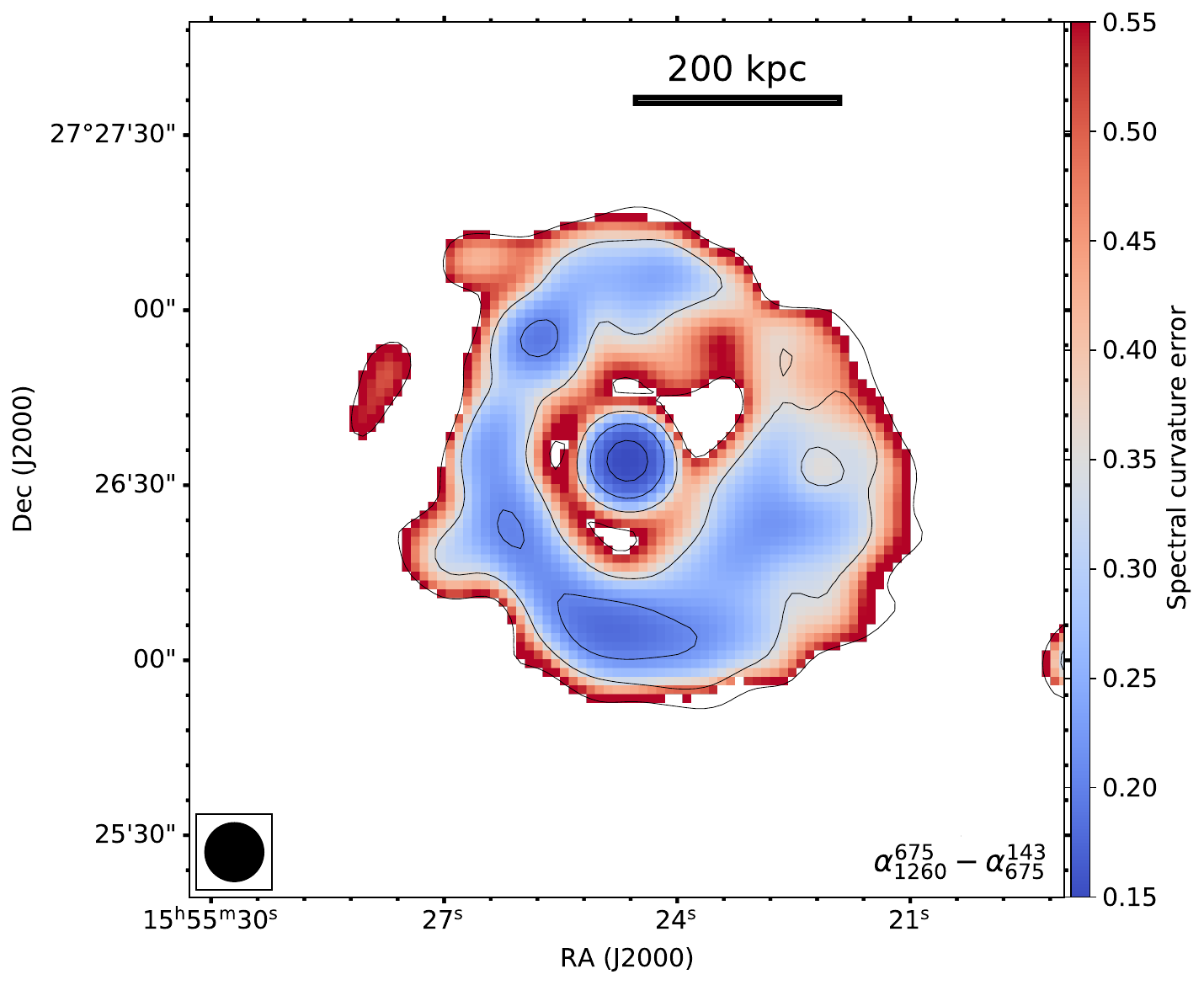}
        \smallskip
        
        \caption{Spectral index and curvature error maps corresponding to the maps shown in Fig. \ref{fig: spixmap}.}
        \label{fig: errspixmap}
\end{figure*}

In Fig. \ref{fig: errspixmap}, we report the error maps corresponding to the spectral index and curvature maps shown in Fig. \ref{fig: spixmap}. These were obtained based on Eqs. \ref{eq: spectralindexformula}, \ref{eq: spectralcurvatureformula}. 

\FloatBarrier


\section{Radio source G}
\label{app: Central AGN spectrum}

In this section we analyse the radio properties of source G. We used images produced with adequate weighting schemes favouring resolution over sensitivity (see Table \ref{tab: flussiagn} and Fig. \ref{fig: img lowres}). We excluded data at 50 MHz as source G is undetected. Instead, we included the GMRT image at 323 MHz (from \citealt{Norris21}) and images from RACS at 1655 MHz and VLASS at 2988 MHz (Appendix~\ref{app: img low res}).

We performed a Gaussian fit to the surface brightness of source G, with results listed in Table \ref{tab: flussiagn}. We obtained integrated to peak flux density ratio $\gtrsim2$ and $\lesssim 1.3$ for images below and above 675 MHz, respectively. This indicates that all flux density measurements at low frequency are contaminated by extended components, which may be associated with the ORC4 filling emission and/or unresolved, small-scale radio jets; some level of contamination may also affect the higher frequency measurements.

\begin{figure*}[htbp]
    \centering
    \begin{minipage}[c]{0.4\textwidth}
        \centering
        \includegraphics[width=\textwidth]{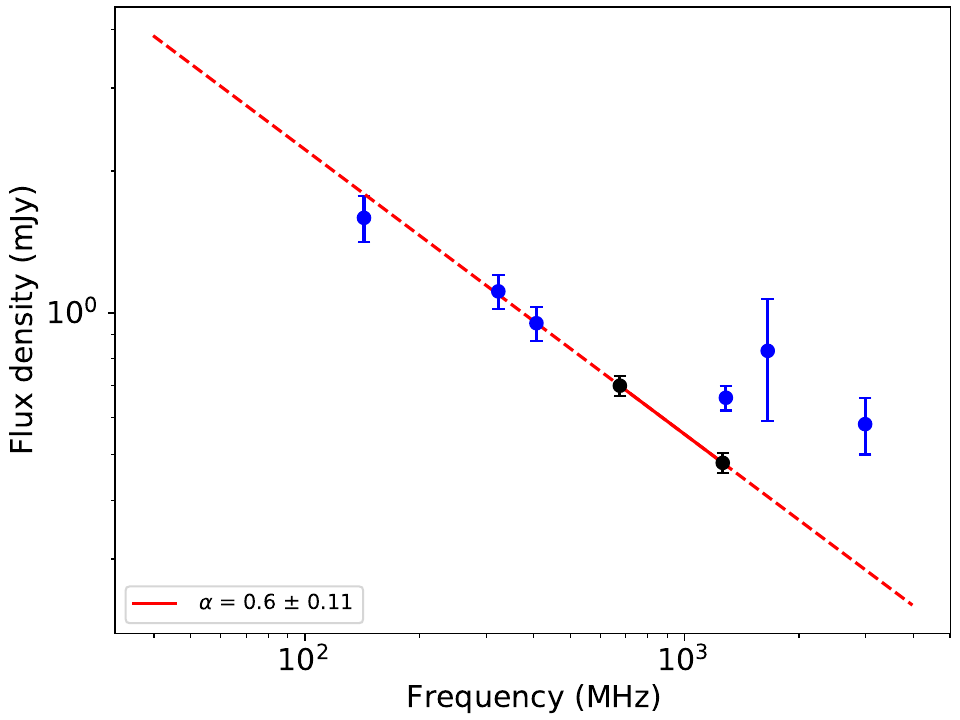}
    \end{minipage}
    \hfill
    \begin{minipage}[c]{0.58\textwidth}
        \caption{Radio spectrum of source G. The reported flux density measurements are $S^{\rm peak}$ from Table \ref{tab: flussiagn}. The black data points were used to compute a two-point spectrum at 675-1260 MHz, which is indicated by the solid line. The dashed line represents the extrapolation of the spectrum.}
        \label{fig: agn spectrum}
    \end{minipage}
\end{figure*}

To minimise the impact of extended emission and determine the spectrum of source G, we initially considered the peak flux density from the highest resolution images at 675, 1260, and 2988 MHz. A single power law cannot fit the three measurements, and we thus obtained a two-point spectrum of $\alpha=0.60 \pm 0.11$ at 675-1260 MHz. In Fig. \ref{fig: agn spectrum} we show this spectrum and its extrapolation with peak flux densities overlaid at higher and lower frequencies. Measurements at 143, 323, and 407 MHz are well predicted by the power law, while measurements at 1284, 1655, and 2988 MHz are significantly higher than the corresponding extrapolated values. By extrapolation, we also predict $S_{50}^{\rm peak}=3.3 \pm 1.2$ mJy, which is $< 3\sigma_{50}$ and consistent with the non-detection by LOFAR LBA.

The observed convex spectrum is suggestive of mixing of multiple components, which cannot be separated with current data. Resolution plays a key role: for instance, by measuring the flux density from the 1260 MHz image at $10''$ (Fig. \ref{fig: img lowres}), we obtained $S_{1260}^{\rm peak}=0.66 \pm 0.04$ mJy and $S_{1260}^{\rm int}=0.71 \pm 0.04$ mJy, consistent with those obtained at 1284 MHz at the same resolution. In conclusion, additional radio data at high (sub-arcsecond) resolution are necessary to properly determine the spectrum of source G, and unveil possible small-scale ($\lesssim10$ kpc) structures.

\begin{table*}[!htbp]
\centering
                \caption{Flux density measurements of source G. }
    \label{tab: flussiagn} 
    \resizebox{0.7\textwidth}{!}{
    \begin{tabular}{c|cccccccc}
        \hline
        \noalign{\smallskip} 
$\nu$ (MHz)  & 143  & 323 & 407 & 675 & 1260 & 1284 & 1655 & 2988 \\  
$\theta$ ($''\times''$)  & $6  \times  4$ & $9  \times  8$ & $8  \times  6$ & $4  \times  4$ & $2  \times  1$ &  $10  \times  10$ & $23  \times  7$  & $3  \times  3$ \\
$\sigma$ (${\rm \mu Jy \; beam^{-1}}$)  & 75  & 70 & 60 & 12 & 10 & 20 & 230 & 70 \\  
$S^{\rm peak}$ (mJy) & $1.59  \pm  0.18$ & $1.11  \pm  0.09$ & $0.95  \pm  0.08$ & $0.70  \pm  0.04$ & $0.48  \pm  0.02$ & $0.66  \pm  0.04$ & $0.83  \pm  0.24$ & $0.58  \pm  0.08$ \\
$S^{\rm int}$ (mJy) & $5.21  \pm  0.59$ & $2.60  \pm  0.21$ & $2.11  \pm  0.18$ & $0.78  \pm  0.04$ & $0.54  \pm  0.02$ & $0.82  \pm  0.05$ & $0.83  \pm  0.24$ & $0.74  \pm  0.10$ \\
$S^{\rm int}/S^{\rm peak}$ & $3.3  \pm  0.2$ & $2.3  \pm  0.1$ & $2.2  \pm  0.1$ & $1.1  \pm  0.1$ & $1.1  \pm  0.1$ & $1.2  \pm  0.1$ & $1.0  \pm  0.4$ & $1.3  \pm  0.1$ \\

        \noalign{\smallskip}
        \hline
        
        \end{tabular}
        }
               \begin{tablenotes}
                \centering
       \item   {\small \textbf{Notes}. Rows 1-3: frequency, resolution, and noise of the considered image. Rows 4-6: peak flux density, integrated flux density, and integrated to peak flux density ratio from the Gaussian fit.   } 
       \end{tablenotes}
   \end{table*}

\FloatBarrier


\section{Comparison with simulations}
\label{app: simulations}

In this section we compare ORC4 with a simulated ORC in the classical DSA framework. Such simulation was performed following the same prescriptions detailed in \cite{Ivleva26} for a galaxy merger shock accelerating particles from the thermal pool.

\begin{figure*}[!htbp]
        \centering
        \includegraphics[width=0.32\textwidth]{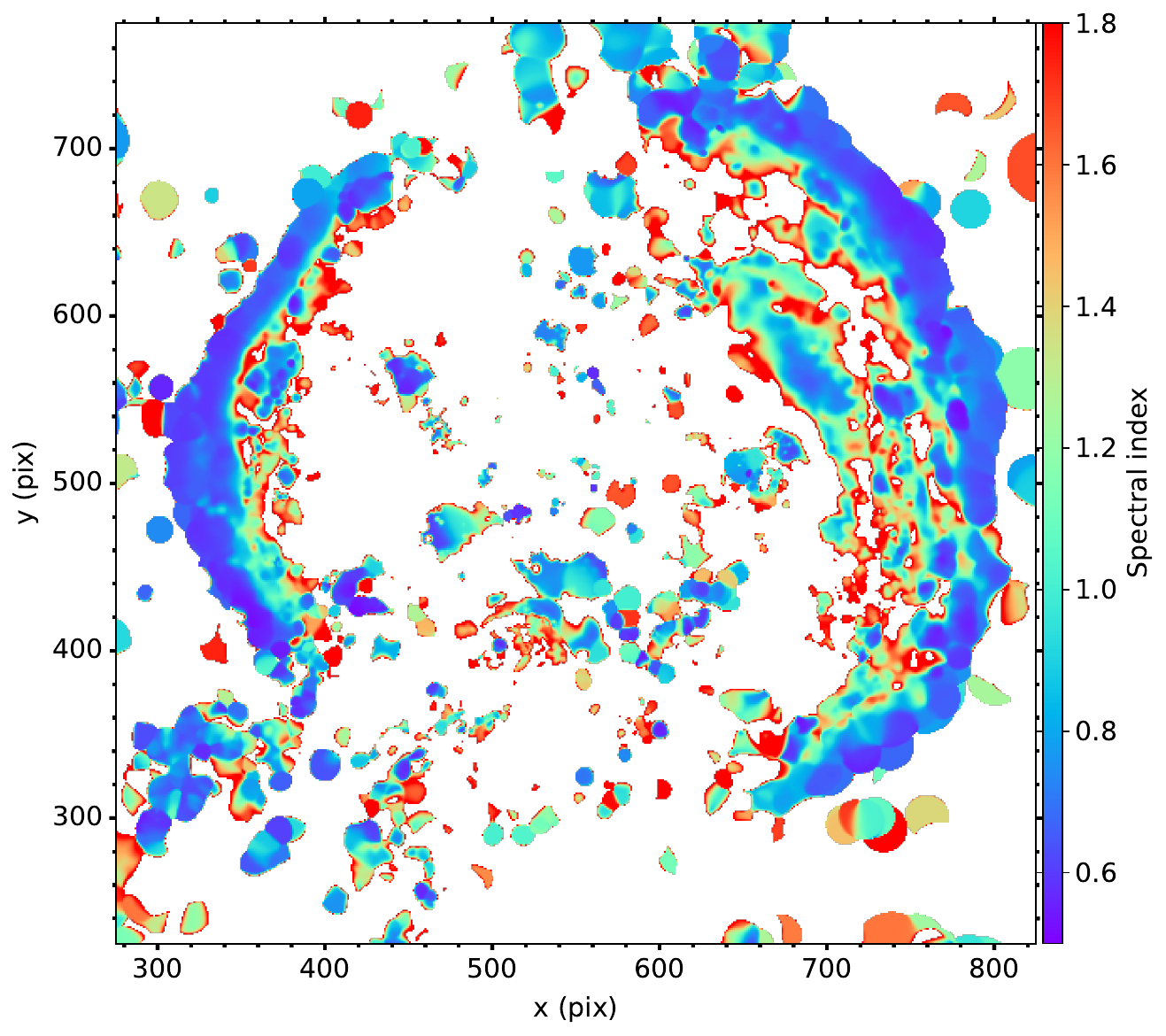}
        \includegraphics[width=0.32\textwidth]{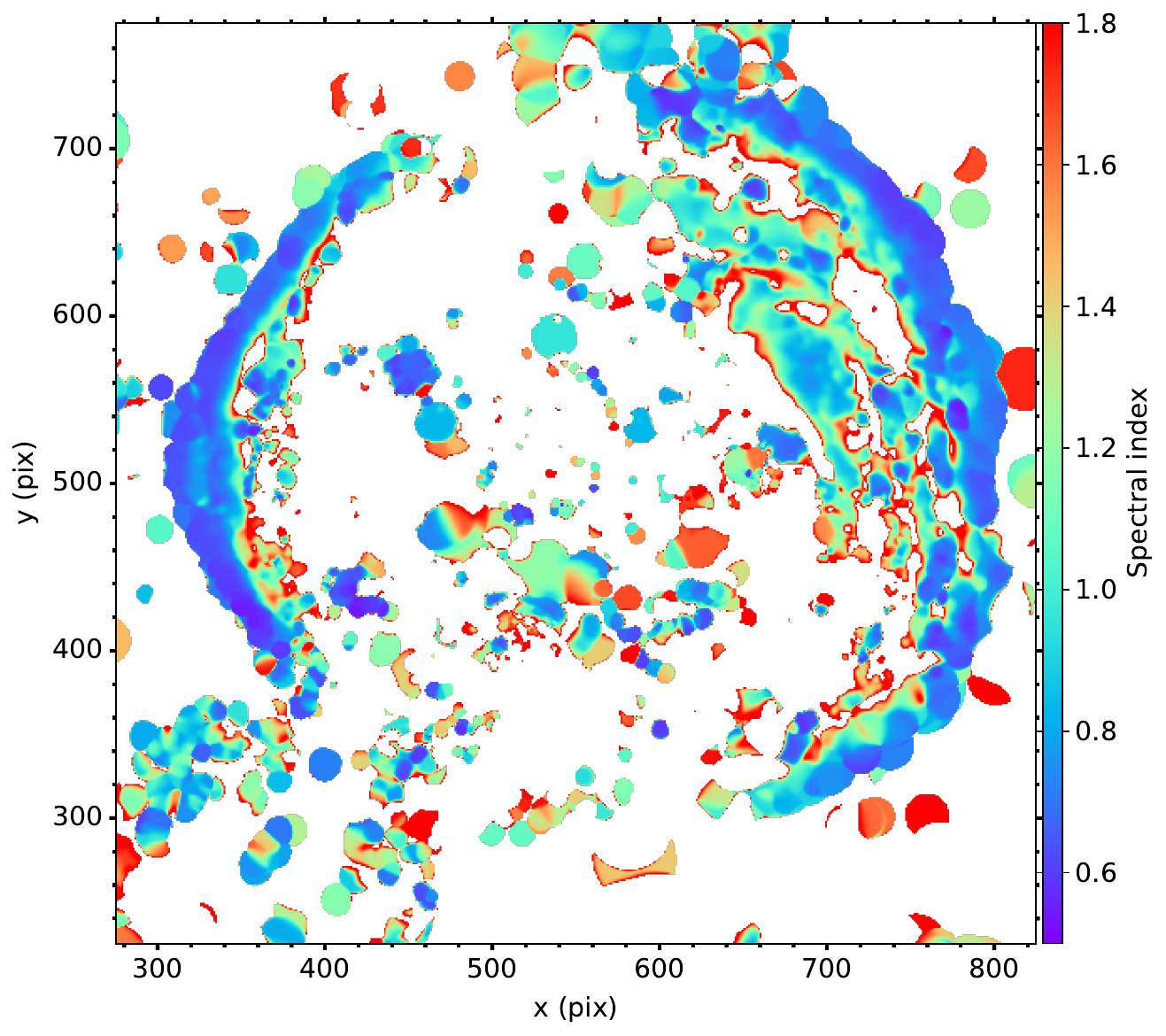}
         \includegraphics[width=0.33\textwidth]{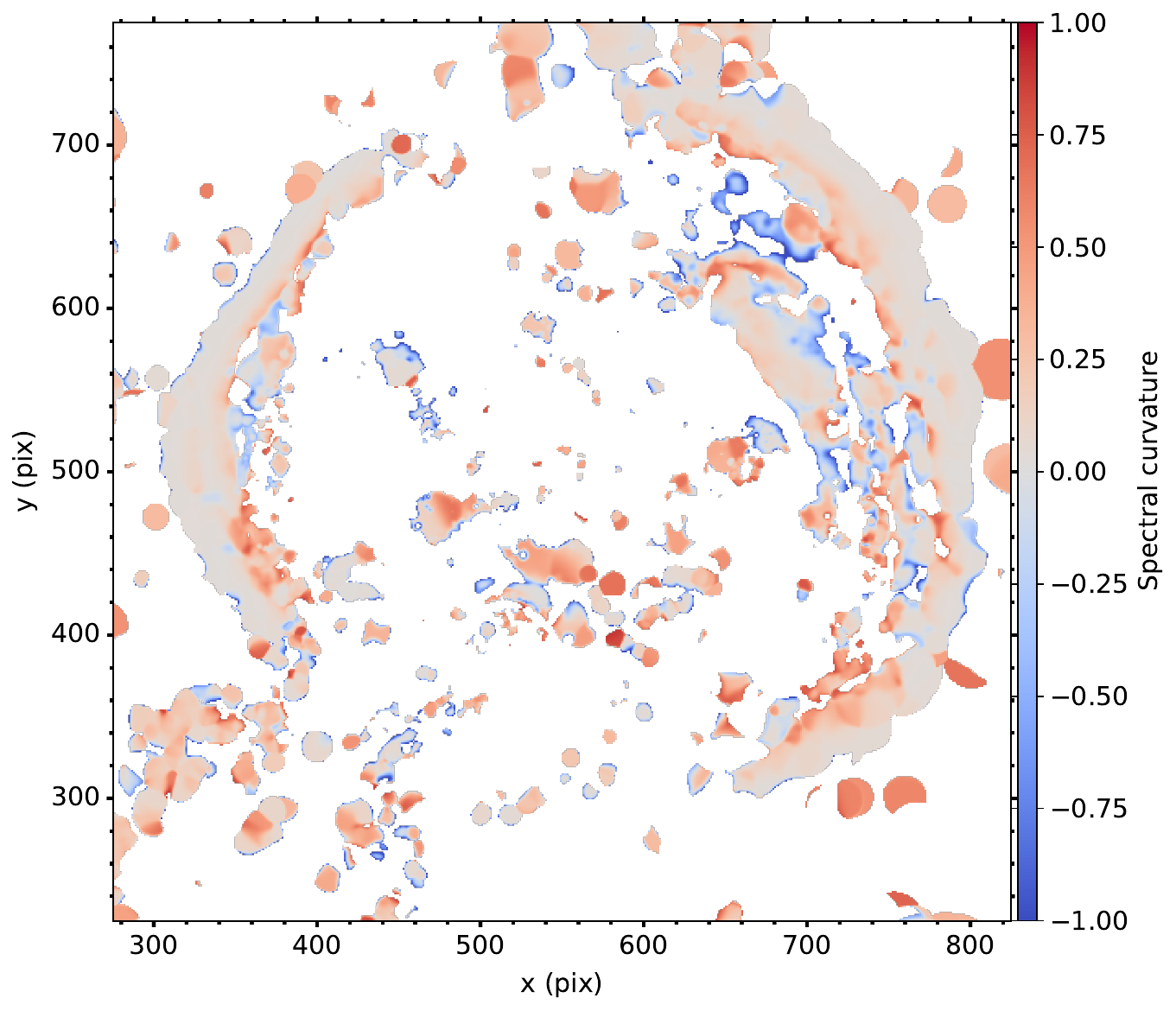}
          \smallskip
    \begin{minipage}[c]{0.4\textwidth}
         \centering
         \includegraphics[width=\textwidth]{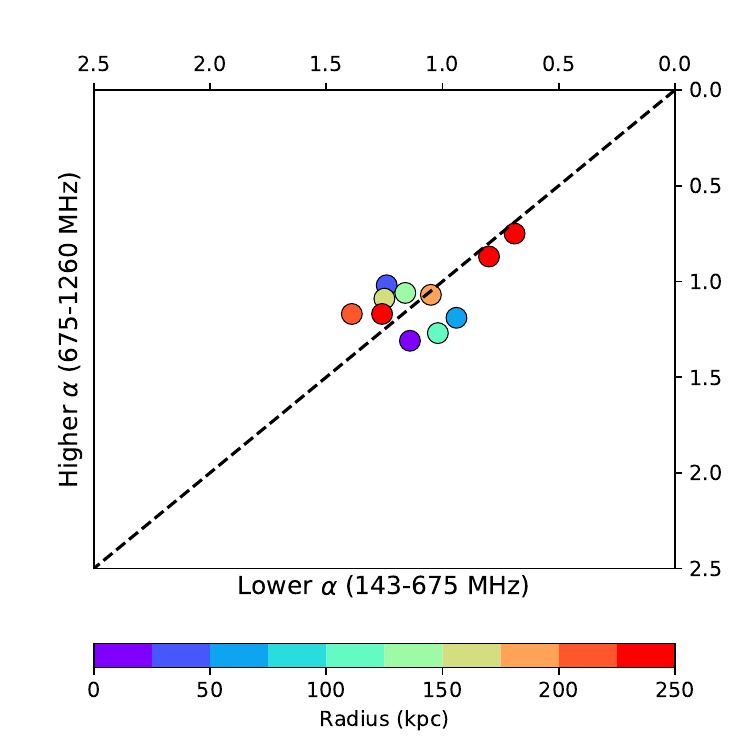}
         \end{minipage}
         \begin{minipage}[c]{0.58\textwidth}
        \caption{Spectral properties of a simulated ORC4-like source as obtained from DSA of thermal electrons in a galaxy merger scenario (see setup in \citealt{Ivleva26}). \textit{Top}: spectral index maps at 143-675 MHz (left) and 675-1260 MHz (centre), and corresponding spectral curvature map (right). In each panel, $160 \; {\rm pix}=200 \; {\rm kpc}$ and the physical resolution of the simulations is $\sim 60$ kpc. \textit{Bottom}: RCCD. Each data point is the median value as measured from the spectral index maps sampled with circular concentric annuli.     }
        \label{fig: simulation}
        \end{minipage}
\end{figure*}

The simulated radio images were obtained at 143, 675, and 1260 MHz with a physical resolution of $\sim 60$ kpc to closely match the observations of ORC4 used in our analysis. In Fig. \ref{fig: simulation}, we report the obtained spectral index maps at 143-675 MHz and 675-1260 MHz, and the resulting spectral curvature map. Pixels with spectral indices $\geq 2$ were masked to reduce the impact of a few pixels dominating the spectral analysis. Afterwards, we extracted the spectral index profiles in concentric annuli, calculating the median across each sampling region. We used these profiles to obtain a RCCD, which is also displayed in Fig. \ref{fig: simulation}. 

The curvature is approximately null or weakly positive. The RCCD shows that the data points are clustered around the injection spectrum ($\alpha_{\rm inj}\sim 0.7$) without any specific spatial trend. This is expected for fresh electrons that are accelerated at the shock front and radiate away energy downstream. These simulations are in contrast with the observed spectral behaviour of ORC4, strengthening the conclusion that DSA from the thermal pool is not a sufficiently adequate scenario for our target.


\FloatBarrier

\end{appendix}

\end{document}